\documentclass[review,12pt]{elsarticle}
\usepackage{xcolor}
\usepackage{graphicx}% Include figure files
\usepackage{dcolumn}% Align table columns on decimal point
\usepackage{bm}% bold math
\usepackage{soul,color}% bold math
\usepackage{amssymb}
\usepackage{amsmath}
\usepackage{lineno,hyperref}
\usepackage{tabularx} % Add this in preamble
\usepackage{booktabs} % For better table lines (optional but recommended)

\begin{document}

\begin{frontmatter}
    
\title{Fundamental concepts of thermal phonon coherence}
\author{Theodore Maranets$^{1}$}
\ead{tedmaranets@gmail.com}
\author{Haoran Cui$^1$}
\author{Milad Nasiri$^1$}
\author{Evan Doe$^1$}
\author{Yan Wang$^1$}
\ead{yanwang@unr.edu}
\address{1. Department of Mechanical Engineering, University of Nevada, Reno, Reno, NV, 89557, USA.}

\begin{abstract}

As scientists seek to transcend traditional material property relationships, novel structures with artificial periodicity, configurational complexity, and disorder are increasingly being explored. For phonons, these architectures can induce significant phase correlations, which strongly shifts lattice heat conduction away from the conventional framework of particle-like scattering. This thermal phonon coherence can develop in space and time separately, resulting in distinct wave-like phenomena. Here we rigorously outline the theories of phonon spatial and temporal coherence. Crucially, this review focuses on demystifying and building conceptual understanding of the fundamental principles linking the various thermal conductivity and phonon property results reported in the literature. Furthermore, we use this understanding to establish a comprehensive physical picture of the wave nature of thermal phonons, unifying the spatial and temporal coherence theories. Altogether, the theoretical dissections in this review provide an extensive knowledge base for understanding and engineering thermal phonons in complex materials. 

\end{abstract}

\begin{graphicalabstract}
\centering
\includegraphics[width=\textwidth]{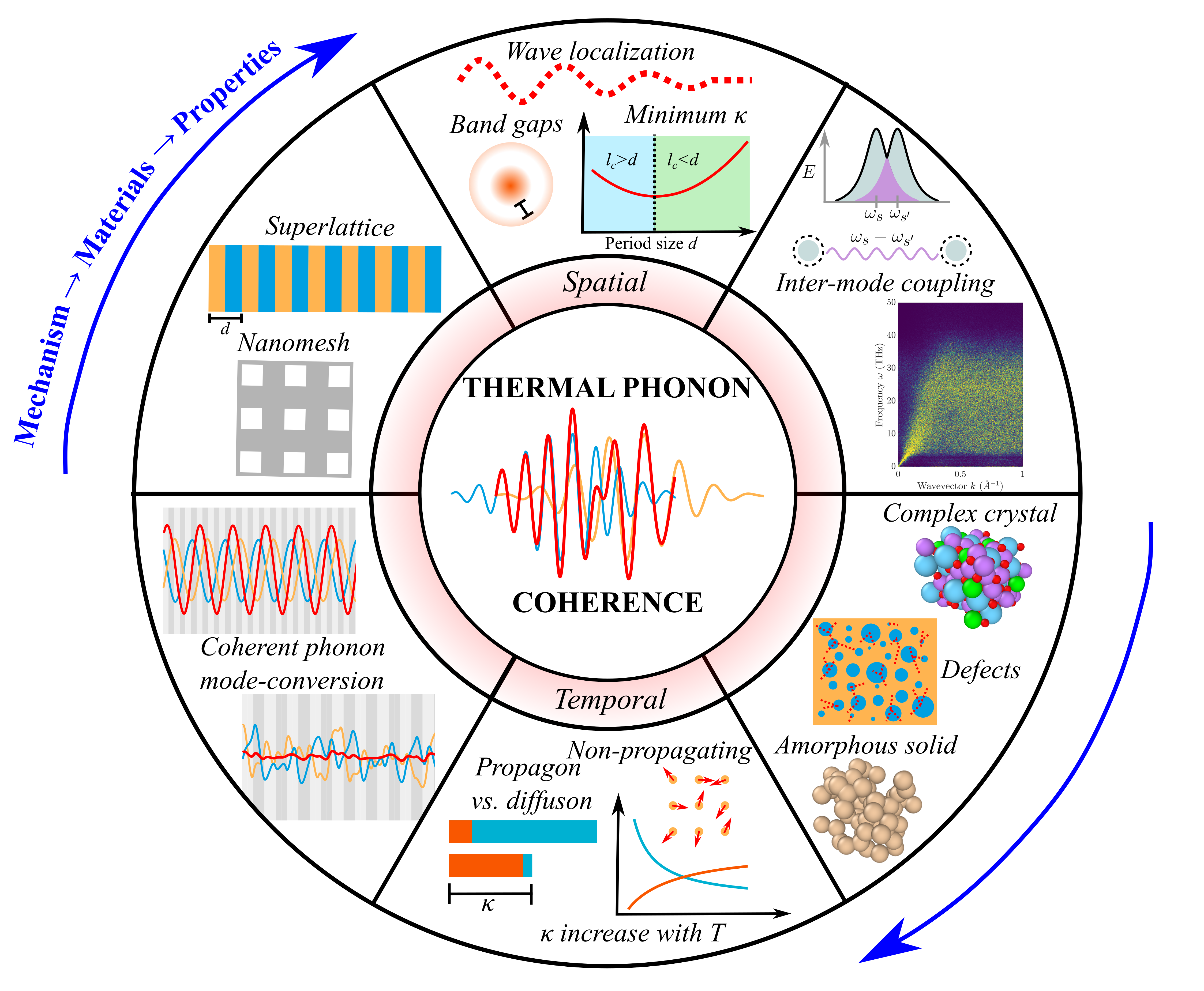}
\end{graphicalabstract}

\begin{highlights}

    \item Phonons possess finite spatial and temporal extensions of their phases.
    \item Spatial extension overlaps with characteristic length enable propagative transport.
    \item Temporal extension overlaps with spectral peaks enable non-propagative transport.
    \item These wave-like phonon behaviors affect how thermal conductivity can be manipulated.
    \item Spatial and temporal coherence are distinct but not mutually exclusive phenomena. 
    
\end{highlights}


\end{frontmatter}

\tableofcontents

\section{Introduction\label{sec:introduction}}

\subsection{Defining thermal phonon coherence\label{sec:coherence_definition}}

Thermal phonon coherence is a phenomenon where thermal phonons---collective lattice vibrations that dominate heat conduction---explicitly exhibit wave-like behaviors. The key consequence of this effect is that the magnitude and dimension of interference---whether with other waves or structural features including interfaces and crystal disorder---determines the amount of heat carried by the phonons. 

Phonon coherence can manifest via spatial or temporal interactions depending on the material structure. Propagative phonon wave transport can evolve in space through artificially periodic metamaterials. Non-propagative phonon heat transfer controlled by temporal phase correlations emerges in structurally complex crystals and disordered solids. 

\subsection{Review synopsis\label{sec:intro_synopsis}}

Manipulating heat conduction in solids is pivotal to increasing thermodynamic efficiencies and improving thermophysical properties in many technologies \cite{henry2020five,qiu2025roadmap}. Exploring exotic behaviors of thermal phonons has long been a key research focus for such a purpose \cite{shi2015evaluating,kim2021strategies,qian2021phonon,ouyang2024advancing}. Over the past three decades, thermal phonon coherence has been among the most studied exotic phonon phenomena, with many insights elucidated from both experiment and computation \cite{chen2021non}. While a plethora of reviews have surveyed the reported results and their isolated conclusions, a comprehensive treatise of the subject has been missing. Moreover, the distinct natures of phonon spatial vs. temporal coherence have been largely overlooked and are frequently conflated. 

In this review, we address these knowledge gaps by conducting an extensive analysis of the theories of thermal phonon coherence. Importantly, we focus on understanding the main physical concepts from which all the observations reported in the literature are fundamentally derived from. The various trends and behaviors associated with thermal phonon coherence are discussed in depth within the contexts of known theoretical frameworks rather than simply being surveyed. When equations are presented, we explore with rigor how their mathematical meanings relate to phonon heat conduction. We extensively provide novel schematic illustrations of key concepts not yet visualized in existing literature. Furthermore, we offer unique analysis of the broader wave nature of thermal phonons and the notion of phonon physical character through dissection of the similarities between spatial and temporal coherence. New perspectives on methodologies, applications, limitations, and potential opportunities in the field are made throughout the review. 

The review is organized as follows: In Sec.~\ref{sec:phonon_transport_overview}, we establish the two main physical characterizations of thermal phonons from which spatial and temporal coherence can be readily understood. Following this, Sec.~\ref{sec:spatial_coherence_theory} and Sec.~\ref{sec:temporal_coherence_theory} expound the theoretical frameworks, observations, and conceptual nuances of phonon spatial and temporal coherence, respectively. In Sec.~\ref{sec:phonon_differences}, we succinctly differentiate thermal phonon coherence from other non-Fourier effects that share some similarities in terminology and observables. With their respective theories having been outlined in the prior sections, the conceptual links between spatial and temporal coherence are scrutinized in Sec.~\ref{sec:spatial_temporal_connection}. Finally, we summarize our discussions in Sec.~\ref{sec:conclusions} and point out future directions in both fundamental and applied materials physics research of thermal phonon coherence.

\section{Phonon transport overview\label{sec:phonon_transport_overview}}

Critical to distinguishing between phonon spatial and temporal coherence phenomena is an understanding of phonon transport regimes and phonon physical character. In this section, we outline a theoretical foundation of phonon transport in regards to how lattice vibrational eigenstates (phonons) are quantized and physically described before initiating our individual discussions of both phonon coherence effects in the following sections.

\subsection{Phonons as propagating wave-packets\label{sec:propagating_wave-packets}}

\begin{figure}
    \centering
    \includegraphics[width=\textwidth]{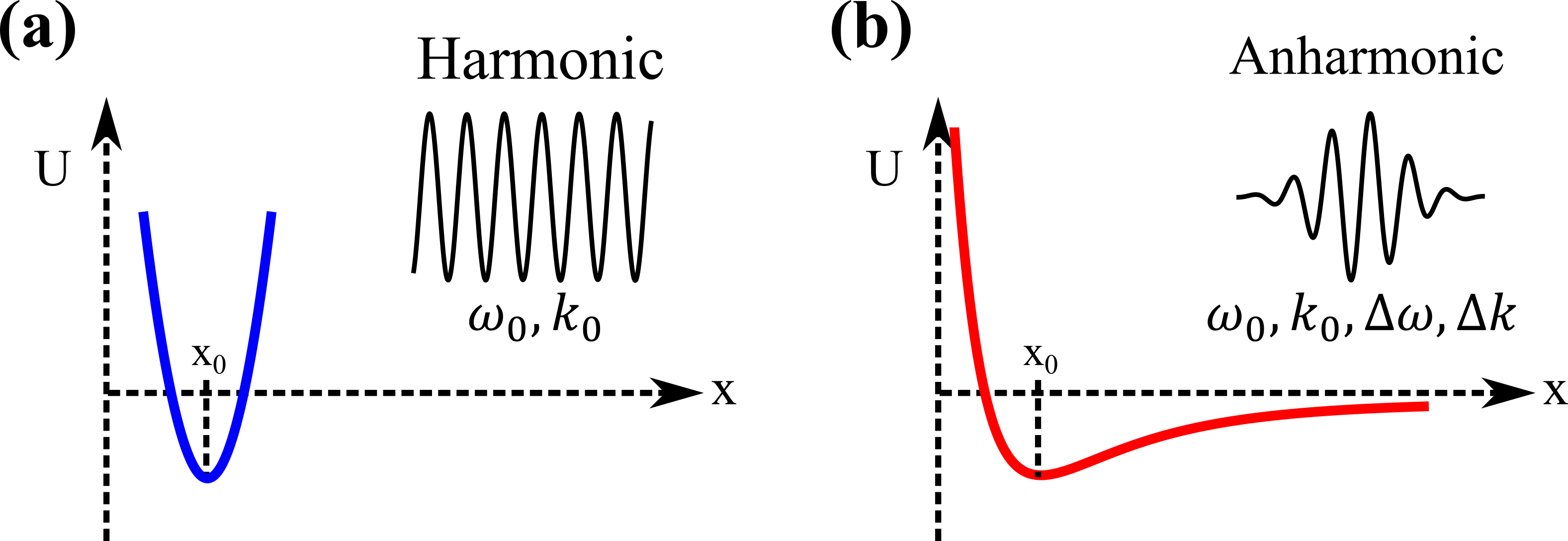}
    \caption{Schematic illustration of the phonon wave representations. Panel (a) plots the harmonic potential in which phonons are quantized as delocalized plane waves and have an exact frequency $\omega_0$ and wavevector $k_0$. Panel (b) plots the Lennard-Jones function, representative of an anharmonic potential, which induces the quantization of phonons as wave-packets possessing frequency $\Delta\omega$ and wavevector $\Delta k$ linewidths centered about $\omega_0$ and $k_0$, respectively. }
    \label{fig:harmonic_anharmonic_wavepacket_potential}
\end{figure}

Broadly, phonons are quanta of the normal modes of lattice vibrations \cite{debye1912theorie}. In the harmonic approximation, for a crystalline structure, the quantization takes the form of delocalized plane wave states described by an exact dispersion relation \cite{georgi1993physics,dove2011introduction}. The dispersion relation defines a precise correspondence between wavevector and frequency in addition to heat capacity, group velocity, and density of states. 

Anharmonicity, corresponding to cubic and higher-order interatomic force constants, introduces broadening of frequency and wavevector characterized by linewidths \cite{ziman2001electrons,srivastava2022physics,latour2017distinguishing}. When the linewidths are smaller than the spacings between modes in the dispersion, there is no correlation of different eigenstates and phonons are quantized as propagating wave-packets, localized plane wave states with well-defined frequency, wavevector, and spatial and temporal extensions that are directly the inverses of the linewidths \cite{latour2017distinguishing}. 

The spatial extension is known as the spatial coherence length and it is definitively greater than the central wavelength of the wave-packet and typically upper-bounded by the mean free path. The temporal extension is exactly the phonon relaxation time or lifetime. This characterization, commonly known as the phonon-gas model, is represented in Peierls' formulation of the phonon BTE \cite{peierls1929kinetischen}. We visualize the contrast between harmonic and anharmonic potential and their respective wave quantizations of phonons in Fig.~\ref{fig:harmonic_anharmonic_wavepacket_potential}.

\begin{figure}
    \centering
    \includegraphics[width=\textwidth]{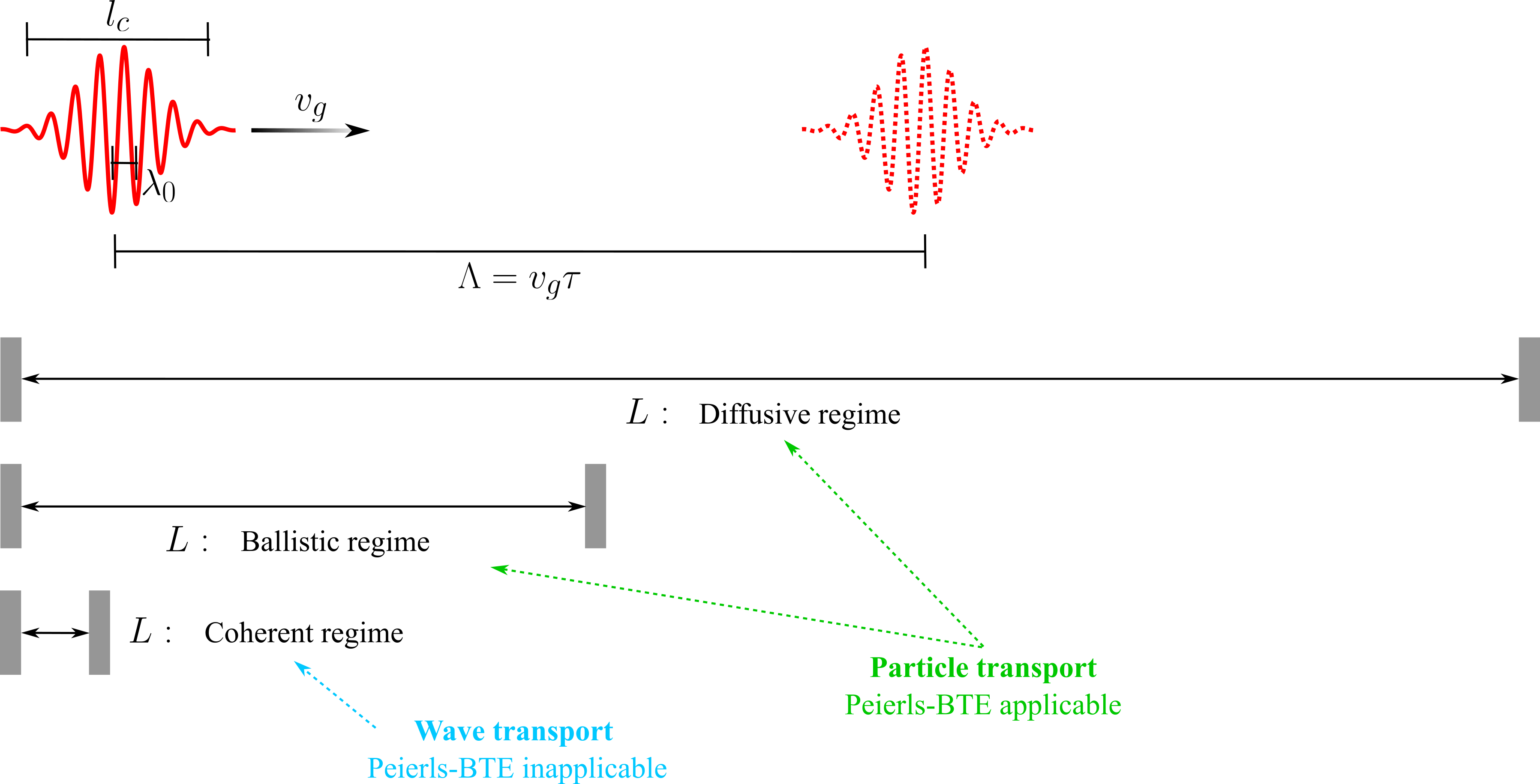}
    \caption{Schematic illustration of the propagation of phonon wave-packets and their three main size regimes in heat conduction. A phonon wave-packet with central wavelength $\lambda_{0}$ and spatial coherence length $l_{c}$ travels a distance $\Lambda$, the mean free path, which is the product of the phonon's group velocity $v_{g}$ and lifetime $\tau$. The comparison of the spatial length scales to the characteristic length $L$ of the material determines the size regime. The Peierls-BTE can accurately describe both diffusive and ballistic regimes, scenarios of particle-like transport, but cannot model the coherent regime ($l_{c}>L$) where wave behaviors predominate.}
    \label{fig:length_regimes}
\end{figure}

\subsubsection{Spatial length scales\label{sec:spatial_length_scales}}

When describing phonons as propagating wave-packets, one must consider the comparison of the characteristic length of the material to the spatial length scales of the phonon wave-packet to assess size effects in thermal transport. The characteristic length of a material is the average distance between heterogeneous features that scatter wave-packets. In a homogeneous medium, the characteristic length would just be the total size of the system. The phonon wave-packet spatial length scales are: 
\begin{enumerate}
    \item Central wavelength $\lambda_{0}$ corresponding to the wavevector value the phonon energy is centered at in reciprocal-space. This value defines the main frequency i.e. energy of the phonon wave-packet per the dispersion relation.
    
    \item Spatial coherence length $l_{c}$ as the inverse of the wavevector linewidth. This value defines the spatial extension or width of the wave-packet.
    
    \item Mean free path $\Lambda$ being the product of the group velocity and phonon lifetime i.e. inverse of the frequency linewidth. This value defines the average distance the wave-packet travels between scattering events.
\end{enumerate} 
Fig.~\ref{fig:length_regimes} visualizes the physical representation of each length scale and illustrates the three different spatial regimes for phonon transport as propagating wave-packets. For the wave-packet to be well-defined, the coherence length must be greater than the central wavelength in order to accommodate a full wave period. For the wave-packet to have a distinct traveling or propagating character, the coherence length must be less than the mean free path. 

There has been considerable conflation of spatial coherence length and mean free path in the literature, most likely owed to the relative equality between the broadening of frequency and wavevector \cite{peierls1955quantum}. However, since the lifetime is multiplied by the group velocity to attain the value of mean free path, mean free path must be greater than the coherence length for a propagating phonon wave-packet. 

One can also intuit this from a physical perspective. Physically, mean free path is the distance the wave-packet travels before dissipating in a scattering event. If the coherence length were equal to or greater than the mean free path, the wave-packet would not travel and transport energy in a propagating manner. Indeed, this can occur for flat bands and modes at the edges of the Brillouin zone when the group velocity approaches zero \cite{latour2017distinguishing}. These modes are typically referred to as standing modes. Distinguishing between spatial coherence length and mean free path is also necessary to classify the ballistic and coherent phonon transport size regimes separately, as will be discussed in the following paragraphs.

\begin{figure}
    \centering
    \includegraphics[width=\textwidth]{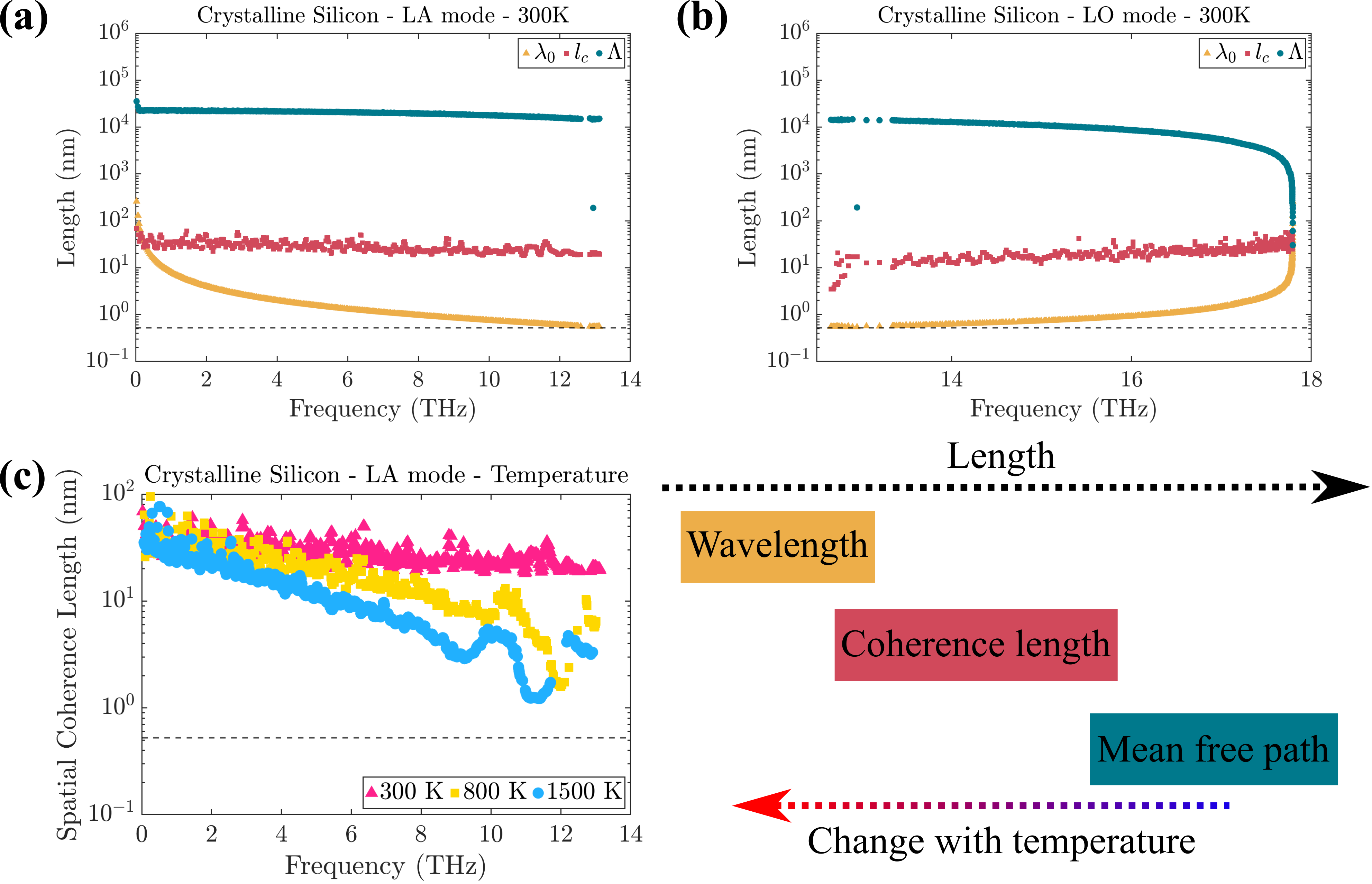}
    \caption{Calculations of the three spatial length scales of phonon wave-packets: central wavelength $\lambda_{0}$, spatial coherence length $l_{c}$, and mean free path $\Lambda$ for several phonon modes. The values for the longitudinal-acoustic (LA) and longitudinal-optical (LO) modes of crystalline silicon modeled with the SW potential \cite{stillinger1985computer} are presented in panels (a) and (b). The $l_{c}$ values for the LA mode of crystalline silicon at 300 K, 800 K, and 1500 K are presented in panel (c). The schematic illustration in panel (d) visualizes the general inequality among the lengths scales and the reduction with increasing temperature.}
    \label{fig:length_calculations}
\end{figure}

For quantitative evidence of these concepts, we present calculations of the spatial length scales for the longitudinal-acoustic and longitudinal-optical modes of crystalline silicon in Figs.~\ref{fig:length_calculations}a-\ref{fig:length_calculations}b. We can clearly see for most frequencies across the spectra: wavelength $\lambda_0$ $<$ spatial coherence length $l_c$ $<$ mean free path $\Lambda$. At the center and edges of the Brillouin zone, the propagating wave-packet characterization can become ill-defined. 

For the plots in Fig.~\ref{fig:length_calculations}, the values were computed from the linewidths of the spectral energy density calculated from equilibrium molecular dynamics simulations conducted in the LAMMPS software \cite{thompson2022lammps}. Spatial coherence length can also be computed from the spatial correlations of the atomic vibrations as demonstrated by Latour et al. \cite{latour2014microscopic} adapting existing mathematical formalisms of coherence applied extensively in optics. Particularly, spatial coherence length, being the spatial extension of a phonon wave-packet, physically quantifies the distance over which the vibrations of a set of atoms are correlated. This property can be captured via a mutual coherence function in equilibrium molecular dynamics simulations that is constructed from the time correlations of the velocity fields between all pairs of atoms along a specified spatial direction. Latour and Chalopin \cite{latour2017distinguishing} proved that calculation of the coherence length in this manner is related to the spectral energy density and that the two approaches give the same values for most frequencies across the dispersion. The deviations occur at the center and edges of the Brillouin zone where the propagating wave-packet description of phonons falls apart due to the group velocity approaching zero and/or the central wavelength exceeding the coherence length.

\subsubsection{Size regimes of propagative transport\label{sec:size_regimes}}

Regarding comparison of the phonon spatial length scales to the characteristic length of the material, there are three different size regimes of transport for propagating phonon wave-packets \cite{chen2002heat,chen2005nanoscale,li2015phonon}, ultimately manifesting in different heat conduction behaviors, as illustrated in Fig.~\ref{fig:length_regimes}. 
\begin{enumerate}
    \item The diffusive regime of phonon transport, modeled by Fourier's law, is defined by the mean free path being less than the characteristic length of the material. In this regime, the phases of the waves are negligible and heat flows diffusely through a stochastic process of multiple scatterings as if the phonons were randomly-moving gas particles. This behavior is referred to classical diffusion and motivates the term ``phonon-gas'' model.
    
    \item The ballistic regime is defined by the mean free path being greater than the characteristic length of the material, ultimately manifesting in size-dependent thermal phonon transport where the wave-packets travel in straight paths and primarily dissipate their momentum and energies through geometric interactions with interfaces and boundaries rather than intrinsic phonon-phonon scattering as is the case for the diffusive regime. The phonon transport is still particle-like due to the inability of the material to facilitate wave-like transport via spatial coherence.
    
    \item The coherent regime is characterized by the spatial coherence length being equal to or greater than the characteristic length of the material. In these scenarios, if the features of the device induce significant specular interactions, the ballistically propagating wave-packets belonging to the constituent materials can constructively interfere with each other to form coherent modes defined by the periodicity and character of the nanostructure in a phenomenon known as phonon spatial coherence. These new coherent phonons also exist as propagating wave-packets and are subject to the ballistic and diffusive transport regimes depending on their new mean free paths, but most importantly, their heat carried is influenced by changes in their wave dynamics, primarily through interference. Thus, their transport is said to be wave-like.
\end{enumerate}  

The diffusive and ballistic regimes, scenarios of particle-like transport, are adequately captured by the Peierls-BTE and Monte-Carlo simulations which effectively approximates solutions to the Peierls-BTE \cite{peraud2011efficient,lindsay2016first,anufriev2020ray}. The Peierls-BTE does not account for the formation of wave-like coherent phonons and thus fails to model the unique trends of thermophysical properties stemming from phonon spatial coherence \cite{xie2018phonon}. 

In Sec.~\ref{sec:spatial_coherence_theory}, we discuss the key concepts in the understanding of phonon spatial coherence, specifically focusing on how it can be stimulated and manipulated in materials. In Sec.~\ref{sec:phonon_differences}, we clarify how spatial coherence differs several other regimes of phonon transport that also don't align with the conventional Peierls-BTE.

\subsection{Phonons as non-propagating oscillators\label{sec:non-propagating_oscillators}}

The description of phonons as propagating wave-packets is not always applicable. When the linewidths exceed the inter-mode spacings in the dispersion relation or when crystal disorder is present, characterizing phonons as propagating wave-packets becomes ill-defined \cite{klemens1951thermal,lv2016examining,seyf2017rethinking,agne2018minimum,hanus2021thermal}. 

For the first case, overlapping linewidths leads to the vibrational eigenstates being quasi-degenerate and clearly delineating spatial and temporal extensions, which parametrize the wave-packet, for different phonon modes becomes challenging. This most often occurs in structurally complex crystals possessing many atoms in the primitive unit cell where an enlarged unit cell consequences in a narrowing of the Brillouin zone and the folding of the band structure. Closer packed bands heightens the possibility for the linewidths of different modes to coincide with one another. This can also occur in situations of high anharmonicity such as high temperatures and weak bonding where linewidths are large. 

Regarding the presence of crystal disorder, a lack of periodicity in the material makes decomposing the normal modes of atomic vibrations into plane wave solutions impossible, thus the propagating wave-packet picture has a weak mathematical foundation. Crystal disorder can be induced by disarrangement of the atom positions in a melt quenching process or by introducing defects via crystal imperfections and/or alloying.

\begin{figure}
    \centering
    \includegraphics[height=0.6\textheight]{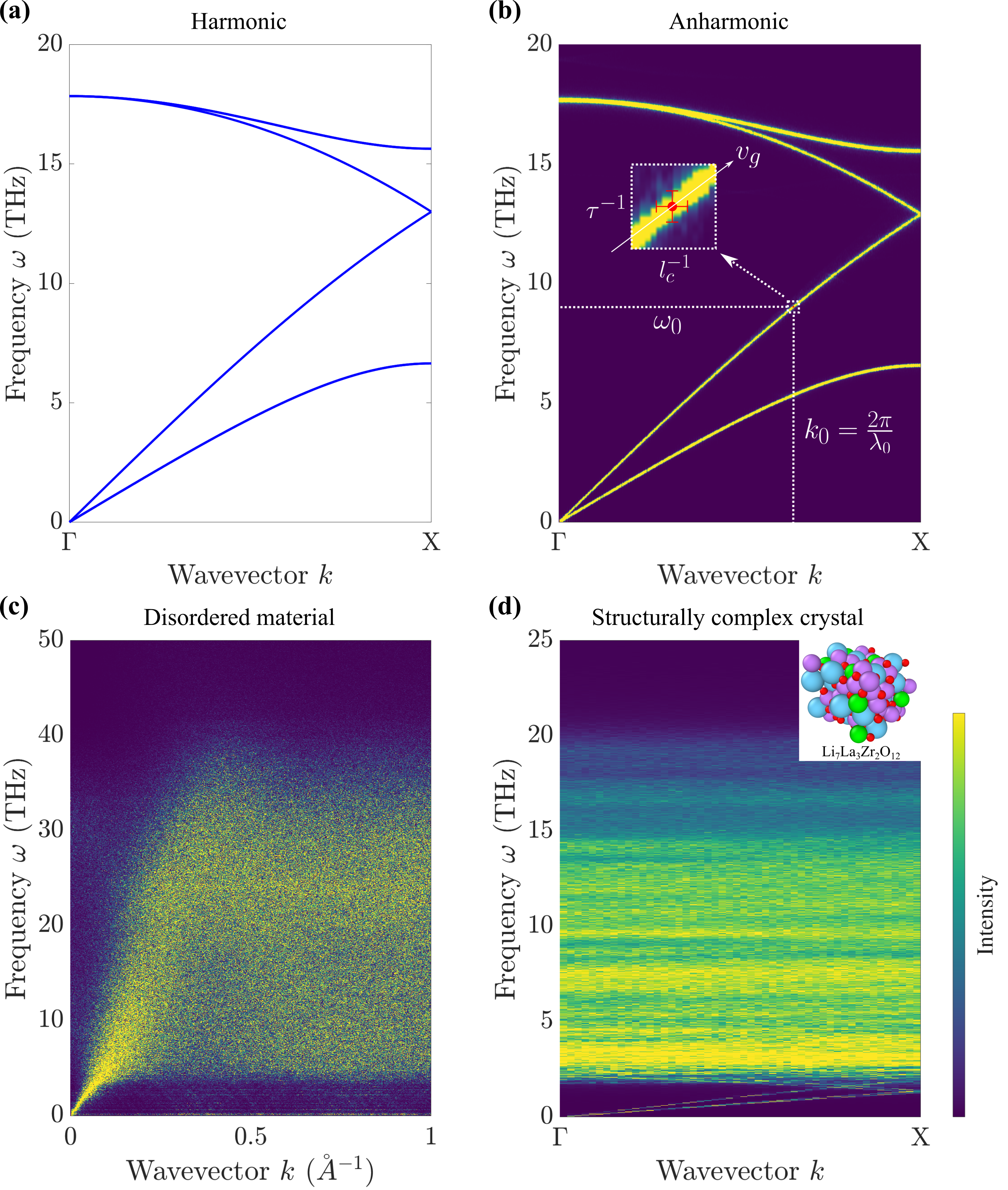}
    \caption{Calculations of the phonon energy distribution in reciprocal-space for several materials. Panel (a) shows the phonon dispersion relation computed from harmonic lattice dynamics of crystalline silicon modeled with the SW potential \cite{stillinger1985computer} as an a example of a structurally simple crystal. Panel (b) shows the spectral energy density computed from equilibrium molecular dynamics, which incorporates anharmonicity, of SW \cite{stillinger1985computer} silicon at 300 K. The frequency $\omega_{0}$ and wavevector $k_{0}$ and their associated group velocity $v_{g}$, spatial coherence length $l_{c}$, and lifetime $\tau$ are marked in the plot. Panel (c) shows the dynamic structure factor computed from equilibrium molecular dynamics of amorphous silicon carbide, modeled with the Tersoff potential \cite{tersoff1990erratum}, as an example of a disordered material. Panel (d) shows the spectral energy density of Li$_7$La$_3$Zr$_2$O$_{12}$ (LLZO), modeled with a machine learning interatomic potential \cite{panneerselvam2024disorder}, as an example of a structurally complex crystal. The atomic structure of LLZO displayed in the inset of panel (d) is visualized with the OVITO software \cite{ovito}.}
    \label{fig:dispersions_all}
\end{figure}

We illustrate these concepts in Figs.~\ref{fig:dispersions_all}c and \ref{fig:dispersions_all}d which show the phonon energy distributions in reciprocal-space, computed from equilibrium molecular dynamics simulations, for amorphous silicon carbide and the solid-state electrolyte Li$_7$La$_3$Zr$_2$O$_{12}$ (LLZO) as an example of a structurally complex crystal. Compared to the plots for crystalline silicon (Fig.~\ref{fig:dispersions_all}a and Fig.~\ref{fig:dispersions_all}b), a structurally simple crystal containing only two atoms in the primitive unit cell, energetically separate dispersion curves from which the three spatial length scales of the propagating phonon wave-packet description are based off of, are absent for the disordered material and structurally complex crystal (besides the low-frequency acoustic branches). Consequently, the Peierls-BTE has failed to accurately predict the thermal conductivities of these types of materials \cite{lindsay2019perspective}.

\subsubsection{Diffusons \& temporal coherence\label{sec:diffuson_modes}}

The desire to quantify thermal transport in these cases has motivated development of a phonon taxonomy where different modes are distinguished by their physical shape and transport mechanism \cite{allen1989thermal,allen1993thermal,allen1999diffusons,lv2016examining,seyf2016method,seyf2017rethinking}. 

When the propagating wave-packet description is invalid, the phonons are primarily theorized to exist as non-propagating oscillators that transport heat by exchanging energy with each other in a hopping process analogous to a random walk. The atoms still move collectively with a temporal phase and frequency, however, the spatial periodicity of the motion is random and thus a distinct phonon wavelength (and consequently spatial coherence length and mean free path) cannot be clearly defined. Per the spatial randomness, the heat conduction is sometimes said to resemble diffusion, though it is physically different from the geometric collisions, creation, and annihilation of particle-like propagating wave-packets; the microscopic mechanisms of scattering in the phonon-gas model. 

\begin{figure}
    \centering
    \includegraphics[width=\textwidth]{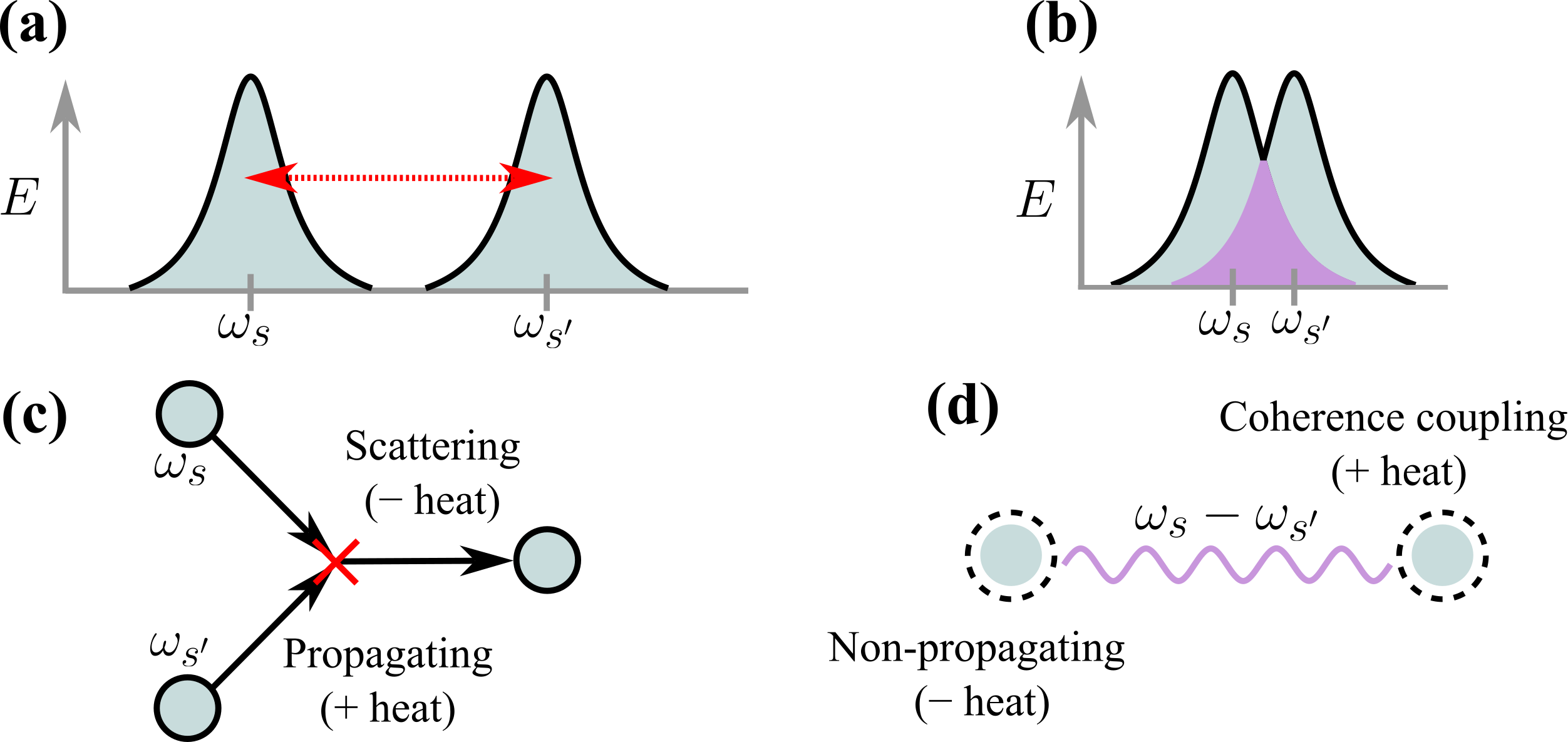}
    \caption{Schematic illustrations of the energy signatures, panels (a) and (b), and the heat conduction pathways, panels (c) and (d) for propagons and diffusons, respectively.}
    \label{fig:propagon_v_diffuson_energy}
\end{figure}

Quantitatively, the non-propagating oscillators, commonly termed ``diffusons'', are represented as the off-diagonal elements ($s\neq s'$) of a lattice thermal conductivity tensor sum over all modes (see Eqn.~\ref{eqn:thermal_conductivity_tensor_sum}) whereas the propagating wave-packets, termed ``propagons'' correspond to the diagonal elements ($s=s'$) \cite{hardy1963energy,allen1999diffusons,sun2010lattice,simoncelli2019unified,isaeva2019modeling,luo2020vibrational,hanus2021uncovering}. This forms a ``two-channel lattice dynamics'' framework.
\begin{equation}
    \kappa = \sum\limits_{\boldsymbol{k},s,s'}C_{ss'}(\boldsymbol{k})v_{ss'}(\boldsymbol{k})v_{s's}(\boldsymbol{k})\tau_{ss'}(\boldsymbol{k})
    \label{eqn:thermal_conductivity_tensor_sum}
\end{equation}
$C_{ss'}(\boldsymbol{k})$, $v_{ss'}(\boldsymbol{k})$, and $\tau_{ss'}(\boldsymbol{k})$ are the heat capacity, velocity, and lifetime matrices for all phonon modes $s$ with wavevector $\boldsymbol{k}$. As the Peierls-BTE only evaluates the propagon contribution, it is usually inadequate in modeling the thermal conductivities of materials with a significant diffuson contribution which typically occurs when a high fraction of the phonon modes are diffusons. 

We refer the reader to an excellent review by DeAngelis et al. \cite{deangelis2019thermal} that comprehensively covers the various methodologies that have been developed to distinguish propagons from diffusons and delineate their relative contributions to the overall thermal conductivity. Additionally, we recommend an article by Hanus et al. \cite{hanus2021uncovering} that offers a compelling physical analogy between lattice heat conduction and shipment of cargo by semi-trucks where the different ways the trucks and the cargo can move along the highway and between each other represent the energy transport mechanisms by propagons and diffusons. We present our own visualizations of the energy signatures and mechanisms of thermal transport for propagons and diffusons in Fig.~\ref{fig:propagon_v_diffuson_energy}, illustrating how overlapping linewidths transforms the character of phonons away from particle-like propagation and scattering.

Generally, the amount of heat carried by diffusons is proportional to the degree of correlation or ``coupling'' between the quasi-degenerate phonon modes \cite{klemens1951thermal,allen1989thermal,allen1993thermal}. In the last decade, theories have emerged suggesting that this relationship is due to the fact that when the diffusons are strongly correlated, they can possess wave-like behaviors in time and enable efficient energy transport \cite{simoncelli2019unified,isaeva2019modeling,zhang2022heat}. This phenomenon is referred to as temporal coherence and it is distinct from the spatial coherence effect of propagons as the wave interference of phonons occurs in time rather than space \cite{latour2017distinguishing}. Consequently, unlike spatial coherence, it does not appear that temporal coherence effectuates a physically wave-like heat conduction in space. In Sec.~\ref{sec:temporal_coherence_theory}, we review the main principles of the phonon temporal coherence, with particular emphasis on how the phenomenon can be quantified and its nuanced differences from various aspects of spatial coherence. 

\section{Spatial coherence theory\label{sec:spatial_coherence_theory}}

Phonon spatial coherence is a profoundly unique size effect that introduces a new type of material architecture, a new framework to model phonon heat conduction in that architecture, and novel pathways to modify thermal conductivity. In this section, we outline the principal theory of spatial coherence and the evolution of coherent phonons in phononic crystals. We also discuss the theoretical subtleties involved and what distinguishes spatial coherence from other propagon transport phenomena. 

\subsection{Theoretical framework\label{sec:spatial_coherence_theoretical_framework}}

\begin{figure}
    \centering
    \includegraphics[width=\textwidth]{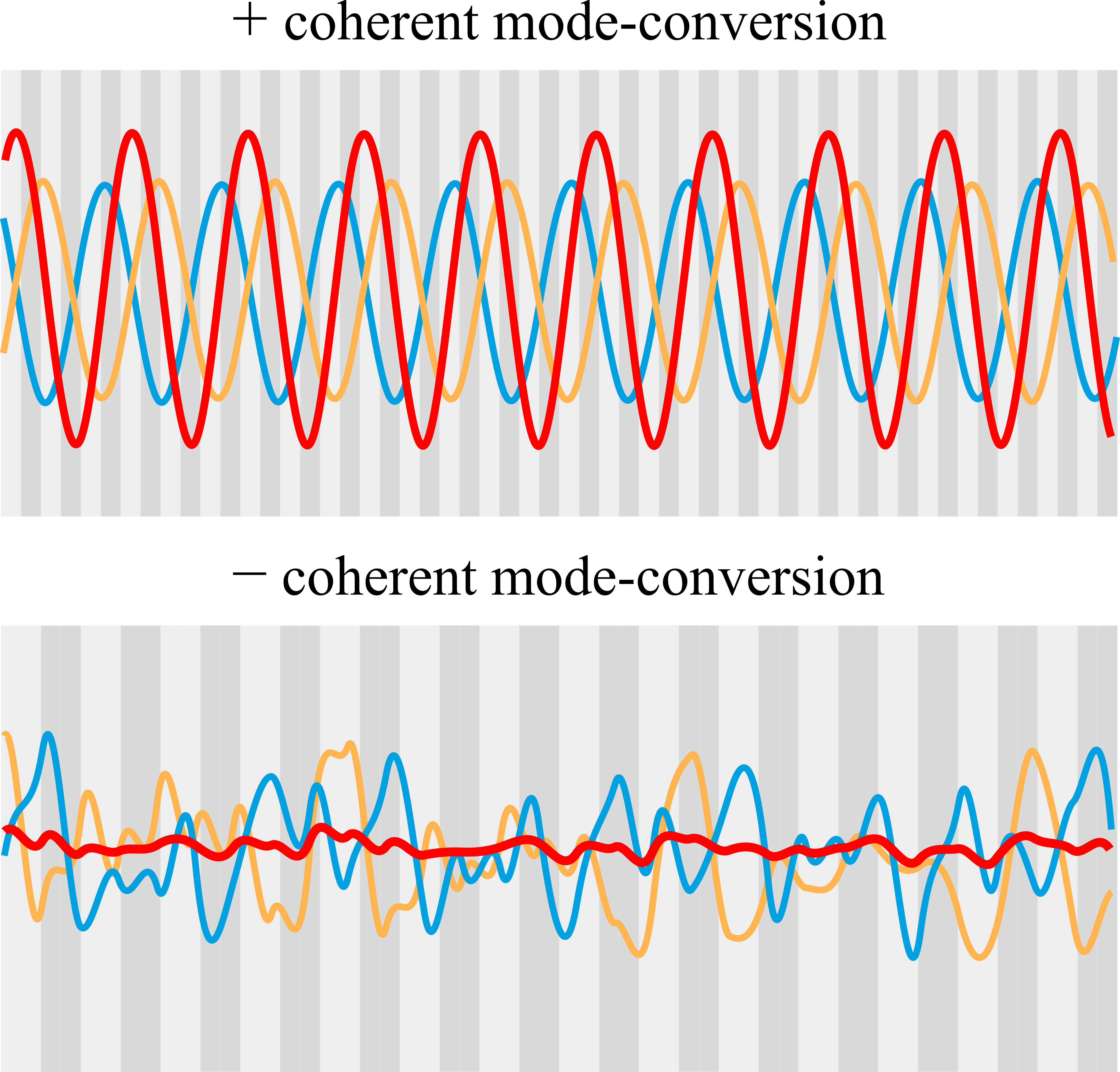}
    \caption{Schematic illustration of coherent mode-conversion. Constructive interference among scattered layer phonons (orange and blue lines) that preserve their phase manifests in formation of coherent modes which transport through the layers with high transmission in a wave-like fashion. Conversely, destroyed phase coherence among scattered phonons inhibits formation of high-transmission coherent modes.}
    \label{fig:coherent_mode-conversion_schematic}
\end{figure}

\begin{figure}
    \centering
    \includegraphics[width=0.9\textwidth]{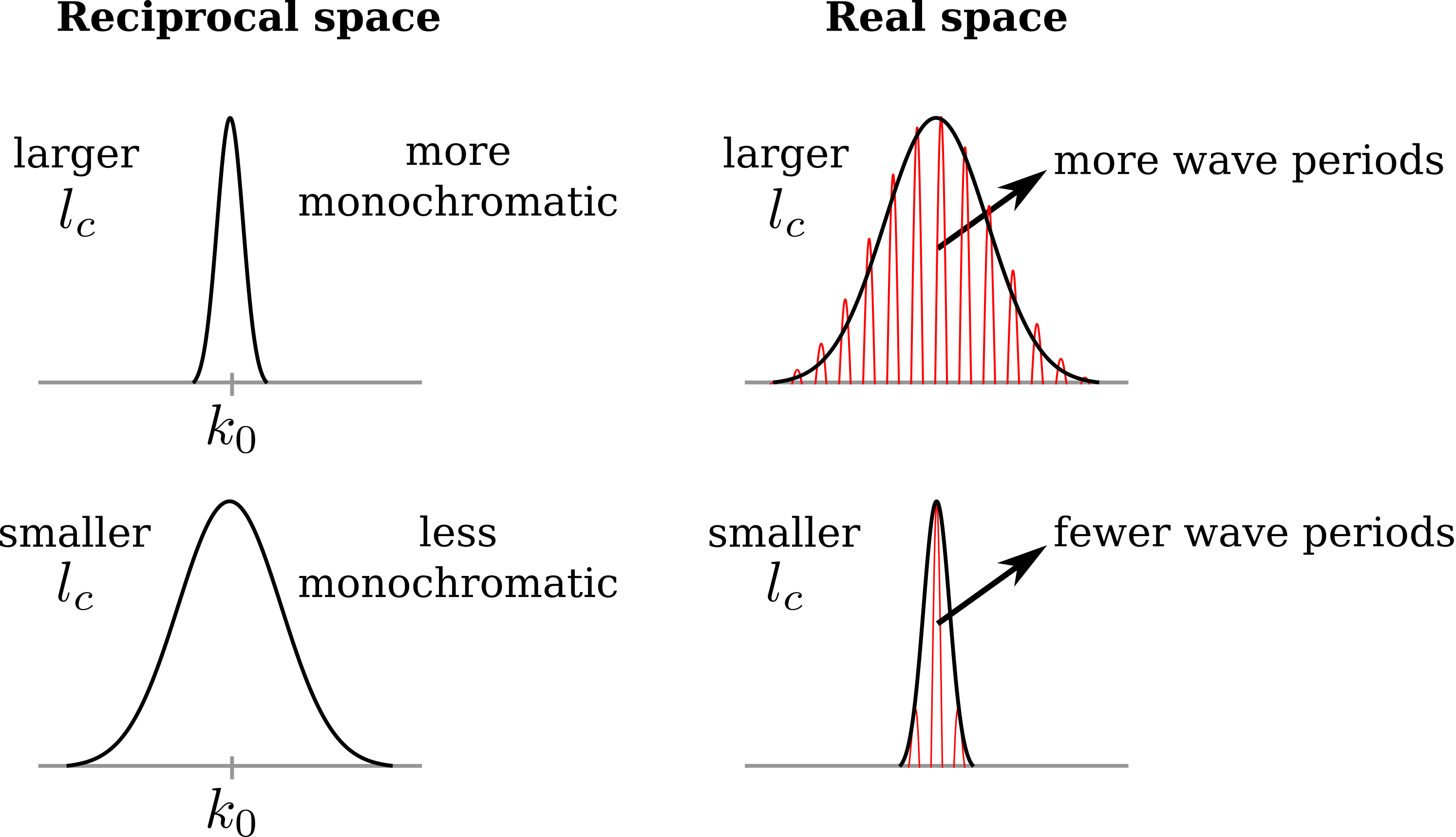}
    \caption{Schematic illustration of the physical meaning of spatial coherence length $l_c$. $l_c$ quantifies the number of wave periods a phonon wave-packet accommodates in real-space. Consequently, wave-packets with larger $l_c$ span a narrower spectrum of phonon wavelengths centered around $k_0$ in reciprocal-space, making the wave-packet more monochromatic and increasing the likelihood of interference among the phonon modes within the wave-packet. Smaller $l_c$ wave-packets span a broader width in reciprocal-space and thus are less susceptible to interference. Furthermore, a greater number of wave periods creates more opportunities for waves to interfere with each other, either constructively or destructively depending on the structure of the material. The figure is reprinted with permission from Ref.~\cite{maranets2024influence}. Copyright (2024) Elsevier.}
    \label{fig:spatial_coherence_length_visualization}
\end{figure}

The relation of the spatial coherence length $l_{c}$ to the characteristic length of the material determines whether a propagating wave-packet transports like a wave or a particle \cite{chen2000phonon,latour2014microscopic,latour2017distinguishing} as illustrated in Fig.~\ref{fig:length_regimes}. Wave transport, often called coherent transport, occurs when the spatial coherence length is equal to or greater than the characteristic length. Particle transport, commonly referred to as incoherent transport, encompassing the ballistic and diffusive regimes, occurs when the spatial coherence length is less than the characteristic length. 

The characteristic length is a measure of the spacing of features in a material that induce scattering. In a homogeneous medium, the characteristic length is just the total size of the system. In materials with heterogeneous features, such as interfaces, defects, or structural boundaries, the characteristic length is the average distance between the features. In the coherent regime, the constructive interference of scattered wave-packets leads to formation of new vibrational modes resonant with the arrangement of features \cite{maranets2024prominent,maranets2025phonon}. These modes, referred to as ``coherent phonons'', can propagate through the material freely without scattering in a wave-like fashion, leading to enhanced heat conduction. In contrast, incoherent transport is characterized by the particle-like behaviors of ``incoherent phonons'' that can't mode-convert. The physical transformation of incoherent phonons to coherent phonons is referred to as ``coherent mode-conversion.''\cite{maranets2025phonon}. We schematically illustrate the mode-conversion process for the reader in Fig.~\ref{fig:coherent_mode-conversion_schematic}. 

The fundamental reason why larger coherence lengths enable or accentuate coherent mode-conversion more than smaller values stems from the physical meaning of spatial coherence length as shown in Fig.~\ref{fig:spatial_coherence_length_visualization}. As a measure of the spatial extension of a phonon wave-packet, $l_c$ quantifies how well-defined the central wavelength $\lambda_0$ of the wave-packet is as well as the number of wave periods the wave-packet contains. Wave-packets with larger $l_c$ have a narrower spectrum of wavelengths about $\lambda_0$, thereby increasing the likelihood of interference among the phonons within the wave-packet. More phonon wave periods also create more opportunities for interference. This principle applies to both constructive and destructive wave interference effects of phonons \cite{latour2014microscopic,maranets2024influence}.   

It is critically important to understand that spatial coherence length, being the literal distance over which the atomic vibrations of a phonon mode are correlated, is the main parameter delineating coherent (wave) from incoherent (particle) transport for propagating wave-packets. Mean free path is a property that conceptually belongs to the particle perspective of phonons formalized in Peierls-BTE and is generally not equal to the spatial coherence length as seen in Figs.~\ref{fig:length_calculations}a-\ref{fig:length_calculations}b and explained in Sec.~\ref{sec:spatial_length_scales}. 

Furthermore, it is imprecise to consider wavelength as the only factor determining the type of transport. Coherent phonon transport has been thought to primarily depend on wavelength based on a notion drawn from analogizing phonons to photons that longer wavelengths scatter specularly, providing more opportunities for wave interference, while shorter wavelengths scatter diffusely in phase-destroying processes \cite{maldovan2015phonon}. Indeed, longer wavelengths tend to possess larger coherence lengths and diffuse scattering often disrupts conditions for spatial coherence; however, specular scattering where the phase of the wave-packet is preserved can have a marked influence on phonon heat conduction in the ballistic size regime of transport where the wave-packets still travel like particles \cite{anufriev2017heat,ravichandran2018spectrally}.

\subsection{Phononic crystals\label{sec:phononic_crystals}}

\begin{figure}
    \centering
    \includegraphics[width=\textwidth]{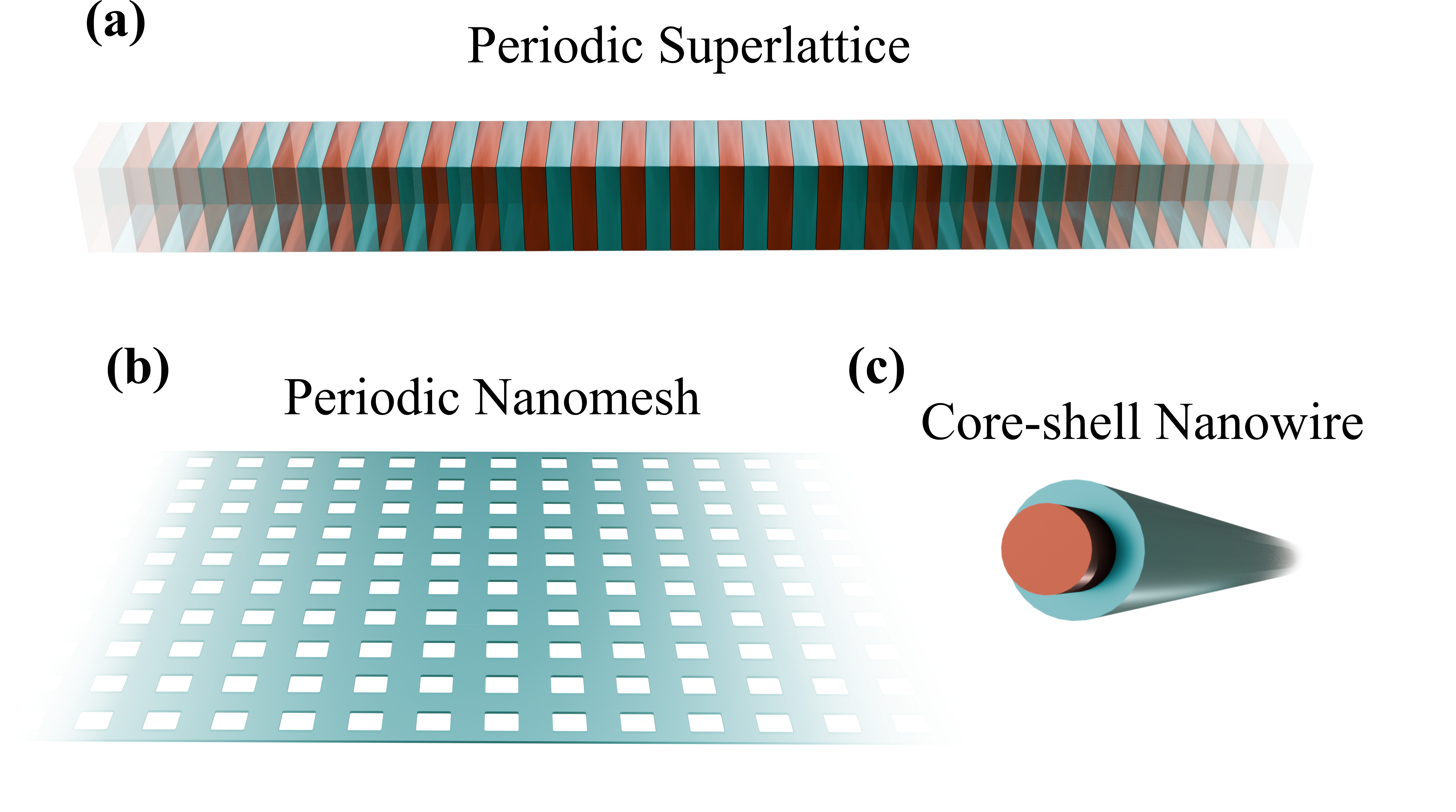}
    \caption{Schematic 3D models of several phononic crystal designs discussed in this review. Panel (a) visualizes the superlattice which has a secondary one-dimensional periodicity perpendicular to the layer interfaces. Panel (b) visualizes the nanomesh which has a secondary two-dimensional periodicity across the plane of the thin film. Panel (c) visualizes the core-shell nanowire which while not strictly possessing secondary periodicity, is a metamaterial that experiences the resonant localization effect (see discussion in Sec.~\ref{sec:resonant_local}) identical to other phononic crystal designs.}
    \label{fig:phononic_crystals_3Dmodels}
\end{figure}

Wave-like heat conduction due to phonon spatial coherence, often referred to as coherent heat conduction, is most prominent in phononic crystals, materials artificially structured with heterogeneous features spaced apart by distances significantly less than the phonon spatial coherence lengths of the constituent materials \cite{volz2016nanophononics,nomura2022review,ma2023phonon}. Typically, the spacing of features is constant such that a new secondary periodicity is created in the system along with the natural periodicity of the constituent materials. Consequently, just as the phonon dispersion relations of the constituent materials in their bulk forms are defined by the crystal structures and lattice constants, a dispersion relation for the phononic crystal can be defined by the geometry of the new artificial unit cell \cite{chen2005nanoscale,latour2014microscopic}. The resonant modes that phonons with $l_{c}$ greater than the size of the phononic unit cell convert to belong to the dispersion relation of the phononic crystal \cite{latour2017distinguishing,maranets2024prominent,maranets2025phonon}. These coherent phonons transport freely without scattering because they essentially perceive the phononic crystal as a pure material \cite{maranets2024influence}. Conversely, phonons with $l_{c}$ less than the phononic unit cell size are scattered by the features as these modes belong to the dispersion relations of the bulk constituent materials instead of the phononic crystal. Clearly, the nomenclature is meant to suggest conceptual similarity to photonic crystals for light \cite{joannopoulos1997photonic}. We visualize several forms of phononic crystals for the reader in Fig.~\ref{fig:phononic_crystals_3Dmodels}.

\begin{figure}
    \centering
    \includegraphics[width=0.75\textwidth]{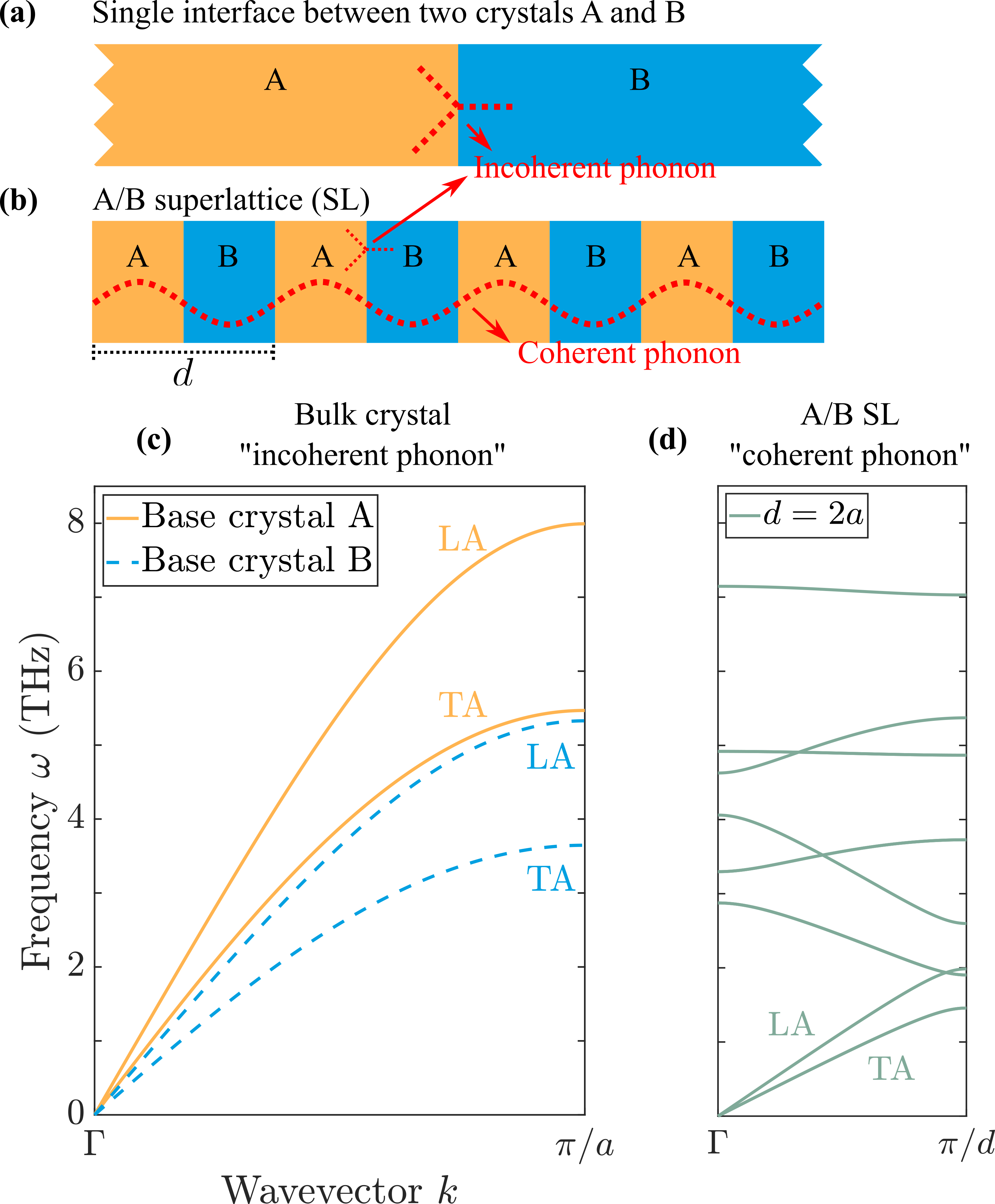}
    \caption{Schematic illustrations of incoherent and coherent phonons within a single interface system (a) and a binary periodic superlattice (b). Particle-like incoherent phonons follow the bulk dispersion relation of the base materials composing the SL (c), while wave-like coherent phonons follow the intrinsic dispersion relation of the superlattice which has secondary periodicity (d). The figure is adapted with permission from Ref.~\cite{maranets2025phonon}. Copright (2024) Elsevier.}
    \label{fig:SL_dispersion_diagram}
\end{figure}

A prominent example of a phononic crystal is the superlattice (SL), a composite structure consisting of periodically alternating nanosized layers of two or more materials as visualized in Fig.~\ref{fig:phononic_crystals_3Dmodels}a. SLs can be constructed through thin film deposition methods like molecular beam epitaxy and metal-organic chemical vapor deposition. For the SL, the secondary periodicity is one-dimensional, occurring along the spatial direction perpendicular to the interfaces. This axis is referred to as the cross-plane direction whereas the axis parallel to the interfaces is referred to as the in-plane direction. Most SLs investigated both experimentally and computationally are binary, meaning they consist of two different materials. Prominent semiconductor examples include Si/Ge and GaAs/AlAs. 

In the SL, the size of the phononic unit cell is the spacing defining the layering pattern known as the period width $d_{SL}$. The layer widths of the two materials composing the SL are often identical meaning $d_{SL}$ would be double the individual layer width. Principally, a phonon transports coherently (wave) if $l_{c}>d_{SL}$ and incoherently (particle) if $l_{c}<d_{SL}$ \cite{latour2014microscopic,latour2017distinguishing}. Physically, one may intuit that incoherent phonons are scattered by the interfaces because they are confined by the interfaces. In contrast, coherent phonons extend past and thus ignore the interfaces. In Fig.~\ref{fig:SL_dispersion_diagram}, we visualize what is meant by coherent and incoherent phonons and their associated dispersion relations in the context of a binary periodic SL.

\subsubsection{Dispersion relations \& Bragg reflection\label{sec:dispersion_Bragg}}

In a bulk natural crystal, the size of the Brillouin zone is determined by the lattice constant $a$. Specifically, a phonon cannot possess a wavelength less than the lattice constant so the maximum wavevector is $\pi/a$. In the SL, the Brillouin zone is now defined by the period width $d_{SL}$ so the maximum wavevector is $\pi/d_{SL}$. Since $\pi/a>\pi/d_{SL}$, the SL dispersion is a zone-folding of the bulk dispersions of the constituent materials. The zone-folding can create ranges of frequencies known as band gaps where phonon transport is forbidden. 

\begin{figure}
    \centering
    \includegraphics[width=\textwidth]{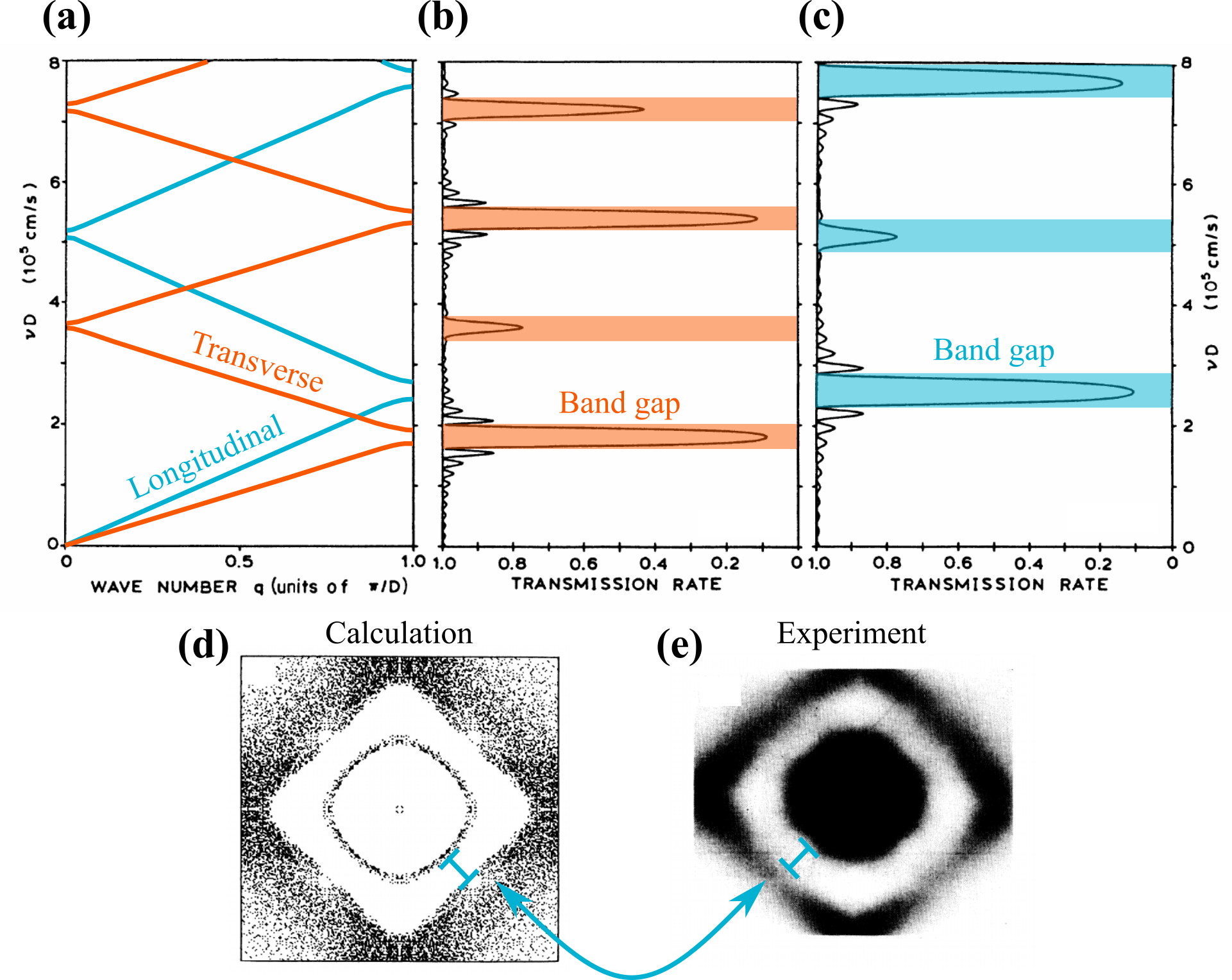}
    \caption{Bragg reflection of coherent phonons in periodic superlattices (SLs). Panel (a) plots a periodic SL dispersion relation for longitudinal and transverse phonons. Panels (b) and (c) presents the calculated phonon transmission spectra for transverse and longitudinal modes, respectively, featuring evenly-spaced dips corresponding to the band gaps. Panel (d) shows the calculated stop-band distribution for longitudinal phonons in a AlAs/GaAs periodic SL. Panel (e) presents the corresponding constant-velocity phonon image for experimental validation. The figures were reproduced with permission from Ref.~\cite{tamura1988acoustic}. Copyright (1988) American Physical Society.}
    \label{fig:Bragg_SL}
\end{figure}

Moreover, akin to optical interference, phonons with wavelengths an integer multiple of the period width (wavevectors at the edges of the Brillouin zone) are totally reflected per the Bragg scattering condition. This effect creates a phonon transmission spectrum consisting of periodically spaced dips when the Bragg condition is satisfied \cite{hurley1987imaging,tamura1988acoustic,tamura1988acousticmulti,tamura1989localized,tanaka1998phonon,maranets2024prominent,maranets2025phonon}. These observations are shown in Fig.~\ref{fig:Bragg_SL}.

The presence of ordered band gaps in the phononic crystal dispersion has attracted interest in engineered waveguiding materials that direct heat flow along specific directions \cite{maldovan2015phonon,laude2021principles}. By blocking the transport of certain frequencies, the phononic crystal can act as an insulating or cladding material that constricts the conduction of those frequencies to a specific geometric pathway dependent on the positioning of the phononic crystal and the heterogenous material it is adjoined to. Experimentally, this principle has been demonstrated in a phonon cavity structure  \cite{trigo2002confinement}. The selective reflection of heat flow is also promising for heat mirror applications \cite{maldovan2015phonon}. 

In addition to creating band gaps, zone-folding, through flattening the bands in the dispersion relation, also decreases the group velocities, reducing the amount of heat carried by the coherent phonons.

\subsection{Signatures of coherent phonon transport\label{sec:signatures_of_coherent_transport}}

In this section, we discuss how phonon spatial coherence is revealed in specific observables.

\subsubsection{Minimum thermal conductivity vs. period size\label{sec:min_kappa_SL}}

\begin{figure}
    \centering
    \includegraphics[width=\textwidth]{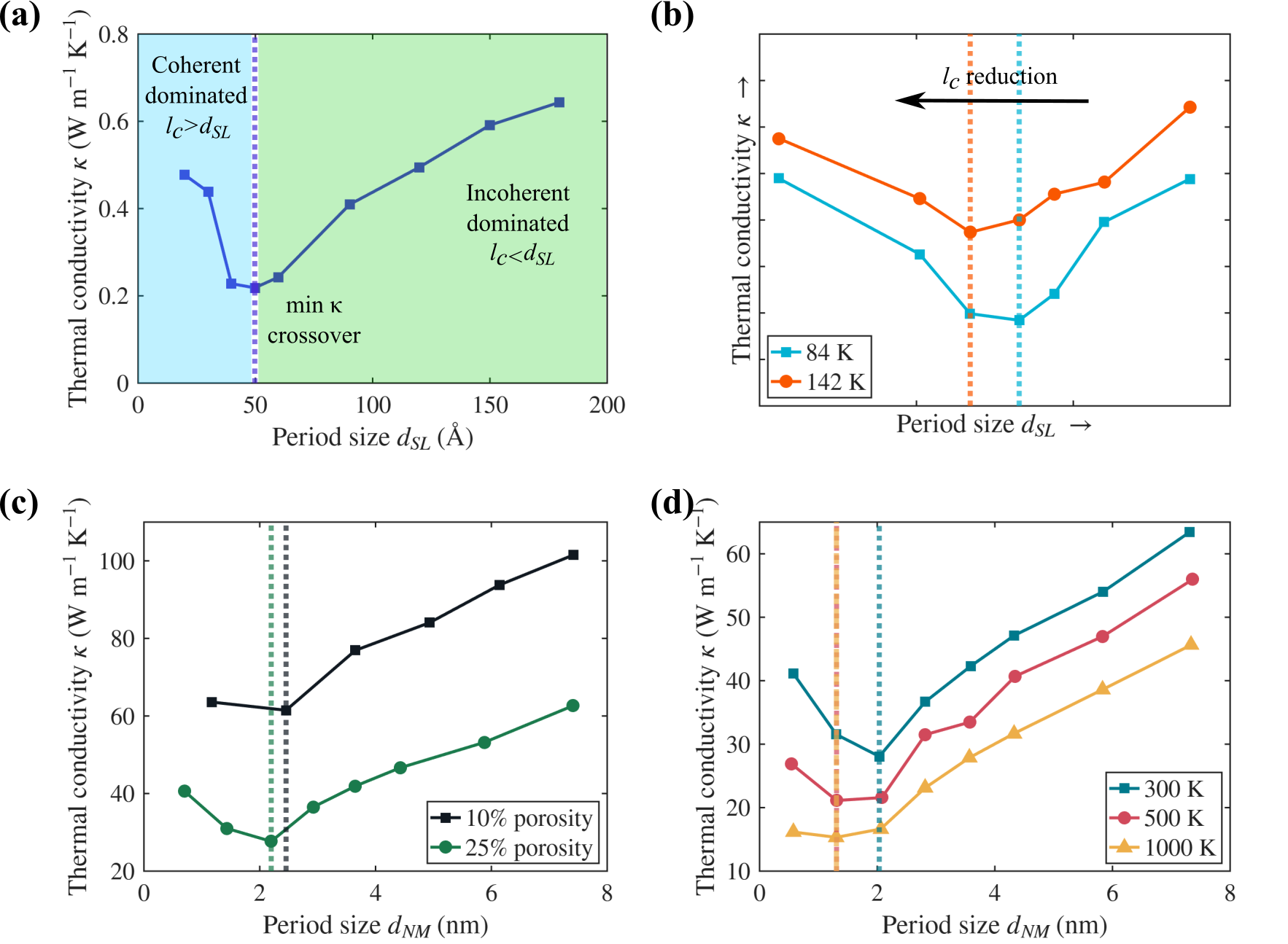}
    \caption{Minimum thermal conductivity with period size for several phononic crystals. Panel (a) shows the seminal experimental result by Venkatasubramanian \cite{venkatasubramanian2000lattice} demonstrating the emergence of phonon spatial coherence in Bi$_2$Te$_3$/Sb$_2$Te$_3$ periodic SLs via a minimum thermal conductivity arising from a crossover between incoherent and coherent dominated phonon transport when period size is reduced. The figure is reproduced with permission Ref.~\cite{venkatasubramanian2000lattice}. Copyright (2000) by the American Physical Society. Panel (b) shows the experimental result by Ravichandran et al. \cite{ravichandran2014crossover}, highlighting that the location of the crossover in periodic SLs changes with temperature due to the associated changes in spatial coherence length (see Fig.~\ref{fig:length_calculations}c). The data is adapted from Ref.~\cite{ravichandran2014crossover}. Copyright (2013) Springer Nature Limited. Panels (c) and (d) show the observations of these behaviors in graphene nanomeshes, another type of phononic crystal structure. The data is adapted from Ref.~\cite{Hu2018CoherentDominant}. Copyright (2018) from the American Chemical Society.}
    \label{fig:min_SL_panels}
\end{figure}

Phononic crystals demonstrate unique thermophysical properties due to the influence of phonon spatial coherence \cite{xie2018phonon}. The most significant phenomenon, which best exemplifies the manifestation of spatially coherent phonon transport, is the presence of a minimum thermal conductivity with the size of the phononic unit cell i.e. the period. Particularly, when the period size is reduced, thermal conductivity does not monotonically decrease as predicted by the magnification of interface scattering by particle-like incoherent phonons when the interface separation is reduced. Instead, past a certain period size, the phonons with spatial coherence lengths greater than the period size begin to predominate energy transport over the incoherent phonons via the coherent transport mechanism, resulting in an increase in thermal conductivity \cite{latour2014microscopic}. This crossover between incoherent and coherent phonon dominated thermal transport is reflected in a minimum thermal conductivity with the variation of period size. 

This minimum thermal conductivity effect has been most extensively observed in SLs \cite{venkatasubramanian2000lattice,simkin2000minimum,chen2005minimum,ravichandran2014crossover}, however, it is also present in many other phononic crystal designs. We display a comparison of results in Fig.~\ref{fig:min_SL_panels}. Importantly, the period size location of the minimum changes with temperature due to the temperature-dependence of spatial coherence length (e.g. Fig.~\ref{fig:length_calculations}c) \cite{latour2014microscopic,maranets2024influence}. We also note that due to inherent similarity of the frequency and wavevector linewidths \cite{peierls1955quantum}, this minimum thermal conductivity behavior and the manifestation of coherent phonon transport can also be assessed through phonon lifetimes \cite{garg2013minimum,nasiri2025evolution}.

\subsubsection{Thermal conductivity dependencies\label{sec:kappa_dependence_spatial}}

Other general markers of phonon spatial coherence include novel dependencies of phonon transmission, thermal conductance, and/or thermal conductivity with system parameters like device length or temperature. Dependencies that both differ from those predicted by the conventional particle analyses of phonons and which align with the theoretical predictions modeling wave-like coherent phonon transport indicate the presence of phonon spatial coherence.

An example dependency characterizing phonon spatial coherence is the marked scaling of thermal conductivity with device length for small period size phononic crystals \cite{chen1998thermal,chen1999phonon,simkin2000minimum,liu2022heat,daly2002molecular,luckyanova2012coherent,wang2014decomposition,wang2015optimization,chakraborty2020complex}. Wave-like coherent phonons are subject to the ballistic size regime of phonon transport (see Sec.~\ref{sec:propagating_wave-packets}) within phononic crystals just like ordinary incoherent phonons are in bulk crystals. In phononic crystals, incoherent phonons do not exhibit significant size effects due severe interface scattering. Thus, any observed size effect and importantly trends consistent with the behavior of incoherent phonons in homogeneous thin films \cite{lukes2000molecular} indicate that length dependence of heat conduction in phononic crystals arises from the coherent phonons that are resonant with the secondary periodicity. This mainly occurs in small period size structures that are dominated by the coherent phonons. For larger period sizes where incoherent phonons are predominant, significant length dependence is absent beyond a few periods \cite{yang2003partially,latour2014microscopic,wang2014decomposition,luckyanova2012coherent,ma2022ex,maranets2024prominent}.

The last prominent signature of phonon spatial coherence is the exotic changes in phonon heat conduction that occur in phononic crystal designs that have been selectively modified to manipulate the wave-like transport of the coherent phonons. This topic is discussed in more detail later in the review in Sec.~\ref{sec:wave_localization} and Sec.~\ref{sec:phononic_character}.

\subsection{Modeling the contributions of coherent and incoherent phonons\label{sec:model_coherent_incoherent}}

The varied signatures of phonon spatial coherence illustrate that heat conduction in phononic crystals stems from the contributions of both wave-like coherent and particle-like incoherent phonons. This has motivated development of heat conduction models that decompose the thermal phonon transport as the summation of the heat carried by coherent and incoherent phonons separately. By delineating two ``channels'', one can describe all heat conduction regimes (coherent dominated, incoherent dominated, or partially coherent) by the relative strengths of the coherent and incoherent phonon contributions. The advantage of this approach stems from the fact that coherent and incoherent phonons derive their properties from separate dispersion relations. This method was formalized from analysis of molecular dynamics simulations of conceptual Lennard-Jones binary SLs by Wang et al. \cite{wang2014decomposition} Drawing on the Landauer framework \cite{datta2005quantum}, the length-dependent thermal conductance of an SL is given as:
\begin{equation}
    G_{SL}(L) = G_{coh}(L)+G_{inc}(L)=G_{coh,0}(L)\frac{\Lambda_{coh}}{\Lambda_{coh}+L} + G_{inc,0}(L)\frac{\Lambda_{inc}}{\Lambda_{inc}+L}
    \label{eqn:coh_inc_equation}
\end{equation}
where the subscripts $coh$ and $inc$ correspond to coherent and incoherent phonon, respectively and $G_{0}$ denotes the ballistic-limit quantity while $\Lambda$ denotes mean free path. The length-dependent thermal conductivity is then $\kappa_{SL}(L) = G_{SL}(L) \cdot L$. This treatment is particularly powerful in determining the thermal conductivity of aperiodic SLs where the aperiodic arrangement of interfaces induces Anderson localization of the wave-like coherent phonons (discussed further in Sec.~\ref{sec:anderson_local}). Particularly, the localization can be captured simply by introducing an exponential decay factor to $G_{coh}(L)$. As the incoherent phonons transport like particles, their amount of heat carried is insensitive to the specific arrangement of interfaces and thus no corrective term to $G_{inc}(L)$ is needed \cite{wang2014decomposition,ma2020dimensionality,chakraborty2025comprehensive}.

Modeling the heat conduction as the summation of coherent and incoherent phonon contributions also conceptually explains why some theoretical methodologies fail to accurately predict the thermophysical properties of phononic crystals across all scenarios. For example, conventional approaches like the Peierls-BTE and Monte-Carlo simulation that solely model the particle behaviors of phonons do not capture the minimum thermal conductivity since the coherent phonon wave effect, intrinsically stemming from spatial coherence length, is not taken into account. Conversely, if only the coherent phonons derived from the phononic crystal's dispersion relation are considered in the Peierls-BTE thermal conductivity calculation, then thermal conductivity would monotonically decrease with increasing period size due to the group velocity reduction of phonon modes in the increasingly folded Brillouin zone. This is inaccurate as the incoherent phonons are less impacted by interface scattering when the period size and consequently the interface separation enlarges. Modifications to these conventional particle-based methods, such as those in Ref.~\cite{simkin2000minimum} and Ref.~\cite{yu2019investigation}, which account for the opposing wave effects, have demonstrated better accuracy.

The atomistic non-equilibrium Green's function method and molecular dynamics simulations are among the most robust approaches to study heat conduction in phononic crystals as both methods natively incorporate the dynamics of wave-like coherent phonons and particle-like incoherent phonons \cite{ma2020first}. We refer the reader to an excellent review by Zhang et al. \cite{zhang2021coherent} that thoroughly examines the different theoretical methodologies that have been utilized to quantify coherent transport and its role in heat conduction in phononic crystals. 

\subsection{Phonon wave localization\label{sec:wave_localization}}

Due to their heat conduction being significantly influenced by spatial coherence, phononic crystals offer a fundamentally new approach to manipulating thermophysical properties through the tailored alteration of phonon wave transport rather than solely the control of phonon particle transport through scattering processes \cite{li2012colloquium,maldovan2013sound,nomura2018thermal}. Specifically, changes to both the arrangement and character of the heterogeneous features defining the phononic crystal can manipulate the wave interference states of coherent phonons, thus affecting the amount of heat carried. 

The most prominent examples of controlling these wave-like phonons are structuring mechanisms that induce wave localization; the suppression of wave transport, characterized by a finite length beyond which the wave energy does not propagate. This localization length, which is often frequency-dependent, is distinct from spatial coherence length and mean free path. Spatial coherence length quantifies the spatial extension of a phonon wave-packet, therefore it is a measure of the range of wavelengths the wave-packet encompasses. Mean free path, being the product of group velocity and relaxation time, quantifies the distance over which the phonon wave-packet travels before being scattered and losing its phase. For localization, phonon waves can be confined and maintain their phase while no longer contributing to energy transport \cite{latour2017distinguishing,maranets2024influence}. Additionally, the scattering phase space for phononic crystals with and without localization can be comparable despite large thermal conductivity differences \cite{ma2020dimensionality}, evidencing that attenuation of heat through localization of coherent phonons is distinct from ordinary phonon scattering. 

The combined inhibition of particle-like incoherent phonons via the intense extrinsic scattering in a typical phononic crystal and wave-like coherent phonons via select variations to the phononic design make localization-modified phononic crystals attractive for applications requiring ultra-low thermal conductivity like thermal barrier coatings and thermoelectric devices. 

There are two types of localization that have been uncovered in phononic crystals. The first is the well-known Anderson localization; an intrinsic wave phenomenon stemming from the coherent backscattering and destructive interference of multiple scattered waves in a disordered medium. The second is resonant localization; an effect arising from the introduction of additional substructures that weaken coherent phonon transport through coupling with the resonant modes of the new substructures. In the following paragraphs, we discuss the theoretical principles and relevant applications of both localization effects.

\subsubsection{Anderson localization\label{sec:anderson_local}}

\begin{figure}
    \centering
    \includegraphics[width=\textwidth]{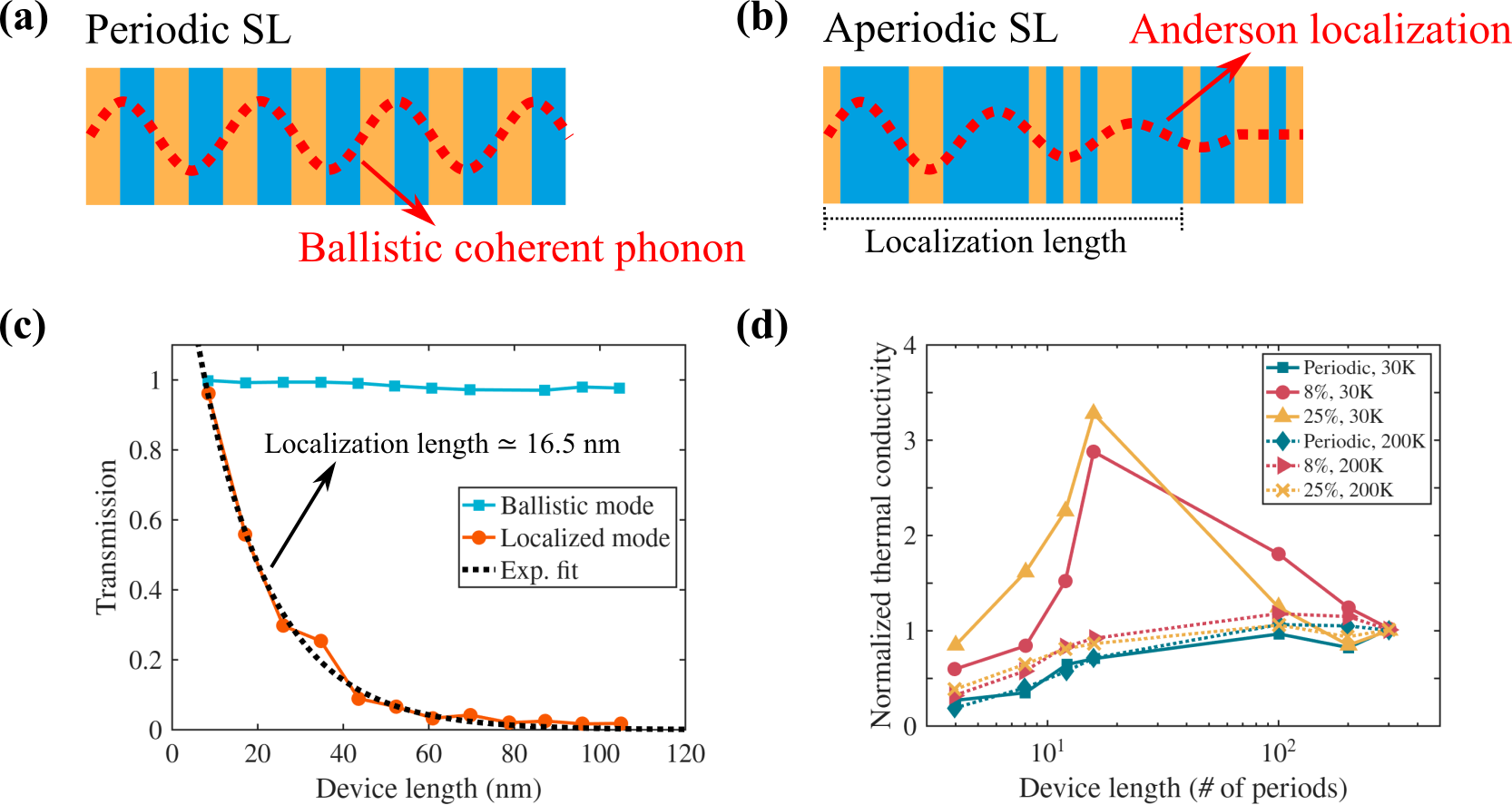}
    \caption{Schematic illustrations of ballistic propagation and Anderson localization of coherent phonons in periodic SLs (a) and aperiodic SLs (b), respectively. Panel (c) compares the near-unity transmission of a ballistic mode and the characteristic exponential decay in transmission of an Anderson localized mode in an aperiodic SL. The data is adapted with permission from \cite{hu2021direct}. Copyright (2021) by the American Physical Society. Panel (d) shows the transition from ballistic to localized transport as revealed by a local maximum in thermal conductivity with device length for SLs whose secondary periodicity was broken by ErAs nanodots embedded at the interfaces. The data is adapted with permission from \cite{luckyanova2018phonon}. Copyright (2018) from The American Association for the Advancement of Science.
}
    \label{fig:Anderson_localization_visual}
\end{figure}

Anderson localization refers the attenuation of wave transport through significant destructive interference events induced by a disordered arrangement of scatterers \cite{anderson1958absence,sheng2007introduction}. For a phononic crystal, the disorder is realized in the disruption of the secondary periodicity. 

Phonon Anderson localization can be estimated or inferred from a variety of measurements \cite{ni2021evidence}. In theoretical studies, Anderson localization is most commonly identified by an exponential decay of phonon transmission (or thermal conductance which is directly proportional to phonon transmission in the Landauer framework \cite{datta2005quantum}) over the length of the system as seen in Fig.~\ref{fig:Anderson_localization_visual}c. The computed rate of decay is the localization length. 

Experimentally, Anderson localization as it pertains to heat conduction has been inferred from a local maximum of thermal conductivity as a function of device length. Luckyanova et al. \cite{luckyanova2018phonon} using time-domain thermoreflectance, measured the thermal conductivities of GaAs/AlAs periodic SLs with and without ErAs nanodots randomly dispersed at the interfaces. We have plotted their data in Fig.~\ref{fig:Anderson_localization_visual}d.

Without the ErAs nanodots, thermal conductivity increases with device length (number of periods) and eventually saturates. As discussed in Sec.~\ref{sec:kappa_dependence_spatial}, the initial increase with length is owed to the ballistic transport of coherent phonons while the saturation signifies the structure being sufficiently longer than the coherent phonon mean free paths such that anharmonic scattering destroys the phonon spatial coherence \cite{luckyanova2012coherent}. Similar results have been shown in molecular dynamics investigations by Wang et al. \cite{wang2014decomposition}. 

In the SLs with the ErAs nanodots, at cryogenic temperatures, the increase in thermal conductivity with length is followed by a non-monotonic reduction, suggesting a transition from ballistic to localized transport. This effect is absent at room temperature due to greater contribution of high-frequency incoherent phonons affected by anharmonic scattering. Additionally, wave localization at low temperatures can also be attributed to the larger spatial coherence lengths which make wave transport behaviors more prominent \cite{latour2014microscopic,maranets2024influence}. Supportive atomistic Green's function calculations show suppression in the transmission of coherent phonons in SLs with ErAs nanodots \cite{mendoza2016anderson,luckyanova2018phonon}.

Since the ErAs nanodots were of comparable size to the SL layer widths, the inclusion of the nanodots at the interfaces effectively broke the periodic layering pattern of the original SL. In a similar vein, much attention has been paid to investigating phonon Anderson localization and heat conduction in aperiodic SLs possessing an aperiodic layering pattern. 

Experimentally, localization was suggested to disrupt coherent phonon transport in aperiodic WSe$_{2}$ SLs as evidenced by an increase in thermal conductivity when the disordered structure was exposed to ion irradiation \cite{chiritescu2007ultralow}. Specifically, through the nucleation of defects in the thin film materials composing the aperiodic SL, irradiation disrupted the specific interface arrangement. A marked increase in thermal conductivity suggests that the aperiodic layering pattern had an attenuating effect on thermal conductivity, possibly through phonon Anderson localization. 

Aperiodic SLs have been investigated more extensively through first-principles calculations and molecular dynamics simulations where Anderson localization can be probed more directly. Given the suppression of both wave-like coherent and particle-like incoherent phonons, several studies have used machine-learning algorithms to optimize the layer widths of aperiodic SLs to best minimize thermal conductivity \cite{chakraborty2020quenching,chowdhury2020machine,hu2020machine,wei2022perspective,tyagi2026machine}.

Still, non-trivial thermal conductivities are observed, suggesting some mechanism facilitates heat conduction in aperiodic SLs. Recent investigations using the molecular dynamics-based wave-packet method have revealed that aperiodic SLs possess their own high-transmission coherent modes defined by an approximate dispersion relation akin to periodic SLs \cite{maranets2024prominent}. However, these coherent phonons are suggested to be of a different character than coherent phonons in the periodic SL, behaving like the non-propagative diffusons and carrying heat diffusely rather than ballistically like propagons, ultimately manifesting in different length dependencies of thermal conductivity \cite{maranets2025phonon}. We discuss these conclusions further in Sec.~\ref{sec:shared_frameworks}. 

How phonon spatial coherence changes and how Anderson localization develops when transitioning from the periodic to the aperiodic SL design are questions of considerable interest. So far, researchers have explored these topics by investigating behaviors in quasi-periodic phononic crystals. Notably, several studies demonstrate partial phonon Anderson localization and strong attenuation of heat conduction in SLs possessing a gradient layering pattern \cite{van2019thermal,ferrando2020beating,guo2021thermal,wu2023suppressed}. A recent study by Doe et al. \cite{doe2026analyzing} has revealed such gradient SLs exhibit coherent mode-conversion behaviors representative of intermediate states between periodic and aperiodic SLs.

\subsubsection{Resonant localization\label{sec:resonant_local}}

\begin{figure}
    \centering
    \includegraphics[width=\textwidth]{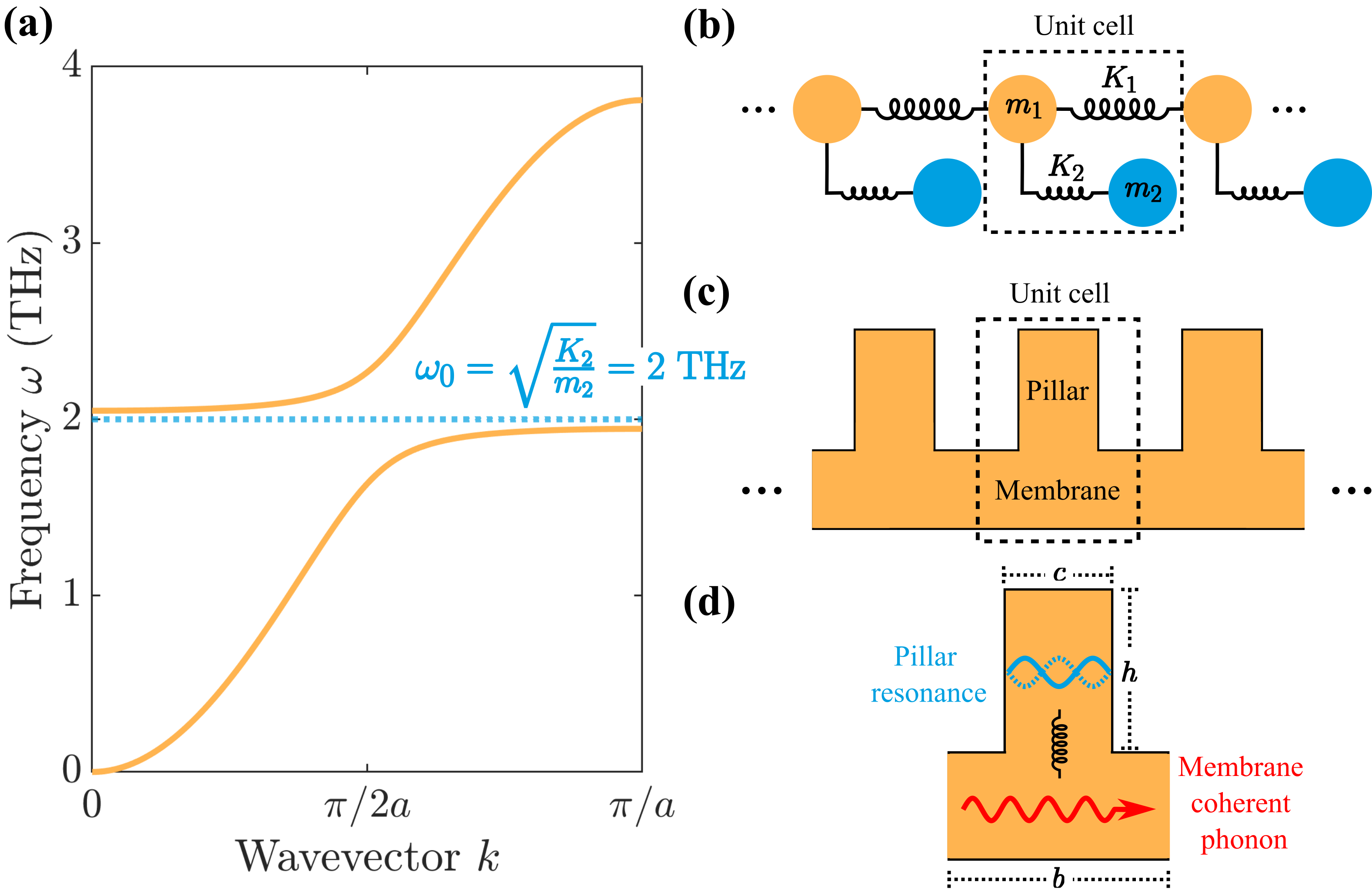}
    \caption{Schematic illustrations of resonant localization. Panel (a) plots the phonon dispersion relation for a periodic 1D mass-spring chain with local resonators as illustrated in panel (b). Panel (c) indicates the unit cell for a phononic metamaterial composed of a thin film membrane with perpendicular pillars. Panel (d) illustrates the coupling between the pillar resonance modes and the in-plane coherent phonons of the membrane. Similar to the 1D chain, the degree of resonance hybridization in the metamaterial depends on the structural parameters of the unit cell including the base membrane width $b$, pillar width $c$, and pillar height $h$ as indicated in panel (d).}
    \label{fig:resonance_localization_visual}
\end{figure}

In contrast to disorder-induced Anderson localization, resonant localization refers to suppression of wave transport through the inclusion of additional substructures to the phononic crystal that possess their own vibrational modes \cite{hussein2020thermal}. These modes, referred to as local resonances or ``vibrons'', can couple with the coherent phonons belonging to the original phononic crystal's dispersion relation. This coupling mechanism, called resonance hybridization, attenuates the transport of coherent phonons through reduction in both group velocities and relaxation times as well as spatial confinement of wave energy inside the substructures \cite{davis2014nanophononic,honarvar2018two,beardo2024resonant}.

Resonant localization is particularly distinct from Anderson localization as the positioning of the introduced substructures does not have to be disordered. The degree of wave localization is dependent on the size and geometry of the substructures in addition to their arrangement. Furthermore, resonant localization takes place across the entire phonon spectrum whereas Anderson localization in a particular phononic crystal device only occurs for phonon frequencies whose localization lengths are less than the length of the device.

Resonant localization in phononic crystals is highly comparable to the effect guest rattler atoms have on host lattice phonons in host-guest crystals like clathrates and skutterudites \cite{wang2009resonant,tadano2015impact,jana2017intrinsic,yang2017unravelling,baggioli2019theory}. Conceptually, resonant localization can be understood through a 1D mass-spring model containing local resonators \cite{ma2023phonon} as illustrated in Fig.~\ref{fig:resonance_localization_visual}b. The local resonance attenuates propagation of phonon modes about the resonant frequency as seen in the dispersion relation plotted in Fig.~\ref{fig:resonance_localization_visual}a.

\begin{figure}
    \centering
    \includegraphics[width=\textwidth]{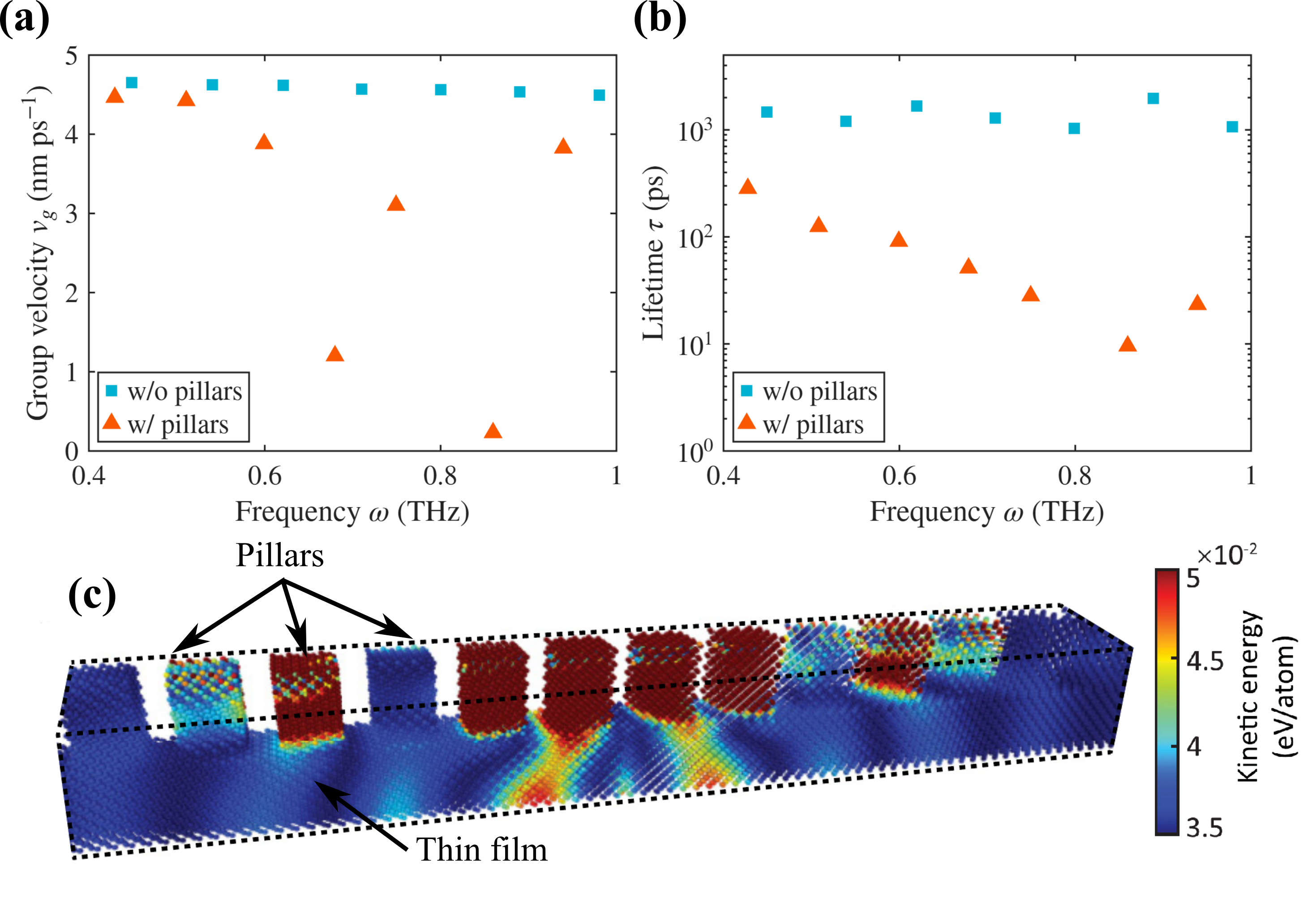}
    \caption{Influence of resonant localization on phonon properties in pillared thin-films. Panel (a) compares group velocity values for in-plane coherent phonons in a silicon thin film with and without pillars. Panel (b) compares phonon lifetime values between the two configurations. Panel (c) shows the kinetic energy in each atom at a representative time step after an excitation at a resonant phonon frequency, demonstrating wave localization of phonons by the pillars. The data in panels (a) and (b) are adapted with permission from Ref.~\cite{honarvar2018two}. Copyright (2018) from the American Physical Society. The image in panel (c) is reprinted with permission from Ref.~\cite{beardo2024resonant}. Copyright (2024) from the American Physical Society. }
    \label{fig:resonant_localization_data}
\end{figure}

Resonant localization has been studied most extensively in semiconductor thin films with pillars adjoined to the surface of the thin film in the cross-plane direction \cite{davis2014nanophononic,wei2015phonon,honarvar2016spectral,honarvar2018two,spann2023semiconductor,beardo2024resonant}. These materials have been referred to as phononic metamaterials, signifying that the base phononic crystal has been artificially modified \cite{hussein2020thermal,jin2021physics}. In this case, the phononic crystal is the thin film membrane unit cell the pillar extends off of as illustrated in Fig.~\ref{fig:resonance_localization_visual}c. To induce wave localization of coherent phonons, the size of the pillar must be less than the spatial coherence lengths. When the pillar is sufficiently small, it modifies the dispersion relation of the phononic crystal along the in-plane direction of the membrane. Specifically, wavevector-independent flat bands are introduced to the dispersion, corresponding to the resonant frequencies of the pillar's vibrational modes \cite{davis2014nanophononic}. The flatness of the bands indicate that these modes exist entirely within the pillar and thus do not contribute to energy transport within the plane of the membrane (see Fig.~\ref{fig:resonance_localization_visual}d). Intersections between the flat bands of the pillar and the dispersive wavevector-dependent modes of the membrane leads to a flattening of the membrane's vibrational modes about the intersection in an effect known as resonance hybridization. Near a resonant frequency of the pillar, the group velocities of the coherent phonons propagating within the plane of the membrane are reduced. At the resonant frequency, the group velocity is zero, meaning these coherent phonons are entirely localized by the pillars and no longer conduct heat within the membrane. 

The influence of the pillars on group velocity can be seen via harmonic lattice dynamics calculations of the dispersion relations. Spectral energy density calculations from equilibrium molecular dynamics simulations incorporating anharmonicity show the resonance hybridization effect also reduces the lifetimes of the coherent phonons in addition to their group velocities  \cite{honarvar2016spectral,honarvar2018two}. The reduction in phonon lifetimes can be attributed to the pillars introducing additional anharmonic and boundary scattering processes. Besides lifetimes, molecular dynamics also reveals that heightened phonon energy is confined to the pillars and that reduction of participation ratios is prominent, particularly about the resonant frequencies of the pillars \cite{wei2015phonon,beardo2024resonant}. All three effects: (1) group velocity reduction near the resonant frequencies, (2) spatial localization of energy at the resonant frequencies, and (3) reduction of phonon lifetimes lead to significant suppression of thermal conductivity from the membrane's base value. We have provided representative data of these effects in Figs.~\ref{fig:resonant_localization_data}a-\ref{fig:resonant_localization_data}c. 

Interestingly, molecular dynamics simulations have shown the thermal conductivity reduction through resonant localization in pillared thin films can be much stronger than reduction through Anderson localization \cite{beardo2024resonant}. Another benefit of the proposed phononic metamaterial design over a typical phononic crystal with engineered Anderson localization is the fact that the structuring mechanism to reduce thermal conductivity, the nanoscale pillar, exists out of the plane of energy transport. This is particularly advantageous for thermoelectric applications as the challenge of introducing heterogeneous features to the plane of energy transport that weaken phonon heat conduction without inducing significant scattering of electrons is avoided.   

The resonant localization phenomenon also occurs in core-shell nanowires (NWs), another type of phononic crystal. The NW structure consists of a pure material, typically a semiconductor crystal, whose geometry is severely elongated in a specific direction like a wire. The extremely high surface-to-volume ratio greatly enhances phonon-boundary scattering, leading to large reduction in thermal conductivity from the bulk form \cite{zou2001phonon,lu2002size,li2003thermal}. Moreover, the narrow dimension can induce coherent phonon transport and resonant localization in core-shell structures (visualized in Fig.~\ref{fig:phononic_crystals_3Dmodels}c) where the NW is coated with a heterogeneous material possessing different vibrational properties. Theoretical studies of this type of structure \cite{pokatilov2005acoustic,yang2005thermal,chen2011phonon,hu2011thermal,chen2012impacts} indicate that coupling between the non-propagating transverse phonon modes in the coating and the propagating longitudinal modes in the NW, the coherent modes, can weaken the transport of the propagating coherent phonons in the NW through reduction of group velocities, ultimately lowering thermal conductivity as experimentally verified for germanium NW coated with silicon by Wingert et al. \cite{wingert2011thermal}. Reduction of participation ratios potentially indicate localization \cite{canisius1985localisation,hafner1993propagating,schober1996low}. 

In line with the minimum thermal conductivity as a function of period size behavior of other phononic crystals discussed in Sec.~\ref{sec:min_kappa_SL}, this resonant localization effect in core-shell NWs only occurs for very small coating thicknesses. Beyond a critical value, the increase in the transverse dimension as result of the coating thickness leads to an increase in thermal conductivity due to the decreased surface-to-volume ratio.  

\subsection{Character of features in phononic crystals\label{sec:phononic_character}}

Our discussion of phononic crystals so far has concerned how the sizes, arrangement, and inclusion of various features composing the devices affect coherent heat conduction. Just as important is how the character of the features influences phonon spatial coherence. Character refers to the shape of a feature which encompasses its geometry as well as its surface roughness. The degree to which these factors affect coherent mode-conversion impacts the overall magnitude of coherent heat conduction in a phononic crystal.

To understand this concept comprehensively, we focus our discussion on the nanomesh (NM), a thin film membrane material possessing holes or geometric boundaries distributed across a two-dimensional plane as visualized in Fig.~\ref{fig:phononic_crystals_3Dmodels}b. For this phononic crystal, the secondary periodicity is two-dimensional with the characteristic lengths mainly being the spacing between the holes and the hole sizes. Computational studies have demonstrated the significant influence of hole size on coherent heat conduction in NMs such as the minimum thermal conductivity phenomenon as seen in Figs.~\ref{fig:min_SL_panels}c-\ref{fig:min_SL_panels}d. Particularly, Hu et al. \cite{Hu2018CoherentDominant} revealed a decrease in the hole spacing i.e. the period size $d_{NM}$ at fixed porosity of graphene NMs with smooth, square holes exhibited a non-monotonic $\kappa$-$d_{NM}$ relation, indicating a transition from incoherent to coherent phonon transport as hole dimensions decrease, similar to that observed in SLs. However, unlike SLs, no experimental study to date has observed the minimum thermal conductivity predicted in computational studies, likely due to the fabrication limitations in hole quality and the requirement for sufficiently small hole sizes below the phonon spatial coherence lengths to observe pronounced coherent transport in NMs. 

Still, the experimental absence of a minimum thermal conductivity in NMs does not necessarily imply the complete absence of phonon spatial coherence. As demonstrated by Cui et al. \cite{cui2024spectral} through spectral heat flux analysis in non-equilibrium molecular dynamics simulations of silicon NMs, long-wavelength phonon modes can still exhibit notable spatial coherence in NMs with relatively large holes (7.2 nm $\times$ 7.2 nm). However, at such hole sizes, there is no observable difference in $\kappa$ between periodic and aperiodic NMs. This underscores the importance of analyzing spatial coherence at the mode level, even in cases where coherent phonon behavior is not reflected in the overall $\kappa$.  

\begin{figure}
    \centering
    \includegraphics[width=\textwidth]{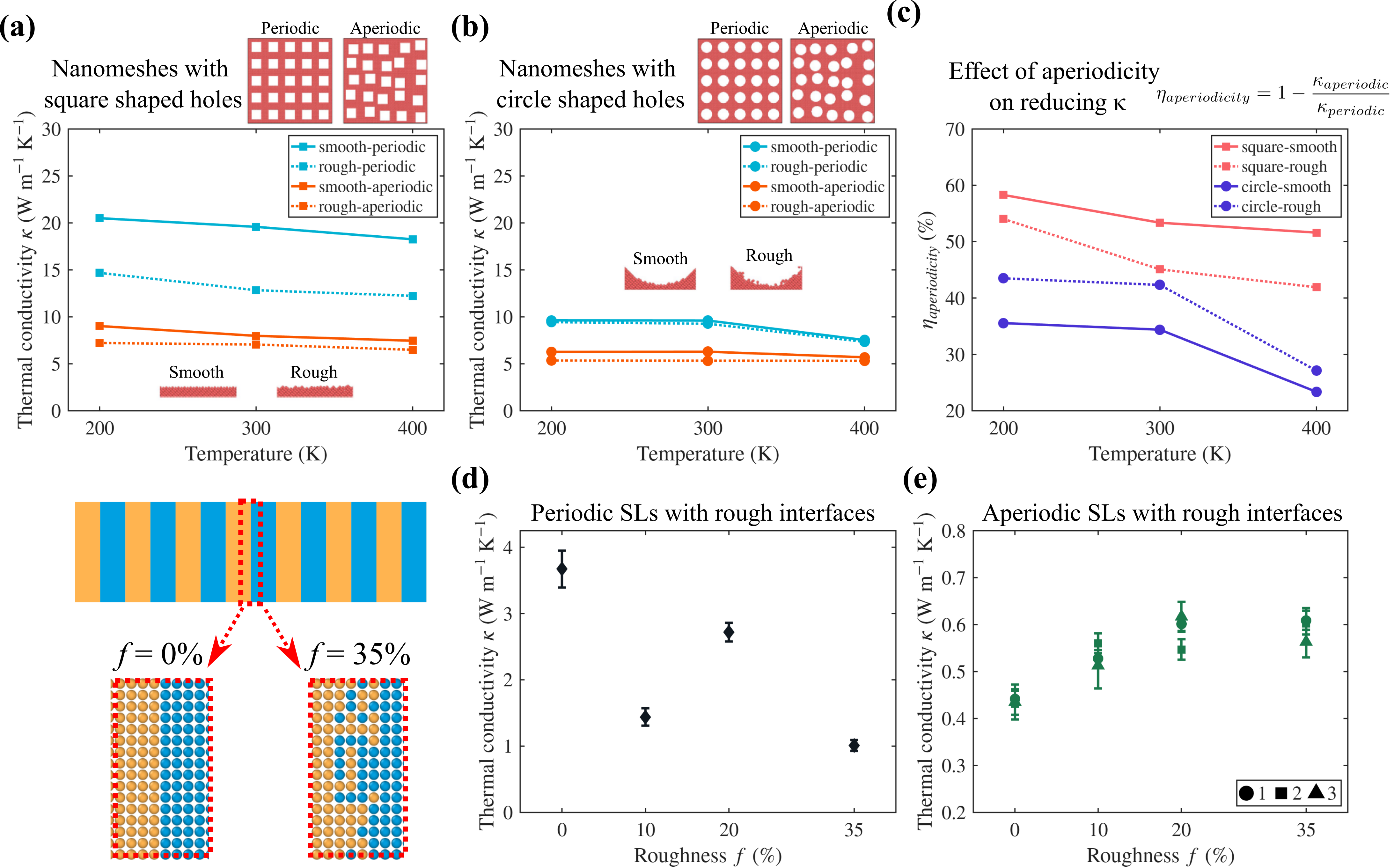}
    \caption{Thermal conductivity $\kappa$ of silicon nanomeshes (NMs) with periodic, aperiodic, smooth hole boundaries, and rough hole boundaries configurations. (a) NMs with squared shaped holes. (b) NMs with circle shaped holes. (c) Effect of aperiodic configuration on $\kappa$ reduction of NMs. The data is adapted with permission from Ref.~\cite{Cui2024NM2D}. Copyright (2024) IOP Publishing Ltd. Thermal conductivity $\kappa$ of Lennard-Jones superlattices (SLs) with varying surface roughness quantified by the mixing fraction $f$ of atomic species at the interfaces. (d) Periodic SLs with rough interfaces. (e) Aperiodic SLs with rough interfaces. The data is adapted with permission from Ref.~\cite{maranets2025role}. Copyright (2025) IOP Publishing Ltd.}
    \label{fig:nanomesh_shape}
\end{figure}

Regarding the hole quality, surface roughness at the hole boundaries has made experimental demonstration of coherent heat conduction in NMs extremely challenging in comparison to the well-documented hallmarks of spatial coherence in SLs. For instance, while an aperiodic layering pattern can markedly reduce $\kappa$ of SLs through Anderson localization, a similar effect was not observed in the experiment by Lee et al. \cite{Lee2017NatureCommu}. Specifically, they reported nearly identical $\kappa$ values for silicon NMs with both periodic and aperiodic hole arrangements over a broad temperature range (14 K to 325 K). The absence of a notable difference in $\kappa$ between periodic and aperiodic NMs in their study can be attributed to the quality of hole boundary surfaces, as existing fabrication techniques are challenged in producing atomically smooth hole surfaces at nanometer length scales \cite{alaie2015thermal}. Rough boundaries introduce more diffuse scattering which destroys the phases of phonons and consequently attenuates coherent mode-conversion (see Fig.~\ref{fig:coherent_mode-conversion_schematic}) and the coherent contribution to $\kappa$ as elucidated in both experimental \cite{wagner2016two} and computational \cite{Cui2024NM2D} studies. Cui et al. \cite{Cui2024NM2D} investigated via molecular dynamics simulations NMs with smooth and rough hole boundaries, as well as square and circular holes. We have plotted their thermal conductivity data in Figs.~\ref{fig:nanomesh_shape}a-\ref{fig:nanomesh_shape}c. Their findings revealed that rough boundaries and circular holes (which effectively act as rough boundaries) significantly attenuate $\kappa$ by suppressing coherent mode-conversion. When rough boundaries are absent, phonon Anderson localization can be realized in NMs \cite{hu2019disorder}.

Interestingly, surface roughness induces different behaviors in SLs than in NMs. As shown in Figs.~\ref{fig:nanomesh_shape}d-\ref{fig:nanomesh_shape}e, surface roughness in SLs, characterized by mixing of atomic species at the interfaces, can enhance lattice heat conduction in aperiodic SLs while decreasing it in periodic SLs. Our own molecular dynamics analyses \cite{maranets2025role} reveal this phenomenon is due to two competing effects of interface mixing on coherent mode-conversion. The first effect is the increased scattering at the mixed interfaces disrupting coherent mode-conversion induced by the interface arrangement. The second effect is the scattering providing new opportunities for constructive interference and coherent mode-conversion. The first effect decreases transmission while the second effect increases it. The second effect enhances transmission of Bragg-reflected modes in the periodic SL and most phonons in the aperiodic SL which exhibit little or no coherent mode-conversion in the structures with smooth interfaces. Conversely, the first effect dominates in the periodic SL as the non-Bragg-reflected modes already have high transmission via significant mode-conversion. 

Contrasting trends of the impact of surface roughness between SLs and NMs exemplifies the variations in thermal behaviors that can exist between different phononic crystal designs. While sharing conceptual similarity in the physical formation of coherent phonons, the particular design of a phononic crystal has a unique influence on the contributions of both coherent and incoherent phonons \cite{lin2013thermal}. Another example of this fact is the discrepancies in thermal conductivity reduction between the periodic and aperiodic forms for SLs and NMs discussed. Owed to their two-dimensional structure, NMs can facilitate enhanced incoherent phonon transport through the neck regions of membrane material separating the holes \cite{xie2018ultra}. Consequently, the impact of aperiodicity may be low or not reported, unlike the SL which possess a weaker incoherent phonon contribution due the interfaces spanning the entire cross-section of the cross-plane direction, thus making the structure one-dimensional and greatly enhancing incoherent phonon scattering.

\subsection{Limitations of phononic crystals\label{sec:nanocomposite}}

\begin{figure}
    \centering
    \includegraphics[width=\textwidth]{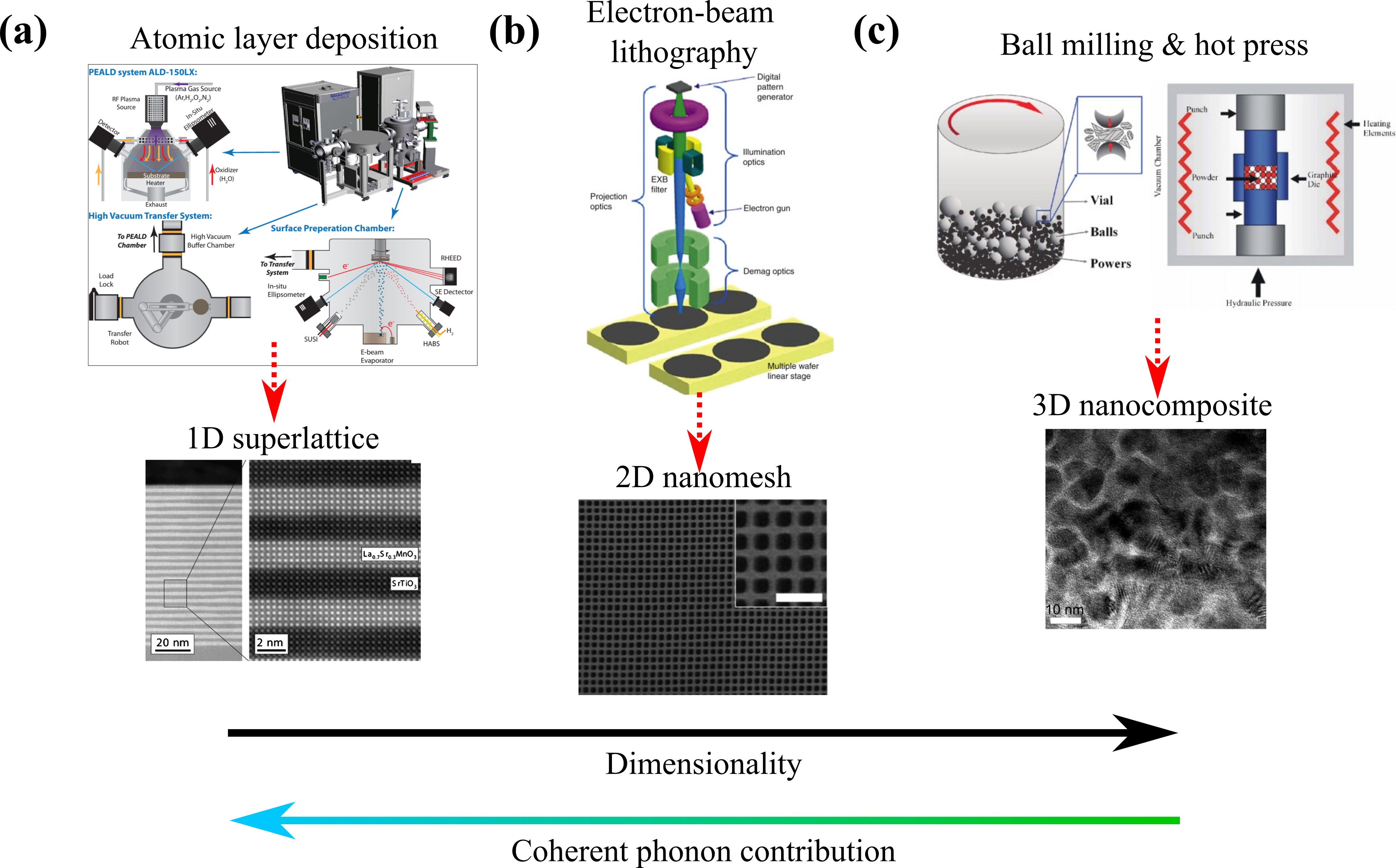}
    \caption{TEM pictures of several nanostructured materials and their associated manufacturing process: (a) 1D periodic superlattice (Image extracted with permission from Ref.~\cite{xiong2016nanostructuration}. Copyright (2016) Elsevier.) fabricated with atomic layer deposition (Schematic illustration extracted with permission from Ref.~\cite{psuAtomicLayer}), (b) 2D periodic nanomesh (image extracted with permission from Ref.~\cite{Lee2017NatureCommu}. Copyright (2017) Springer Nature.) fabricated with electron-beam lithography (Schematic illustration extracted with permission from Ref.~\cite{figueiro2015process}. Copyright (2015) Thiago Rosa Figueiro.), and (c) 3D nanocomposite (Image extracted with permission from Ref.~\cite{wang2008enhanced}. Copyright (2008) AIP Publishing.) fabricated with ball-milling and hot press sintering (Schematic illustrations extracted with permission from Ref.~\cite{li2017processing} and Ref.~\cite{moustafa2011hot}, respectively. Copyright (2017), Science China Press and Springer-Verlag GmbH Germany. Copyright (2011) Scientific Research.). The phononic crystals like superlattices and nanomeshes require expensive and laborious fabrication processes to manufacture the precise nanoscale features, while nanocomposites can be fabricated in bulk processes that can be scaled to mass production more easily.}
    \label{fig:TEM_structure_compile}
\end{figure}

When discussing the heat conduction theories and applications of nanostructured materials, phononic crystals are often placed in sharp relief to nanocomposites \cite{chen2000phonon,dresselhaus2007new,minnich2009bulk,kim2021strategies,xie2023brief}. Nanocomposites are nanoscale composite structures containing features like grain boundaries, embedded nanoparticles, or defects spaced apart by distances less than the anharmonic mean free paths of thermal phonons. Generally, these features are randomly dispersed throughout the nanocomposite and possess disordered geometries and rough surfaces. In contrast, the features in phononic crystals have an ordered character and/or arrangement and there are minimal amounts of defects and imperfections, allowing for the stimulation of wave-like coherent phonon transport. 

Importantly, coherent mode-conversion requires the scattered phonon wave-packets along a specific direction to be in phase with each other \cite{maranets2024prominent,maranets2025phonon} as visualized in Fig.~\ref{fig:coherent_mode-conversion_schematic}. By scattering phonons in many different directions, often in phase-destroying events, rough surfaces and disordered geometries lessen the opportunities for coherent mode-conversion. Consequently, the coherent phonon contribution is negligible or nonexistent in the thermal conductivities of nanocomposites. Furthermore, the impact of rough surfaces and disordered geometries explains why most phononic crystals are low-dimensional as control of the nanostructure to facilitate clean wave interference is easier with fewer dimensions and boundary scattering to stimulate interference is intensified with heightened surface-to-volume ratio \cite{maire2017heat}. 

To achieve the precise material structures necessary for phonon spatial coherence, phononic crystals are manufactured in expensive and laborious nanoscale fabrication processes, whereas nanocomposites can be more easily manufactured in bulk fabrication processes, as visualized in Fig.~\ref{fig:TEM_structure_compile} \cite{minnich2009bulk,li2017processing}. Thus, there is a significant practical cost associated with achieving the unique thermophysical properties associated phonon spatial coherence in phononic crystals such as phonon band gaps, large thermal conductivity anisotropy, and below alloy thermal conductivity through suppression of both particle-like and wave-like phonons.

Developing manufacturing processes to more easily fabricate phononic crystals is necessary for advancing the proliferation of phononic crystal devices. Progress may lie with self-assembled SLs consisting of multiple different structural phases, rather than different chemical compositions with the same crystal structure (requiring atomic layer deposition techniques) as has been conventionally studied e.g. Si/Ge and AlAs/GaAs. A recent molecular dynamics study by Lotfpour et al. \cite{lotfpour2026ultrafast} demonstrates this concept in the self-assembly of 1D multi-phase wurtzite-GaN/zincblende-GaN SLs via ultrafast melt quenching simulation. Similarly, colloidal nanocrystals consisting of inorganic crystal cores and connective organic ligands can self-assemble from solution into 2D and 3D periodic arrangements and possess phononic crystal properties \cite{weidman2015interparticle,poyser2016coherent,sadat2016colloidal,yazdani2019nanocrystal}. 

\section{Temporal coherence theory\label{sec:temporal_coherence_theory}}

Phonon temporal coherence has a rich history of theory and experimental measurements, yet its conceptual understanding is lacking compared to propagons and spatial coherence. In this section, we amend this issue by outlining the most advanced theoretical frameworks quantifying temporal coherence and the pathways for diffusons to conduct heat. Following our discussion of frameworks, we review how phonon temporal coherence manifests in measurable properties and highlight the distinguishing features between propagon-dominated and diffuson-dominated heat conduction.

\subsection{Theoretical frameworks\label{sec:temporal_coherence_theoretical_framework}}

\begin{figure}
    \centering
    \includegraphics[width=\textwidth]{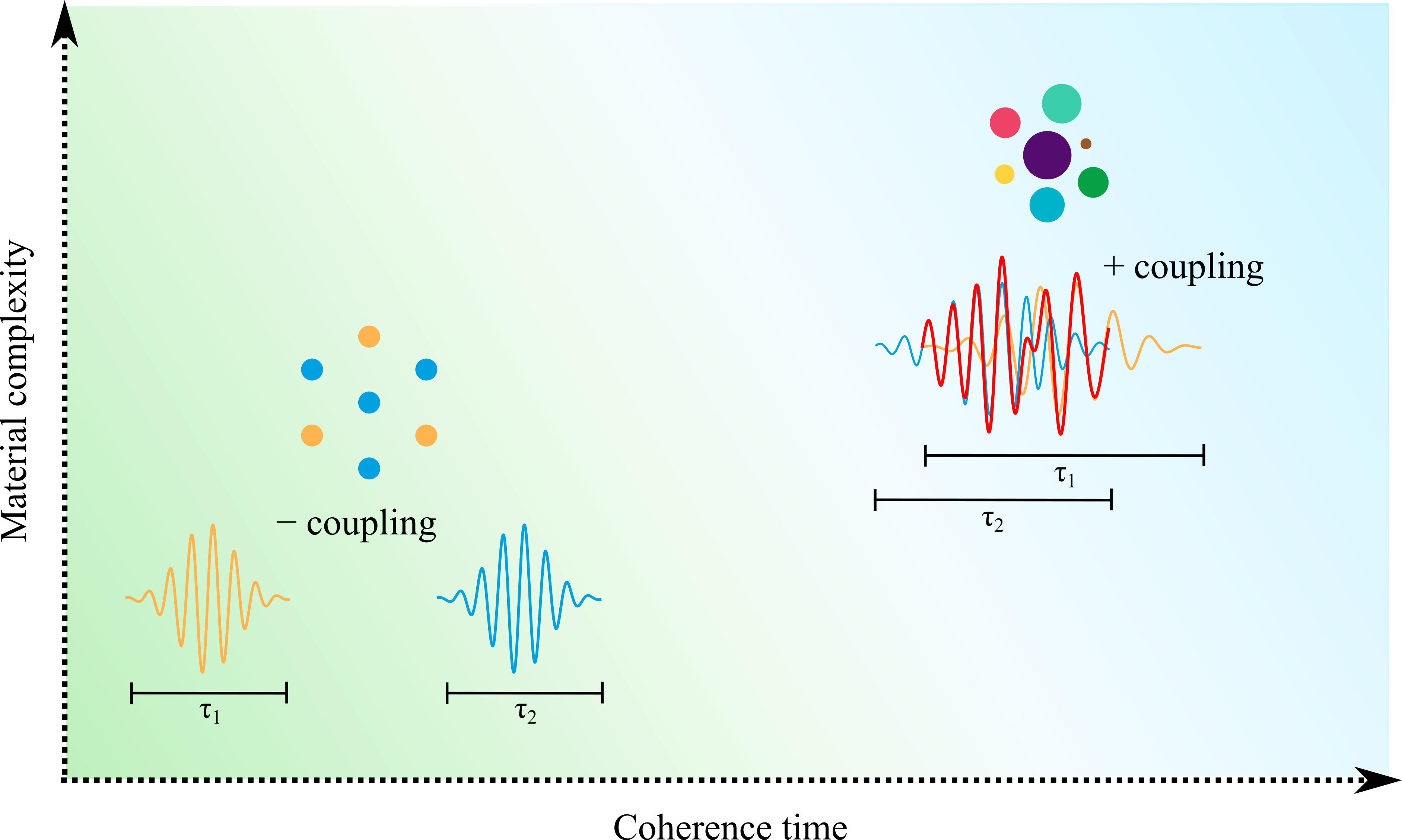}
    \caption{Schematic illustration of phonon temporal coherence. Phonon modes whose temporal phases overlap beyond their individual lifetimes can constructively interfere and conduct heat. This is the primary energy transfer mechanism of diffusons, which become more predominant with higher material complexity. Material complexity can result from: (1) crystal unit cell sophistication, (2) crystal disorder and defects, or (3) larger chemical composition variance.}
    \label{fig:temporal_coherence_visual}
\end{figure}

As discussed in Sec.~\ref{sec:non-propagating_oscillators}, diffuson-type phonons lack periodicity in space, but they do possess periodicity in time, allowing for well-defined phases and frequencies. Consequently, just as propagons can transport like waves when they are close enough to each other in space such that their spatial periodicity i.e. wavelength overlaps and they constructively interfere (spatial coherence), diffusons can exhibit wave-like behaviors for similar events in time (temporal coherence) as visualized in Fig.~\ref{fig:temporal_coherence_visual}. 

The requirement for the spatial coherence length of the propagon to be equal to or exceed the characteristic length in order to transport coherently has been the condition for phonon spatial coherence considered most extensively in existing literature, mainly because it is physically intuitive and consistent with existing optical coherence theory \cite{chen2000phonon,latour2014microscopic}. The larger the coherence length, the more wave periods there are and consequently the more opportunities exist for interference. Additionally, the larger coherence length means the central wavelength of the wave-packet is more well-defined and there is heightened possibility for constructive interference that facilitates coherent mode-conversion, the process in which a particle-like incoherent phonon transforms into a wave-like coherent phonon \cite{maranets2024influence,maranets2024prominent}. These concepts are illustrated in Fig.~\ref{fig:spatial_coherence_length_visualization}.

Unlike spatial coherence, a similar logical reasoning for the determining factor(s) of temporal coherence is lacking due to the presently poor understanding of the physical picture of diffuson thermal transport. This can be attributed to the fact that the interactions of diffusons are primarily occurring in time and so it is harder to visualize the energy transfer processes in comparison to the spatial mechanisms occurring for propagons. As such, there have been several different theories that have been developed to accurately quantify when the diffusons behave like waves and conduct significant heat through this temporal coherence effect. In this section, we review the following three distinct theoretical frameworks to model phonon temporal coherence:
\begin{enumerate}
    \item A wavelet decomposition of the normal mode atomic velocities from molecular dynamics that describes the temporal signature of a phonon into a particle-like lifetime and wave-like coherence time. The coherence time, like the spatial coherence length for propagons, quantifies the extent of temporal correlation between phonon modes and consequently determines whether a phonon behaves like a wave or a particle. The thermal conductivity calculation incorporating temporal coherence effects depends on coherence time.
    
    \item A Wigner transport equation, built from a Wigner phase-space formulation of the dynamics of quantum harmonic oscillators, that quantifies an inter-mode tunneling strength determining the magnitude of heat carried by the correlated or coupled vibrational eigenstates (phonons). The tunneling strength is reflected in a conventional phonon lifetime used to compute thermal conductivity akin to the Peierls-BTE. The relative magnitude of the lifetime dictates a wave-like or particle-like behavior in heat conduction.
    
    \item A Green-Kubo approach that unlike the Peierls-BTE, does not presume the physical character of the phonon modes in the thermal conductivity calculation. The particular technique is a sophistication of the seminal lattice dynamics theory by Allen and Feldman, introducing corrections for anharmonicity, thus properly accounting for both propagon and diffuson contributions to thermal transport.
\end{enumerate}
Though different in their mathematical forms, all approaches reduce to the Peierls-BTE calculation of thermal conductivity for the case of structurally simple crystals where the diffuson presence is minimal. Additionally, they are accurate in diffuson-dominated regimes where the Peierls-BTE typically fails.

\subsubsection{Wavelet transform\label{sec:wavelet}}

We first discuss the molecular dynamics-based wavelet approach, developed by Zhang and co-authors \cite{zhang2021generalized,zhang2022heat}, as it is, in our opinion, a conceptual stepping stone from the conventional quantization of phonon transport in crystals. 

Reviewed in Sec.~\ref{sec:propagating_wave-packets}, the description of phonons as plane waves in crystals stems from a Fourier decomposition of the atomic vibrations that are mathematically modeled as harmonic oscillators. Non-linear interatomic forces associated with anharmonicity results in the plane waves being localized in space and time, leading to the propagating wave-packet characterization discussed heavily throughout this review. Following this understanding, Zhang and co-authors proposed revisiting the initial step in the lattice dynamics analysis where a wavelet decomposition is applied instead of a Fourier one. Where the Fourier transform solely utilizes a delocalized sine wave $e^{-i\omega t}$ as its basis function, the wavelet transform incorporates an envelope expression that turns the sine wave into a wave-packet, also referred to as a wavelet, reflecting the time localization of phonons. In the approach by Zhang et al. \cite{zhang2021generalized}, a Gaussian function is used as the envelope in the following normalized wavelet basis function: 
\begin{equation}
    \psi_{\omega_{\textbf{k}s},\tau_{\textbf{k}s}^{c},t_{0}}(t) = \pi^{-1/4}\Delta_{\textbf{k}s}^{-1/2}e^{[i\omega_{\textbf{k}s}(t-t_{0})]}e^{\left[\frac{-(t-t_{0})^{2}}{2\Delta_{\textbf{k}s}^{2}}\right]}
    \label{eqn:wavelet}
\end{equation}
Eqn.~\ref{eqn:wavelet} is then used in the following integral transform on the normal mode velocities, which are computed from equilibrium molecular dynamics, to obtain the wavelet transform.
\begin{equation}
    \Lambda(\omega_{\textbf{k}s},\tau_{\textbf{k}s}^{c},t_{0}) = \int\psi_{\omega_{\textbf{k}s},\tau_{\textbf{k}s}^{c},t_{0}}(t)\dot{q}_{\textbf{k}s}(t) dt
    \label{eqn:wavelet_transform}
\end{equation}
$\omega_{\textbf{k}s}$ and $\Delta_{\textbf{k}s}$ are the angular frequency and wave-packet duration, respectively of phonon mode $s$ at wavevector \textbf{k} whose time-dependent normal mode velocity is specified by the variable $\dot{q}_{\textbf{k}s}$. $t$ is time while $t_{0}$ is the location of the wave-packet center. The full width at half maximum of the wave-packet, $\tau_{\textbf{k}s}^{c}=2\sqrt{2ln2}\Delta_{\textbf{k}s}$, is known as the coherence time and it measures the degree of phonon temporal coherence in a very similar way to spatial coherence length for the spatial coherence of propagons. An important distinction is that propagons are phonon modes that physically transport in real-space as wave-packets whereas Eqn.~\ref{eqn:wavelet} is simply a mathematical object used to describe the time signal of a phonon, be it a propagon or diffuson, with a characteristic frequency and duration, both of which are captured in a wave-packet form. Following the wavelet transform calculation in Eqn.~\ref{eqn:wavelet_transform}, the phonon number density is computed as:
\begin{equation}
    N(\omega_{\textbf{k}s},\tau_{\textbf{k}s}^{c},t_{0}) = \frac{1}{2}|\Lambda(\omega_{\textbf{k}s},\tau_{\textbf{k}s}^{c},t_{0})|^{2}/\hbar\omega_{\textbf{k}s}
    \label{eqn:phonon_number_density}
\end{equation}

\begin{figure}
    \centering
    \includegraphics[width=\textwidth]{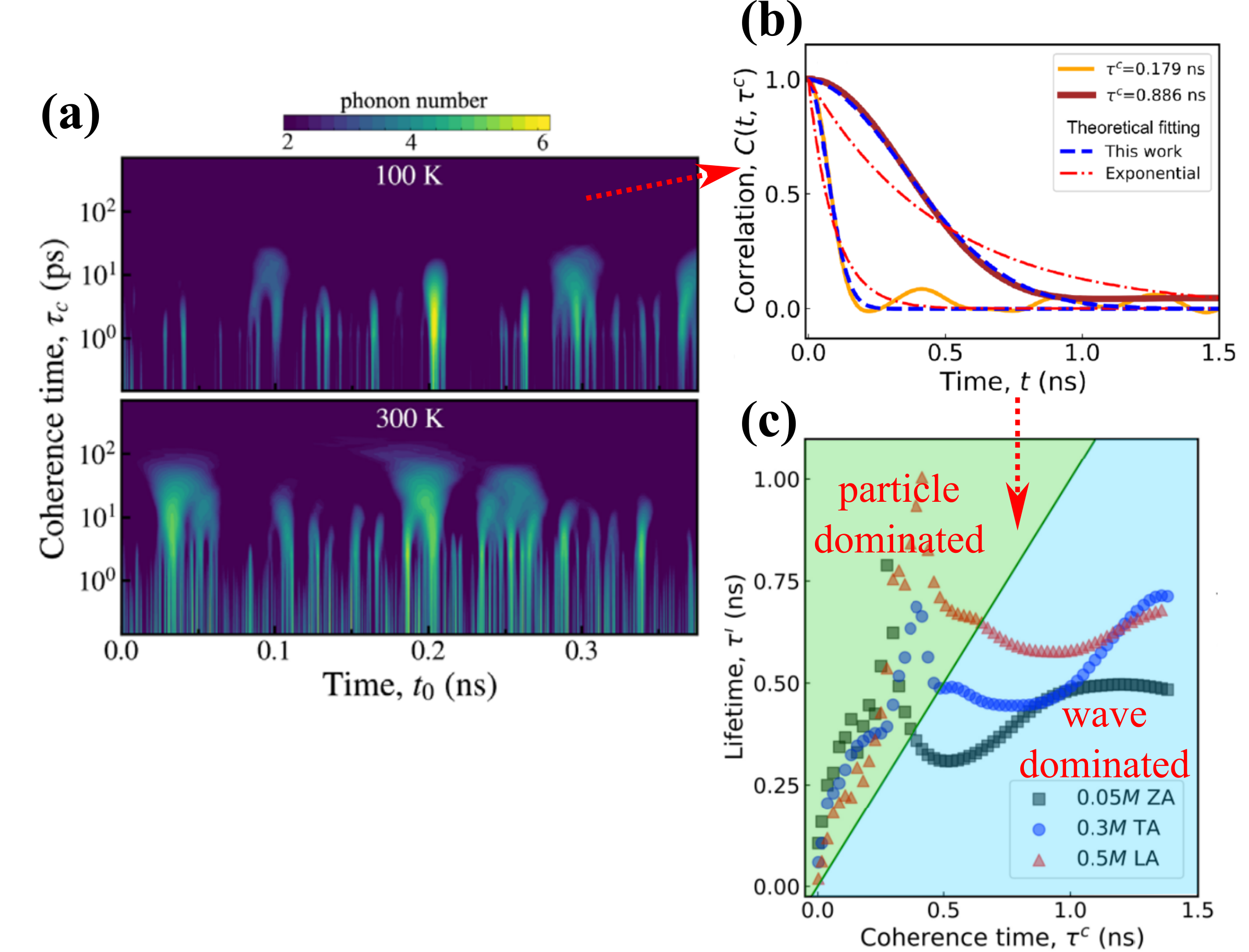}
    \caption{Wavelet transform calculations workflow to quantify phonon temporal coherence. Panel (a) shows the phonon number density heat maps, computed with Eqn.~\ref{eqn:phonon_number_density}, of an optical mode in Tl$_3$VSe$_4$ at 100 K and 300 K. The plots are reprinted with permission from Ref.~\cite{zhang2022heat}. Copyright (2022) from the American Physical Society. Panel (b) shows example results for the subsequent step in the workflow, calculation of the normalized autocorrelation of phonon number, revealing the differences in energetic decay between particle-like phonons with small coherence times $\tau^c$ and wave-like phonons with large coherence times. Lastly, panel (c) shows the distribution of lifetimes $\tau^l$ and coherence times for several modes. A green line corresponding to $\tau^l=\tau^c$ is plotted to show the crossover between particle-like and wave-like contributions to thermal conductivity. The data in panels (b) and (c) are results for suspended single-layer graphene with the plots  reprinted with permission from Ref.~\cite{zhang2021generalized}. Copyright (2021) from the American Physical Society.}
    \label{fig:wavelet_transform_panels}
\end{figure}

Eqn.~\ref{eqn:phonon_number_density} quantifies how the energy of phonon mode (\textbf{k},s) at time $t_{0}$ is distributed in coherence time, effectively indicating which wave-packet sizes are the most prominent for mode (\textbf{k},s). When plotted as a function of both real time and coherence time, one can obtain a heatmap like that shown in Fig.~\ref{fig:wavelet_transform_panels}a which visually indicates the degree to which temporal coherence manifests for a phonon mode. When Eqn.~\ref{eqn:phonon_number_density} is summed over all coherence times, one obtains the time-dependent phonon number $N(\omega_{\textbf{k}s},t_{0})$ which is the total energy of mode (\textbf{k},s) at time $t_{0}$. 

In the case of a Fourier transform in Eqn.~\ref{eqn:wavelet_transform}, the normalized autocorrelation of the phonon number is a simple decaying exponential function whose time constant is the particle-like phonon lifetime $\tau^{l}_{\textbf{k}s}$ that enters into the Peierls-BTE \cite{ladd1986lattice,mcgaughey2004quantitative,turney2009predicting}. When a wavelet transform is used, the autocorrelation function now follows the form:
\begin{equation}
    C(t,\tau_{\textbf{k}s}^{c}) = e^{-t/2\tau^{l}_{\textbf{k}s}}\cdot e^{-4ln2[t^{2}/\tau_{\textbf{k}s}^{c}{}^{2}]}
\end{equation}
where an additional decay term that quantifies the temporal coherence through the phonon coherence time $\tau_{\textbf{k}s}^{c}$ is present \cite{zhang2021generalized}. Based on this new relation, larger values of $\tau_{\textbf{k}s}^{c}$ prolong the phonon decay, creating an increased opportunity for the interference in time of different phonon modes. This difference in this phonon decay is illustrated by the wavelet data and fitting functions plotted in Fig.~\ref{fig:wavelet_transform_panels}b. Increased inter-mode coupling leads to heightened heat conduction and so a thermal conductivity calculation that accounts for this phonon temporal coherence effect increases with coherence time.
\begin{equation}
    \kappa_{p+w} = \frac{1}{3}\sum_{\alpha}\sum_{\textbf{k}s}C_{\textbf{k}s}v^{2}_{\textbf{k}s,\alpha}\sqrt{\frac{\pi}{4ln2}}\tau_{\textbf{k}s}^{c}\exp{\left[\frac{\tau_{\textbf{k}s}^{c}{}^{2}}{128ln2\tau_{\textbf{k}s}^{l}{}^{2}}\right]}
    \label{eqn:kappa_coherence_time}
\end{equation}
$\kappa_{p+w}$ is the total thermal conductivity that incorporates the particle-like contribution ($p$) and wave-like contribution ($w$) through inclusion of both phonon lifetime and coherence time terms \cite{zhang2022heat}. When the two properties are equivalent i.e. $\tau_{\textbf{k}s}^{l} = \tau_{\textbf{k}s}^{c}$, Eqn.~\ref{eqn:kappa_coherence_time} reduces to the Peierls-BTE form. This and the wavelet-computed autocorrelation function of the time-dependent phonon number indicates that coherence time must exceed lifetime in order to have significant wave-like contribution (see Fig.~\ref{fig:wavelet_transform_panels}c). This condition is very similar to the spatial coherence length criterion for phonon spatial coherence of propagons discussed in Sec.~\ref{sec:spatial_coherence_theoretical_framework}. Both coherence time and spatial coherence length quantify the sizes of the phonon wave-packets in their respective domains. Like the characteristic length which specifies the spatial distance between geometric scattering events, lifetime measures the time ``distance'' between phonon-phonon scattering events or more broadly, interactions that destroy the phases and modifies energies of the phonons. Diffusons can exchange energy with each other and facilitate enhanced heat conduction through constructive interference in time when the wave-packet extends beyond the lifetime. By capturing the coherence mechanisms of diffusons in time, the wavelet transform approach is shown to agree well with experimental measurements and molecular dynamics data for systems where diffusons are prominent \cite{zhang2022heat,zhang2022coherence,zhang2023assessing}. 

Furthermore, by quantifying distributions of phonon lifetimes and coherence times, the wavelet transform method reveals that the heat transfer pathway of a phonon is not fixed. If this were so, there would be correspondence of a single lifetime and single coherence time to a single mode. Instead, the lifetime and coherence time distributions such as that shown in Fig.~\ref{fig:wavelet_transform_panels}c indicate that the mechanism of energy transport for a phonon mode changes in time dynamically, likely due to local non-equilibrium fluctuations that always exist at finite temperatures.

\subsubsection{Wigner transport equation\label{sec:WTE}}

The Wigner transport equation (WTE), developed by Simoncelli and co-authors \cite{simoncelli2019unified,simoncelli2022wigner,simoncelli2023thermal}, is an analytical model for phonon heat conduction akin to the Peierls-BTE in that both methods relate thermal conductivity to the changes in a scalar distribution function describing the spread of phonon energy across real-space, phase-space, and time. 

The primary distinction between the two methods is that the Peierls-BTE predefines the phonons as particles such that the distribution function is a literal measure of the population of phonon particles whereas the WTE makes no assumption of phonon character, ultimately enabling the distribution function to encompass the energetic contributions of individual vibrational eigenstates, the particle-like phonons or propagons, and the correlations between pairs of eigenstates, the wave-like phonon behavior characterizing the temporal coherence effect of diffusons. The WTE achieves this through a second-quantization formalism for quantum harmonic oscillators that allows calculation of the impact of both wave-like and particle-like interactions on the overall evolution of vibrational energy in a system. A Wigner-Weyl transform is then used to describe these dynamics through phase-space functions which can subsequently be used to construct a transport equation in the language of classical mechanics similar to the Peierls-BTE. 

A concise derivation of the WTE can be found in Ref.~\cite{simoncelli2019unified} and a more in-depth discussion of the theoretical framework is made in Ref.~\cite{simoncelli2022wigner}. Here, we summarize some key aspects of model, highlighting the concepts that make the WTE a generalization of the Peierls-BTE and improving our understanding of phonon temporal coherence in the process.

We start by recalling the form of the Peierls-BTE \cite{peierls1929kinetischen,ziman2001electrons,lindsay2016first,sparavigna2016boltzmann}.
\begin{equation}
    \frac{\partial\Psi_{\textbf{k}s}}{\partial t}+\mathbf{v}_{\textbf{k}s}\cdot\nabla_{\mathbf{R}}\Psi_{\textbf{k}s} = \left(\frac{\partial\Psi_{\textbf{k}s}}{\partial t}\right)_{\text{scattering}}
    \label{eqn:BTE}
\end{equation}
$\Psi_{\textbf{k}s}$ is the probability distribution for the population of phonon particles of wavevector $\textbf{k}$ and mode $s$ at locations in real-space and time. The left-hand side of Eqn.~\ref{eqn:BTE} is the material derivative of $\Psi_{\textbf{k}s}$ while the right-hand side is the rate of change due to phonon scattering which involves both interactions with other phonons (intrinsic scattering) and interactions with the environment (extrinsic scattering). The WTE has a qualitatively similar structure to the Peierls-BTE with several modifications to quantify phonon temporal coherence.
\begin{equation}
\begin{split}
    \frac{\partial\Psi_{\textbf{k},s,s'}}{\partial t}+i[\omega_{\textbf{ks}}\Psi_{\textbf{k},s,s'}-\Psi_{\textbf{k},s,s'}\omega_{\textbf{k}s'}] 
   \\ +\frac{1}{2}\{\sum_{s''}\mathbf{v}_{\textbf{k},s,s''}\cdot\nabla_{\mathbf{R}}\Psi_{\textbf{k},s'',s'} + \sum_{s''}\nabla_{\mathbf{R}}\Psi_{\textbf{k},s,s''}\cdot\mathbf{v}_{\textbf{k},s'',s'} \}  \\ = \left(\frac{\partial\Psi_{\textbf{k},s,s'}}{\partial t}\right)_{\text{scattering}}
    \label{eqn:WTE}
\end{split}
\end{equation}
$\Psi_{\textbf{k},s,s'}$ and $\mathbf{v}_{\textbf{k},s,s'}$ are general forms of the phonon population and group velocity, referred to as the Wigner distribution and velocity operator, respectively. One can see that Eqn.~\ref{eqn:WTE} is essentially the same as Eqn.~\ref{eqn:BTE} but with a modified expression for the directional derivative of the phonon distribution function and an additional correction term. When only the particle-like phonons, represented by the eigenstates $s=s'$, are considered, Eqn.~\ref{eqn:WTE} reduces to Eqn.~\ref{eqn:BTE}. The scattering operator, sometimes referred to as the collision term, on the right-hand side of the transport equations follows a similar trend. Not shown here, the linearized form of the Wigner scattering operator matches the scattering operator in the linearized BTE when only the $s=s'$ elements are considered \cite{simoncelli2022wigner}. These relationships reveal that the scattering of the two phonon types, $s=s'$ (propagons) and $s\neq s'$ (diffusons), are decoupled from each other, meaning the energies of the wave coherences are not affected by the energies of the particles and vice versa. 

Before continuing, we emphasize several points. Again, we note that scattering broadly refers to interactions where the phases and energies of the phonons are not preserved. Additionally, while the propagons are the phonon modes that physically transport in real-space like particles, the diffusons are not physically wave-like, but have a wave character in time due to their well-defined phases and frequencies. Diffusons can conduct heat through constructive interference in time, which is quantified in the WTE. 

Lastly, the BTE and WTE align well due to sharing many similarities in their representations of phonons \cite{simoncelli2022wigner}. In both frameworks, the phonon frequencies and polarizations are taken from the eigenvalues and eigenvectors, respectively of a dynamical matrix determined from the force constants and crystal structure. Furthermore, phonon scattering is quantified through linewidths $\Gamma_{\textbf{k}s}$ determined from Fermi's Golden Rule and the phonons are well-defined quasiparticle excitations where $\hbar\omega_{\textbf{k}s} > \hbar\Gamma_{\textbf{k}s}$.

Following Eqn.~\ref{eqn:WTE}, the vibrational energy field and heat flux as functions of $\Psi_{\textbf{k},s,s'}$ and $\mathbf{v}_{\textbf{k},s,s'}$ are derived to obtain a solution for the steady-state thermal conductivity which ends up being a summation of the two heat conduction pathways $\kappa_{Tot} = \kappa_{P} + \kappa_{C}$: particle ($P$) transport of propagons and wave coherences ($C$) of diffusons. The works of Simoncelli and co-authors and others show many examples of the $\kappa_{Tot}$ computed via the WTE matching well with experimental measurements for amorphous materials and structurally complex crystals, cases where the Peierls-BTE, which only calculates the particle contribution $\kappa_{P}$, is typically inaccurate. 

\begin{figure}
    \centering
    \includegraphics[width=\textwidth]{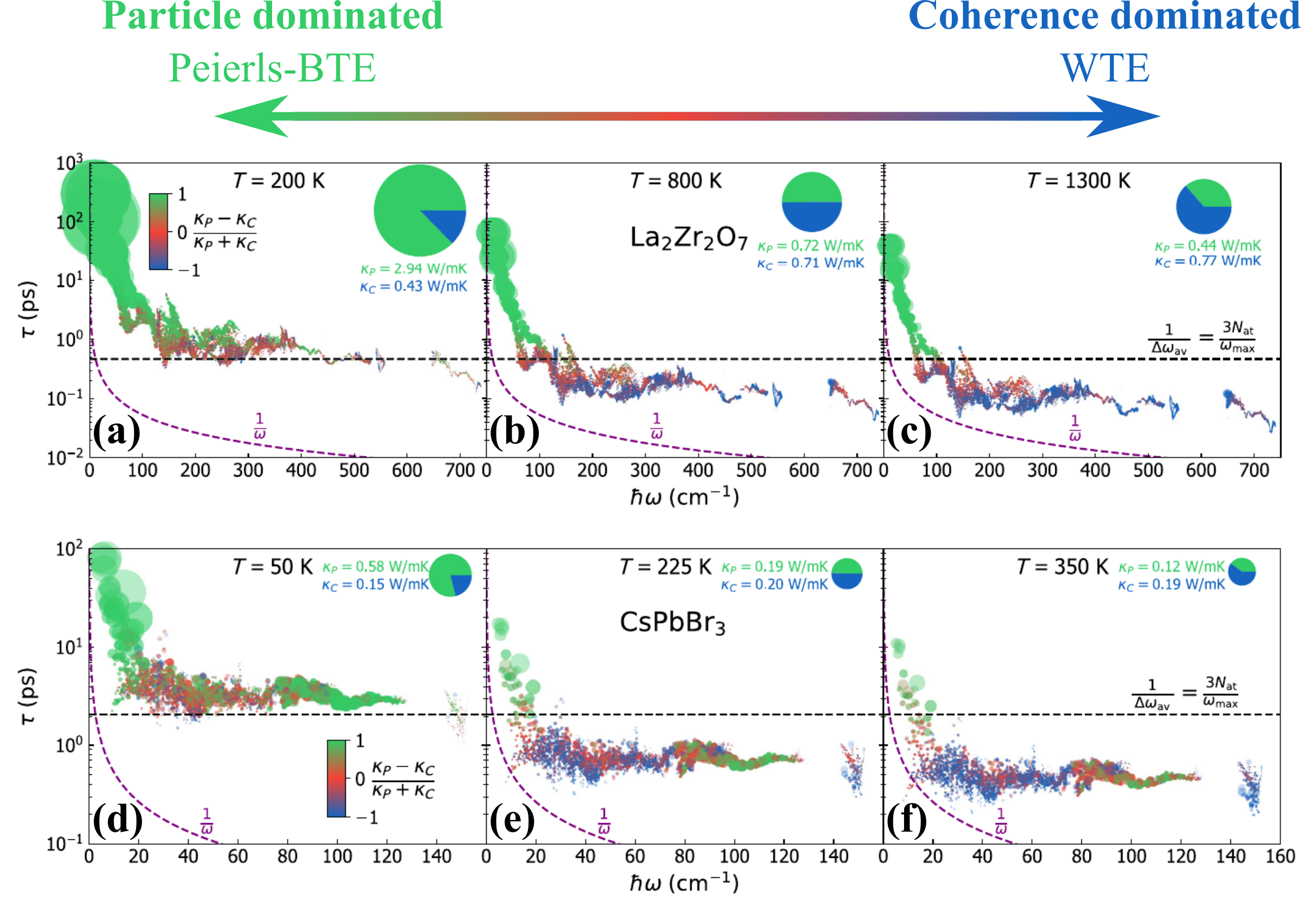}
    \caption{Phonon lifetimes and modal thermal conductivity computed via WTE for La$_2$Zr$_2$O$_7$ and CsPbBr$_3$. Panels (a), (b), and (c) show the quantities as a function of phonon mode energy $\hbar\omega$ for La$_2$Zr$_2$O$_7$ at temperatures $T$ 200 K, 800 K, and 1300 K, respectively. Panels (d), (e), and (f) show the quantities for CsPbBr$_3$ at temperatures $T$ 50 K, 225 K, and 350 K, respectively. The area of each circle is proportional to the total modal thermal conductivity and the color of each circle reflects whether the modes contributes more through the particle-like transport of propagons $\kappa_P$ or through the wave-like temporal coherence of diffusons $\kappa_C$. The pie charts in the top right of each panel have an area proportional to the total thermal conductivity and the slices delineate the relative contributions of the particle and wave pathways. The dashed black horizontal line corresponds to the phonon lifetime equal to the inverse of the average phonon inter-band spacing $\tau = [\Delta\omega_{avg}]^{-1}$ as defined in Eqn.~\ref{eqn:wigner_limit}. The dashed purple hyperbola corresponds to the Ioffe-Regel limit in time $\tau=1/\omega$. The Wigner framework is applicable to phonon modes with lifetimes greater than the Ioffe-Regel limit. The plots are reprinted with permission from Ref.~\cite{simoncelli2022wigner}. Copyright (2022) from the American Physical Society.}
    \label{fig:Wigner_plots}
\end{figure}

The Wigner formalism also provides a metric for evaluating whether a phonon contributes more to the particle-like conductivity or more to the wave-like thermal conductivity. In Ref.~\cite{simoncelli2022wigner}, Simoncelli and co-authors highlight the following relationship:
\begin{equation}
    \frac{\kappa_{C,\textbf{k}s}}{\kappa_{P,\textbf{k}s}} \simeq \frac{\Gamma_{\textbf{k}s}}{\Delta\omega_{avg}} = \frac{[\Delta\omega_{avg}]^{-1}}{\tau_{\textbf{k}s}}
    \label{eqn:wigner_limit}
\end{equation}
where $\Delta\omega_{avg} = \frac{\omega_{max}}{3N_{at}}$ is the average phonon inter-band spacing defined as the ratio between the maximum phonon frequency and the total number of phonon bands in the dispersion relation (three times the number of atoms in the primitive unit cell $N_{at}$). Eqn.~\ref{eqn:wigner_limit} shows that when the linewidth $\Gamma_{\textbf{k}s}$ exceeds the average spacing, a phonon carries more heat through the temporal coherence mechanism than through particle-like propagation. Eqn.~\ref{eqn:wigner_limit} also indicates that the transition between particle-like and wave-like heat conduction is not discrete, but rather occurs continuously as seen by the color scales in Fig.~\ref{fig:Wigner_plots}. Thus, in our discussion of propagons and diffusons throughout this review, we stress to the reader that not all phonon modes can be classified as distinctly a propagon or distinctly a diffuson. However, distinguishing between the two phonon characters is critical to differentiating spatial and temporal coherence effects. Spatial coherence is a wave effect in real-space for propagons, while temporal coherence is a wave effect in time for diffusons. Fundamentally, a phonon mode's heat transfer mechanism generally differs depending on its physical character.

Qualitatively, the criterion for the regime of thermal transport established in Eqn.~\ref{eqn:wigner_limit} is similar to that for the wavelet approach in Sec.~\ref{sec:wavelet} in that it is an inequality between the central parameter in the respective thermal conductivity calculation (phonon lifetime $\tau_{\textbf{k}s}$ for the WTE and phonon coherence time $\tau_{\textbf{k}s}^{c}$ for the wavelet approach) and a fundamental time scale of the material.

\subsubsection{Quasi-harmonic Green-Kubo\label{sec:Isaeva_GK}}

\begin{figure}
    \centering
    \includegraphics[width=\textwidth]{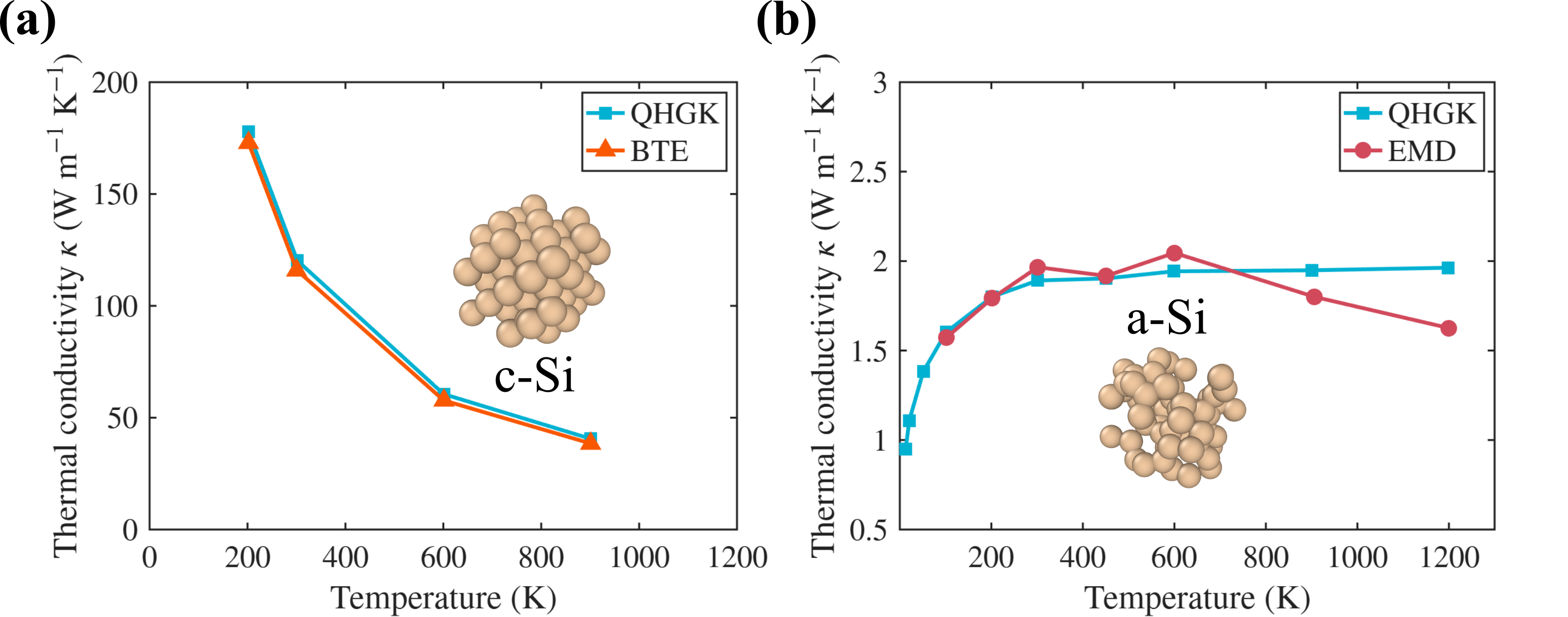}
    \caption{Quasi-harmonic Green-Kubo (QHGK) thermal conductivity plots. Panel (a) compares QHGK to values calculated with Peierls-BTE for crystalline silicon (c-Si). Panel (b) compares QHGK to values calculated from equilibrium molecular dynamics (EMD) for amorphous silicon (a-Si). The data is adapted with permission from Ref.~\cite{isaeva2019modeling}. Copyright (2019) Springer Nature.}
    \label{fig:Isaeva_plots}
\end{figure}

The final distinct method for quantifying phonon temporal coherence that we discuss in this review is the quasi-harmonic Green-Kubo (QHGK) analysis developed by Isaeva and co-authors \cite{isaeva2019modeling}. This approach is a sophistication of the seminal theory by Allen and Feldman \cite{allen1989thermal,allen1993thermal} for quantifying phonon heat conduction in disordered solids like glasses. In that theory, thermal conductivity is evaluated via the Green-Kubo formalism \cite{green1954markoff,kubo1957statistical} and Hardy's harmonic heat flux operator \cite{hardy1963energy}. In the Green-Kubo formalism, thermal conductivity is a function of the autocorrelation of heat flux:
\begin{equation}
    \kappa_{\alpha\beta} = \frac{1}{Vk_{B}T^{2}}\int_{0}^{\infty}\langle J_{\alpha}(t)J_{\beta}(0)\rangle dt
    \label{eqn:GK-kappa}
\end{equation}
where $V$ is the material volume, $k_{B}$ is the Boltzmann constant, $T$ is temperature and $J_{\alpha}$ is the heat flux component in direction $\alpha$. Hardy expressed the harmonic heat flux vector as a function of the atomic energies and positions \cite{hardy1963energy}:
\begin{equation}
    \boldsymbol{J} = \sum_{i}\left(\frac{1}{2}mv_{i}^{2} + U_{i}\right)\boldsymbol{v}_{i} + \sum_{i}\sum_{j}\left(\frac{\partial U_{j}}{\partial r_{i}}\cdot\boldsymbol{v}_{i}\right)\boldsymbol{r}_{ij}
    \label{eqn:hardy-heatflux}
\end{equation}
where $m_{i}$, $r_{i}$, $v_{i}$, and $U_{i}$ are the mass, position, velocity, and potential energy of atom $i$. The first term in Eqn.~\ref{eqn:hardy-heatflux} is associated with the energy transfer stemming from the motion of individual atoms while the second term corresponds to energy transfer arising from the interactions between atoms \cite{boone2019heat,maranets2023ballistic}. In a solid, we can associate atomic motion directly to phonons since the atoms are coupled oscillators \cite{georgi1993physics}. 

Unlike the Peierls-BTE, where the heat flux (and consequently thermal conductivity) is explicitly a function of the phonon population distribution function, the mathematical forms of Eqn.~\ref{eqn:GK-kappa} and Eqn.~\ref{eqn:hardy-heatflux} do not presume the physical character of the phonons and so can properly account for the diffuson contribution to thermal conductivity. Specifically, the autocorrelation function in Eqn.~\ref{eqn:GK-kappa} exactly captures the correlations of thermal vibrations that enable diffusons to conduct heat.

The Allen-Feldman theory was introduced as a harmonic lattice dynamics calculation of supercells of amorphous solids where anharmonic scattering processes are neglected, assuming thermal transport is wholly due to the coupling interactions of diffusons (temporal coherence). As a consequence, the theory fails in accurately modeling materials featuring high anharmonicity and/or significant propagon contributions e.g. crystals. 

The primary achievement of the QHGK analysis is the inclusion of anharmonic scattering effects to the Allen-Feldman theory by factoring in the normal-mode lifetimes \cite{isaeva2019modeling}. The lifetimes are obtained from linewidths calculated from Fermi's Golden Rule as is conventional for lattice dynamics calculations. At the completion of the QHGK derivation, Isaeva and co-authors achieve a unified result similar to the WTE framework \cite{simoncelli2019unified} in that the total lattice thermal conductivity is decomposed into the particle-like ($s=s'$) and wave-like temporal coherence ($s\neq s'$) components. The QHGK technique was found to be in good agreement with equilibrium molecular dynamics simulations (which naturally incorporates all possible phonon interactions and atomic-scale features \cite{mcgaughey2006phonon,xu2023quantifying}) for both crystals and amorphous materials \cite{isaeva2019modeling,barbalinardo2020efficient} as seen in Fig.~\ref{fig:Isaeva_plots}.

\subsection{Signatures of temporal coherence\label{sec:signatures_of_temporal}}

In this section, we discuss the unique properties of phonon temporal coherence and its observables.

\subsubsection{Thermal conductivity dependencies\label{sec:kappa_dependence_temporal}}
\begin{figure}
    \centering
    \includegraphics[width=\textwidth]{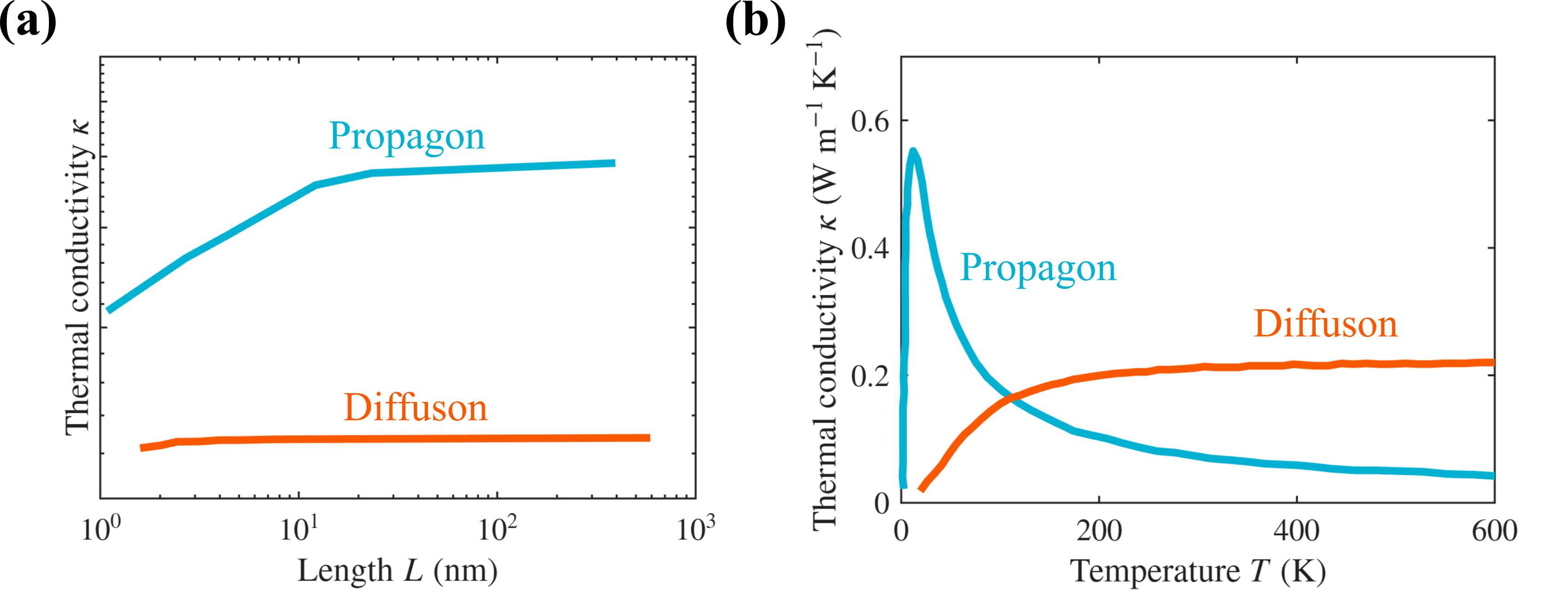}
    \caption{Differences in thermal conductivity dependencies for propagons and diffusons. Panel (a) shows the contrast in temperature. Panel (b) shows the contrast in device length or system size. The example data in the plots are adapted with permission from Ref.~\cite{larkin2014thermal} and Ref.~\cite{bernges2023analytical}. Copyright (2014) from the American Physical Society. Copyright (2023) Elsevier.}
    \label{fig:length_temperature_propagon_diffuson}
\end{figure}

Macroscopically, phonon temporal coherence is characterized by thermal conductivity values which deviate from the standard predictions of the Peierls-BTE \cite{zhou2020thermal}. Moreover, deviations in the dependencies of thermal conductivity, as shown in Fig.~\ref{fig:length_temperature_propagon_diffuson}, further evidence that the physical process of heat conduction for diffusons is different from that of particle-like propagons. The key differences to mention are: (1) a lack of observed size effects and (2) a contrasting trend with temperature. 

Regarding size effects, because diffusons are physically non-propagating, they cannot facilitate any significant variation in their heat carried with device length in the way propagons do \cite{li2015phonon}. Diffusons do not have any intrinsic ballistic transport behavior like propagons. Instead, they interfere with the local disorder and become insensitive to the size of the device \cite{bodapati2006vibrations,zhang2022coherence,maranets2025phonon}. This results in an apparent diffuse transmission of heat.

For temperature, the diffuson contribution to thermal conductivity notably increases with temperature and eventually saturates at high-temperature \cite{auerbach1984universal}. The initial increase is due to larger linewidths heightening the coupling between diffusons while the saturation corresponds to this effect eventually reaching a diminishing point. The temperature-dependence of diffuson thermal conductivity strongly deviates from the propagon thermal conductivity which scales with the inverse of temperature to some power as determined by extent of anharmonic phonon scattering \cite{ma2020first}. A consistent $1/T$ dependence indicates the predominance of traditional three-phonon scattering while deviations correspond to prominent four-phonon scattering \cite{feng2017four}. In fact, consideration of higher-order phonon scattering can be critical to accurate thermal conductivity modeling of materials which exhibit significant phonon temporal coherence \cite{wei2026phonon}. With the introduction of four-phonon effects, phonon lifetimes can drop below the Wigner limit (see Eqn.~\ref{eqn:wigner_limit}) and contribute more heat through the coherence mechanism than the particle transport pathway, thus increasing thermal conductivity.

Other signatures of temporal coherence are revealed in analyses of microscopic phonon properties. The aforementioned techniques in Sec.~\ref{sec:temporal_coherence_theoretical_framework} can elucidate the thermal conductivity contributions of the particle and coherence pathways of propagons and diffusons, respectively. These quantities can also be estimated analytically \cite{agne2018minimum,bernges2023analytical} and the total thermal conductivity contributions of individual phonon modes, regardless of their physical character, can be assessed by applying the Green-Kubo calculation in Eqn.~\ref{eqn:GK-kappa} on a modal basis in equilibrium molecular dynamics simulations \cite{lv2016direct}. The wavelet transform \cite{zhang2021generalized} and WTE \cite{simoncelli2022wigner} methods specifically provide criteria for distinguishing whether an individual phonon mode will contribute more to the particle or coherence components of thermal conductivity.

\subsubsection{Distinguishing phonon physical character\label{sec:phonon_character}}

\begin{figure}
    \centering
    \includegraphics[width=\textwidth]{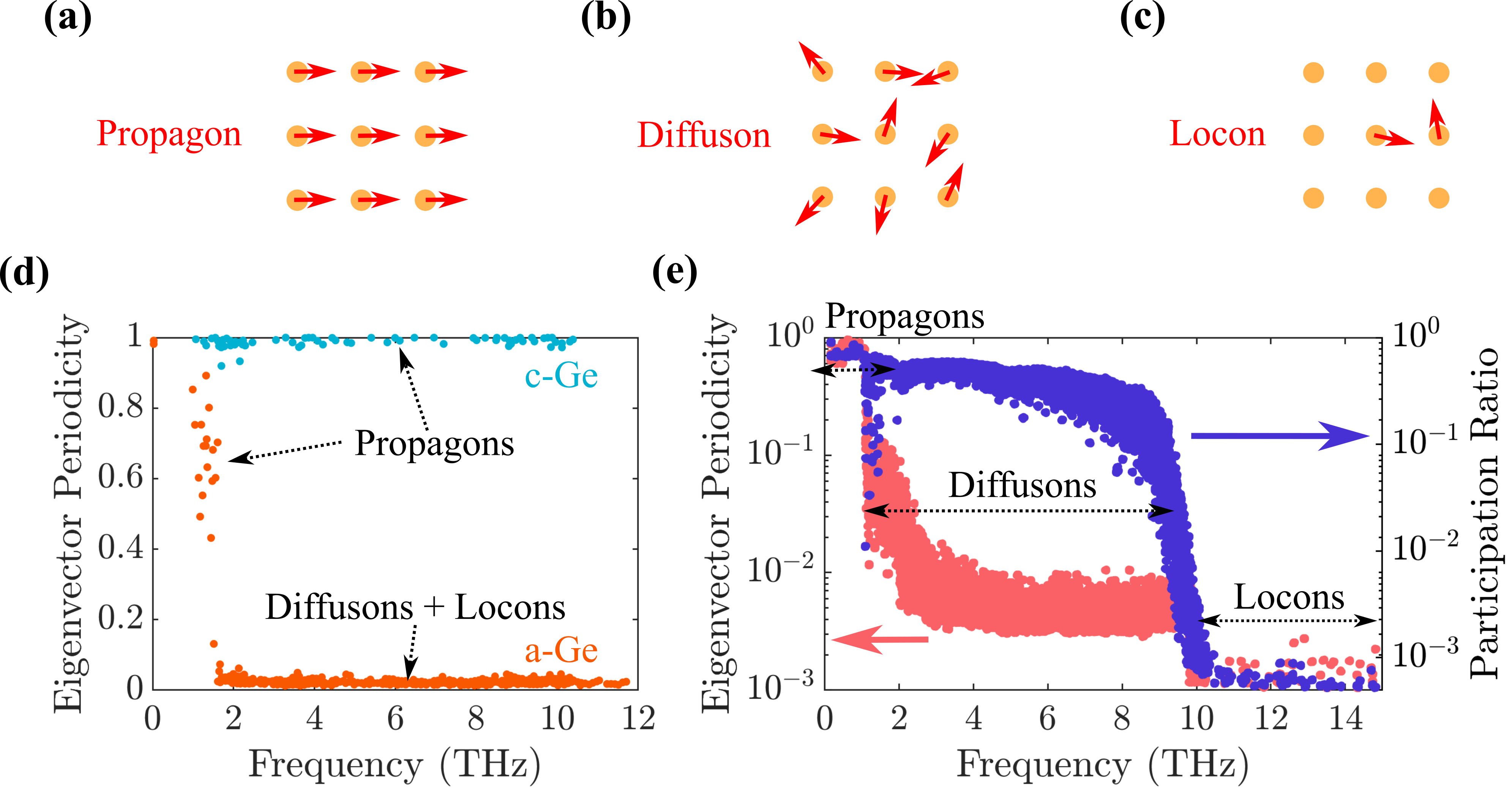}
    \caption{Schematic illustrations of the normal mode velocity fields for propagons (a), diffusons (b), and locons (c). Panel (d) shows a comparison of eigenvector periodicity across the phonon spectrum for crystalline germanium (c-Ge) and amorphous germanium (a-Ge), indicating the contrast between propagons and diffusons. Panel (e) shows a comparison of eigenvector periodicity and participation ratio for phonons in a-Ge, indicating how participation ratio can elucidate locons, but does not clearly distinguish between propagons and diffusons unlike eigenvector periodicity. The data in panels (d) and (e) is adapted with permission from Ref.~\cite{seyf2016method}. Copyright (2016) AIP Publishing.}
    \label{fig:EP_PR_plot}
\end{figure}

Besides thermal conductivity modeling, several analyses have been developed systematically classify the spectrum of phonon modes for a given material into the propagon and diffuson descriptions \cite{deangelis2019thermal}. When a large fraction of modes in a material possess a diffuson character, heat conduction will most often have a significant contribution from the phonon temporal coherence effect. In Figs.~\ref{fig:dispersions_all}c-\ref{fig:dispersions_all}d, we applied one such analysis of computing the dynamic structure factor for an amorphous solid and spectral energy density for a complex crystal which visually reveals for which frequencies is a well-defined phonon dispersion nonexistent and the propagon description invalid. The transition frequency region in this analysis is commonly known as the Ioffe-Regel crossover \cite{taraskin2000ioffe}, however, it should be reiterated that the crossover is not sharp and does not predetermine the heat conduction pathway. As discussed in Sec.~\ref{sec:WTE} and exemplified by several data points in Figs.~\ref{fig:Wigner_plots}d-\ref{fig:Wigner_plots}f, propagons and diffusons can sometimes conduct heat through the nominally opposing transport mechanism. 

More quantitative assessments of phonon character involve calculation of participation ratio which can be used to elucidate locons which are diffuson-type phonon modes that are spatially localized to one or a few atoms \cite{biswas1988vibrational} unlike propagons and diffusons that are spatially extended across multiple atoms. Propagons, diffusons, and locons can be distinguished more generally through eigenvector periodicity analysis \cite{seyf2016method} which assesses how spatially periodic a phonon's mode shape, as revealed by its normal mode's velocity field, is. We visualize the technical differences between propagons, diffusons, and locons in Figs.~\ref{fig:EP_PR_plot}a-\ref{fig:EP_PR_plot}c. Calculations of eigenvector periodicity and participation ratio in Fig.~\ref{fig:EP_PR_plot}d and Fig.~\ref{fig:EP_PR_plot}e show how the two quantities differ in distinguishing between propagons, diffusons, and locons.

\subsection{Manipulating temporal coherence\label{sec:manipulate_temporal_coherence}}

Unlike phonon spatial coherence, the temporal coherence effect does not effectuate the phonons to physically transport in real-space like waves. Rather, constructive wave interference in time serves as a suitable and rigorous explanation for how non-propagating vibrational modes, diffusons, can conduct heat. Being purely temporal, it is presently unclear how to artificially structure a material to significantly manipulate the thermal transport by these diffusons like it is for propagons whose heat can be controlled via size effects. In this section, we discuss the current research into manipulating temporal coherence and the thermal conductivity contributions of diffusons.

\subsubsection{Propagon vs. diffuson competition\label{sec:propagon_vs_diffuson_competition}}

\begin{figure}
    \centering
    \includegraphics[width=\textwidth]{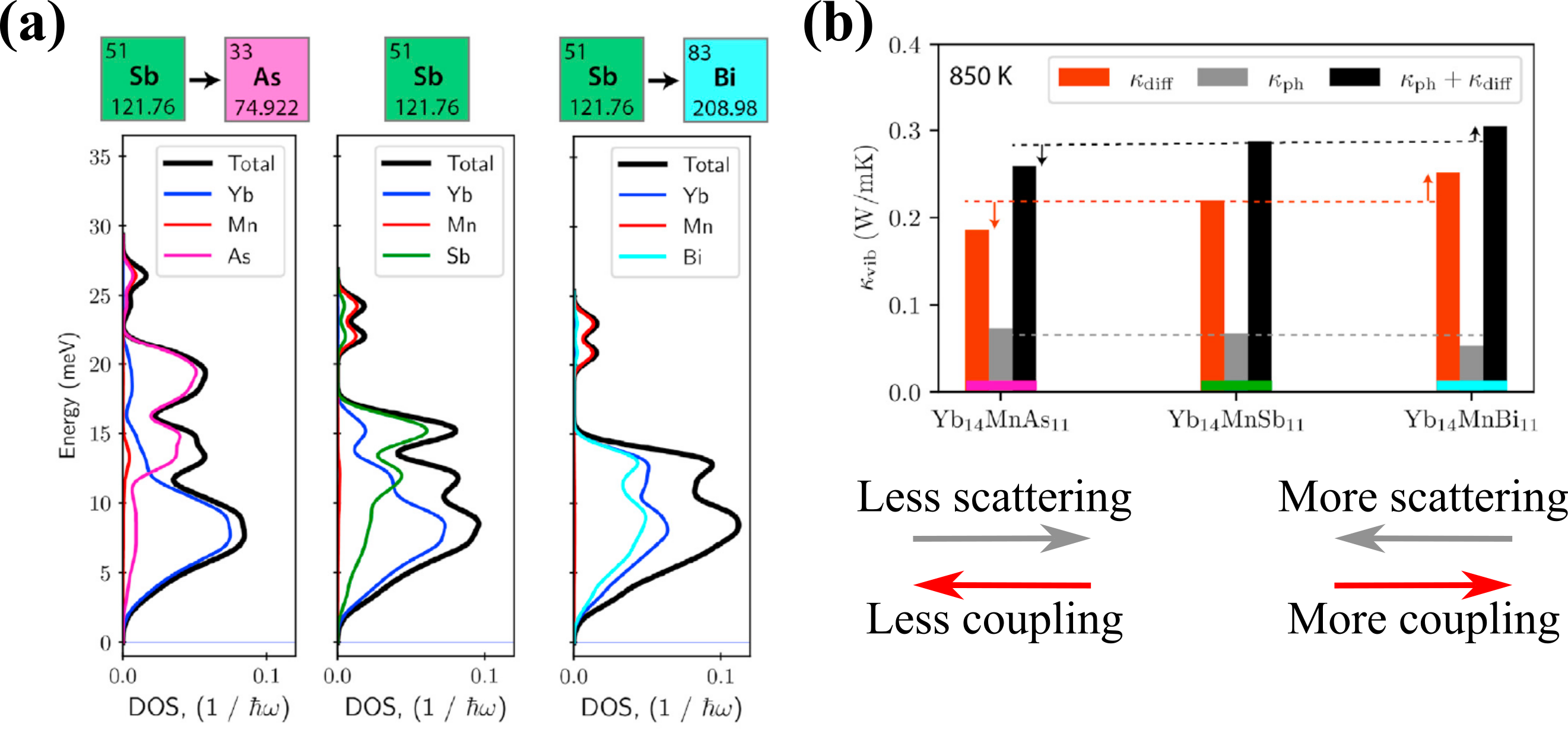}
    \caption{Influence of Sb-site mass on phonons and thermal conductivity in Yb$_{14}$MnSb$_11$. Panel (a) shows the phonon density of states decomposed for each atomic species. Panel (b) reveals the effect of changing Sb-site mass on the propagon and diffuson contributions in addition to the total thermal conductivity. The plots are reprinted with permission from Ref.~\cite{hanus2021uncovering}. Copyright (2021) Elsevier.}
    \label{fig:Hanus_data_panels}
\end{figure}

Hanus et al. \cite{hanus2021uncovering} investigated via first-principles calculations and experimental measurement the influence of atomic mass on the thermal conductivity of Yb$_{14}$MnSb$_{11}$, a complex crystal whose heat conduction is dominated by diffusons. We present their key results in Fig.~\ref{fig:Hanus_data_panels}.  When the mass of the Sb-site was lowered, the diffuson contribution decreased while the propagon contribution increased. Overall, there was a minor reduction in thermal conductivity. Increasing the mass of the Sb-site showed opposing behavior with an increase in the diffuson contribution, a decrease in the propagon contribution, and an overall minor increase in thermal conductivity as seen in Fig.~\ref{fig:Hanus_data_panels}b. Changing the mass of the Sb-site affected the linewidths and spacing between the branches in the dispersion relation which can be assessed by atomic species decomposition of the vibrational density of states as shown in Fig.~\ref{fig:Hanus_data_panels}a. Such changes to the phonon band structure alter the degree of overlap/coupling between different phonon modes and consequently the amount of heat carried via the temporal coherence effect, which is reflected in the diffuson contribution to thermal conductivity. 

The fact that both the propagon and diffuson contributions change simultaneously and in opposing directions, ultimately leading to only slight changes in the total thermal conductivity, highlights a fundamental challenge in manipulating phonon temporal coherence towards desired thermophysical properties. The mechanism to decrease the diffuson contribution found by Hanus et al. \cite{hanus2021uncovering}, energetic isolation of different phonon branches whereby the spacings between branches is increased and the linewidths are reduced, directly raises the propagon contribution by decreasing the phonon scattering phase space \cite{peierls1955quantum,ziman2001electrons,lundstrom2002fundamentals,srivastava2022physics}.

In general, as the thermal conductivity of a solid is decreased, a greater fraction of the heat is conducted by diffusons due to the increased scattering of propagons effectuating larger coupling among diffusons \cite{cahill1992lower,agne2018minimum,hanus2021uncovering}. The opposing heat conduction pathways are visualized in Figs.~\ref{fig:propagon_v_diffuson_energy}c and \ref{fig:propagon_v_diffuson_energy}d, and the effects on thermal conductivity are visualized by the illustration of arrows in Fig.~\ref{fig:Hanus_data_panels}.  A general materials engineering design to simultaneously decrease or increase both diffuson and propagon contributions to thermal conductivity has yet to be found and would be a paradigm shift in engineering phonon heat conduction for extreme thermal conductivity materials if discovered. Several materials have demonstrated ultra-low thermal conductivity via combined low propagon and low diffuson contributions \cite{zeng2024pushing,xiong2025decoupling,wu2025pushing}, but a systematic strategy for direct decoupling of propagons and diffusons applicable to a wide range of structures is still lacking.

\subsubsection{Optimizing contributions\label{sec:materials_design_temporal}}

Presently, alteration of the thermal conductivity of materials with significant diffuson contribution is often made by influencing any non-trivial propagon contribution. As an example, size effects associated with the ballistic regime of propagon transport (discussed in Sec.~\ref{sec:propagating_wave-packets}) are reported in thin film amorphous Si \cite{regner2013broadband,larkin2014thermal,braun2016size,kwon2017unusually} which has a significant propagon presence \cite{moon2018propagating,moon2019thermal} unlike amorphous SiO$_{2}$ \cite{larkin2014thermal,zhu2018thermal} or amorphous Nb$_{2}$O$_{5}$ \cite{cheng2019diffuson} where the size effects are not observed in the thin film forms. Cai et al. \cite{cai2024diffuson} similarly found vacancy point defects, another source of extrinsic scattering, to suppress the propagon contribution in CH$_{3}$NH$_{3}$PbI$_{3}$ while hardly affecting the diffusons. Variation in the short-range order of amorphous Si structures was found to impact the propagon contribution far more than the diffusons \cite{zhang2026decoding}. Ishibe et al. \cite{ishibe2021heat} observed that thermal transport across amorphous-crystal interfaces, a common feature in electronics devices, can be influenced by tuning the propagon contribution. Large propagon contributions can even effectuate thermal conductivity behaviors consistent with phonon spatial coherence in periodic SLs composed of defective and/or disordered materials \cite{giri2018localization,perez2022incoherent}.

Though it is currently unclear as to how heat conduction by diffusons can be directly controlled and decoupled from propagons, it is our opinion that a solution may be found through more studies like that by Hanus et al. \cite{hanus2021uncovering} where close examination is paid to how diffuson dynamics and thermal conductivity contributions vary with different structural parameters of diffuson-dominated materials. Existing studies of these materials have mostly just examined total thermal conductivity trends with some works including microscopic analyses of the propagons. For instance, Nie et al. \cite{nie2017structural} found structural disorder in SiGe alloys attenuates thermal conductivity more effectively than compositional disorder and assessed the differences in the broadening of propagon linewidths. Another case is our own molecular dynamics study \cite{maranets2023lattice} that observed particle number density to modulate thermal conductivity of semiconductor nanocomposites far more strongly than the particle size distribution. These examples highlight nuanced variance in diffuson thermal transport with the material configuration, particularly its level of disorder, which if probed further, could elucidate novel mechanisms to fine tune the diffuson and propagon thermal conductivities. 

Furthermore, from an application standpoint, the task of optimizing the propagon and diffuson contributions is often associated with a broader objective of extreme maximization or minimization of the total thermal conductivity. Consequently, high-throughput frameworks for thermal conductivity calculation coupled with machine learning algorithms for efficient prediction when applied to large materials databases becomes a promising approach \cite{ohnishi2026database}. In recent years, the emergence of high-throughput workflows for accurately calculating thermal conductivity directly from first-principles, incorporating diffuson contributions \cite{zhu2024high,li2025high,rodriguez2026approaching}, and the development of machine learning models for thermal conductivity prediction \cite{luo2023predicting} make tackling this sophisticated optimization problem in a robust data-driven manner now feasible.

\section{Differences between thermal phonon coherence and other exotic phonon phenomena\label{sec:phonon_differences}}

Our last discussion of phonon theories in this review concerns distinguishing both phonon spatial and temporal coherence from other phonon physics, specifically those involving fundamental changes in physical character or transport behavior, thereby impacting phonon heat conduction. In this section, we provide a cursory discussion of the fundamental principles of several exotic phonon effects with a unique emphasis on their conceptual differences from both phonon coherence phenomena to clarify any possible confusion.

\subsection{Phonon hydrodynamics\label{sec:hydrodynamics}}

Phonon hydrodynamics is physical regime where phonons propagate similar to fluid flow \cite{ghosh2022phonon}. This occurs when the Normal phonon scattering processes that conserve momentum and do not contribute to thermal resistance dominate over the Umklapp phonon scattering processes which do not conserve momentum and effectuate thermal resistance \cite{peierls1955quantum,ziman2001electrons}. This leads to the phonons possessing a high drift velocity and a physical transport behavior akin to the motion of a fluid since the energetic behaviors of the phonons and molecules in fluid flow are now comparable \cite{chen2005nanoscale}. In the case of a heat pulse applied to a material in the hydrodynamic regime, the heat propagates like a pressure wave in a fluid and is termed phonon second sound \cite{ward1952iii,prohofsky1964second}.

While bearing some resemblance, phonon hydrodynamics is distinct from phonon spatial coherence because spatial coherence is a size effect occurring at a length scale much less than the anharmonic phonon scattering mean free path (illustrated in Fig.~\ref{fig:length_regimes}) whereas phonon hydrodynamics can contain many intrinsic scattering events and is not necessarily limited to a specific length scale. 

Temporal coherence is different from phonon hydrodynamics because the coupling interactions of diffusons is physically different from both Normal and Umklapp scattering processes. Additionally, the existing material systems and environmental conditions (very low temperatures) in which phonon hydrodynamic effects are observed are usually dominated by propagons \cite{ghosh2022phonon,machida2024phonon}.

\subsection{Anomalous phonon transport\label{sec:anomalous_transport}}

Several theoretical models and experimental measurements of various low-dimensional materials demonstrate anomalous phonon transport characterized by divergent or infinite thermal conductivity with system length \cite{zhang2020size}. This is a size effect attributed to the classical confinement of vibrational eigenstates and restricted scattering phase space when the crystal dimension is reduced \cite{schutz2018thermal}. As an example, the mirror symmetry of suspended single-layer graphene severely restricts the three-phonon scattering phase space of the out-of-plane modes in comparison to the in-plane modes, thereby effectuating a much larger thermal conductivity contribution \cite{lindsay2010flexural}. This is similar to how phonon spatial coherence can realize substantial heat conduction through high transmission wave-like coherent phonons manifested at low dimensions. However, several factors require a clear distinction to be made between anomalous transport and spatial coherence. 

Firstly, while phononic crystals such as SLs can possess low-dimensional artificial periodicity to stimulate spatial coherence, the natural periodicity of the solid remains three-dimensional as visualized in Fig.~\ref{fig:phononic_crystals_3Dmodels}, unlike the truly one- and two-dimensional crystal structures exhibiting anomalous transport. As mentioned in Sec.~\ref{sec:wave_localization}, attenuation of wave-like coherent phonons through Anderson localization in aperiodic SLs was shown to not impact the scattering phase space \cite{ma2020dimensionality}, evidencing a fundamental difference between anomalous transport affected by scattering and spatial coherence affected by wave interference. 

Secondly, as discussed in Sec.~\ref{sec:model_coherent_incoherent}, heat conduction in phononic crystals features the coexistence of two phonon types: coherent and incoherent phonons that propagate like waves and particles, respectively. In contrast, anomalous transport is characterized by a singular phonon type whose physical character and thermal transport magnitude is altered by the reduced dimension. Owed to the fact that phonon temporal coherence is not associated with size effects, there is no physical relation to anomalous phonon transport. However, owed to their unique combinations of crystal order and disorder \cite{cui2026lattice}, polymer chains are shown to exhibit both phonon temporal coherence and anomalous transport across different structural dimensions \cite{henry20101d,wang2017phonon,cai2024anomalous}.

\subsection{Ultrafast spectroscopy\label{sec:coherent_spectroscopy}}

The term "coherent phonon" is widely used in condensed matter physics, ultrafast spectroscopy, and nanoscale heat transfer, but its meaning depends strongly on the research context. The common element is the existence of a well-defined phase relationship in lattice vibrations. What differs is which phase relationship is relevant, how the phonons are generated, and which observable is used to establish coherence. In thermal transport studies, as discussed so far in this review, coherent phonon transport generally refers to the preservation of vibrational phase during propagation and scattering over a sufficient distance for wave interference to influence phonon transmission, phonon dispersion relations, or heat conduction. In ultrafast laser spectroscopy, by contrast, the same term generally denotes vibrational wave packets that are impulsively excited by femtosecond optical pulses. While both usages highlight the importance of phase relationships in lattice vibrations, the underlying physical mechanisms and observables differ.  

In ultrafast spectroscopy, coherent phonons emerge from sudden perturbations of the lattice potential induced by electronic excitation, such as displacive excitation \cite{zeiger1992theory} or impulsive stimulated Raman scattering \cite{merlin1997generating,stevens2002coherent,dhamija2022revisit}. The resulting oscillation is therefore a driven nonequilibrium state rather than a statistical thermal population of phonons. They are temporally phase-locked across the excited volume, giving rise to oscillatory modulations in reflectivity, transmission, or diffraction. These modes are inherently non-thermal and typically decay within a few picoseconds due to anharmonic interactions and scattering with carriers or defects. Consequently, the decay time of an ultrafast coherent-phonon signal does not necessarily represent a conventional phonon transport mean free path, a pure population lifetime, or even a temporal coherence time. Coherent optical phonons near the Brillouin-zone center are common examples, although femtosecond excitation can also generate propagating coherent acoustic wave packets and zone-folded acoustic phonons in multilayers and superlattices \cite{thomsen1986surface,bartels1998coherent,ye2025coherent}.

Importantly, the existence of an ultrafast coherent phonon does not by itself imply a significant contribution to heat conduction. A zone-center optical phonon, for example, may exhibit an exceptionally clear coherent oscillation while possessing a small group velocity and carrying little net heat. Here, the word ``coherent'' primarily describes the phase-defined dynamical state produced by the excitation. 

In contrast, many thermal transport studies, as discussed in this review, employed the term coherent phonon to describe propagating vibrational states that retain their wave character over extended distances. In artificial periodic structures such as superlattices or phononic crystals, spatial phase coherence enables constructive or destructive interference, phonon bandgap formation, and altered group velocities, thereby reshaping phonon transmission and heat conduction. In disordered or amorphous systems, coherence debates focus on whether certain vibrational modes (e.g., propagons) can traverse large distances without losing temporal phase memory, as opposed to diffusons or localized modes. In the previous sections of this review, we have provided a comprehensive discussions of coherent phonons in this context.

Despite these distinctions, the two concepts can converge in experiments that use ultrafast optical excitation to prepare a coherent phonon state and then exploit the propagation or interference of that state in a nanostructure. Semiconductor superlattices provide an early example: femtosecond laser pulses can excite coherent zone-folded longitudinal acoustic phonons whose frequencies and propagation are determined by superlattice periodicity; appropriately timed pulse sequences can selectively enhance or suppress particular coherent acoustic modes \cite{bartels1998coherent,bartels1999coherent}. Ultrafast measurements of coherent acoustic phonons have also been used to investigate phonon propagation and scattering in Bi$_2$Te$_3$/Sb$_2$Te$_3$ thermoelectric superlattices, directly connecting time-domain coherent-phonon dynamics with mechanisms relevant to their low thermal conductivity \cite{wang2010acoustic}.

More recently, layer-selective femtosecond optical excitation has been demonstrated to create a transient coherent-phonon flat band in GaAs/AlAs superlattices through coupling between optically generated lattice motion and the superlattice acoustic branches \cite{ye2025coherent}. Such approaches illustrate a particularly interesting intersection of ultrafast phononics and thermal transport: rather than using structural periodicity only to passively engineer the equilibrium phonon spectrum, ultrafast excitation may provide a route to dynamically create, select, amplify, suppress, or reshape coherent vibrational states. This emerging direction points toward actively reconfigurable phonon transport and, ultimately, time-dependent control of nanoscale heat flow. 

\subsection{Topological phononics\label{sec:topological_phonon}}

The highly geometric nature of phonon spatial coherence suggests some equivalence to topological phononics. Since the wave description of phonons fundamentally derives from Bloch's theorem, topological concepts such as as Berry connection and Berry curvature, Chern number, nodal lines and rings, and Weyl points, associated with Bloch wave functions are proposed to apply to phonons \cite{liu2020topological}. The emerging field of topological phononics explores how phonon modes are influenced by the topology of the phonon dispersion relation (band structure), which can be modified artificially in metamaterials \cite{zhu2023topological}. This dynamic is quite similar to the manifestation of coherent phonons and their dispersion relations in phononic crystals, which are selectively fabricated. Thus, phonon spatial coherence can be considered, in a way, a sub-phenomenon of topological phononics wherein new vibrational states (coherent phonons) emerge when new (secondary) periodicity is introduced to the material. However, there are several points which presently distinguish spatial coherence from most other topological phonon effects. 

Firstly, as previously discussed in Sec.~\ref{sec:model_coherent_incoherent}, heat conduction due to phonon spatial coherence is dictated by the interplay of two phonon types: coherent and incoherent phonons described by two dispersion relations. In contrast, topological phononics underscores the topological properties of a singular band structure.

Secondly, the topological phonon states appear highly unique to specific structures whereas spatial coherence is far less structure dependent. Phenomena such as minimum thermal conductivity with period size, wave localization, and Bragg transmission are fundamentally equivalent both conceptually and physically across different phononic crystal designs, be they superlattices, nanomeshes, pillared thin films, etc. For topological phononics, phenomena can be conceptually comparable across different materials, but physically manifest in entirely different ways. For example, phonon chirality can originate from chiral molecular helices \cite{ishito2023truly,abraham2024quantifying} or pseudoangular momentum in 2D lattices \cite{zhu2018observation}. 

Lastly, we emphasize that the spatial coherence of thermal phonons is well established and observed, unlike topological phononics which so far has demonstrated only minimal influence on thermal transport. The topic of topological phonons within the context of heat conduction is very nascent and is, in our opinion, ripe for advancement.

\section{Relationships between phonon spatial and temporal coherence\label{sec:spatial_temporal_connection}}

A key objective of this review was to clearly delineate the difference between phonon spatial and temporal coherence effects since both phenomena are frequently conflated in the literature by the general label ``phonon coherence." However, while being distinct physical processes, wave interference in space versus wave interference in time, phonon spatial and temporal coherence possess some relationships and similarities corresponding to a broader wave nature of phonons. In this section, we review these conceptual connections between spatial and temporal coherence.

\subsection{Wave-nature of thermal phonons\label{sec:spatiotemporal_thermal_phonon}}

We first emphasize that while our discussion of propagons and diffusons was necessary to distinguish between phonon spatial and temporal coherence, it simultaneously demonstrates that the two coherence effects are both linked to the concept of a phonon's physical character. How a phonon physically manifests, as determined by the dispersion relation and linewidths, controls the mechanisms by which that phonon carries heat. 

The conventional characterization of phonons as particle-like propagating wave-packets fails to accurately quantify thermal transport when the frequency and wavevector linewidths, which delineate the phonon's phase coherence in time and space, respectively, become non-negligible relative to the time and length scales of the material, respectively. The length scale is logically the characteristic length measuring the average distance between heterogeneous features. The time scale is less obvious; the spacing between spectral peaks in the frequency domain, which can be assessed through the average separation between branches in the dispersion relation (see Eqn.~\ref{eqn:wigner_limit}) or a new parameter termed coherence time that describes temporal phase extensions in a comparable fashion to spatial coherence length (see Eqn.~\ref{eqn:wavelet}). 

Since the time scale measures the phase correlations of multiple phonon modes, Zhang and co-authors \cite{zhang2022heat} referred to the temporal coherence effect as ``mutual coherence." In contrast, spatial coherence was termed ``intrinsic coherence," as it pertains to the phase correlations of a single phonon mode and its scattered variants. Ultimately, spatial and temporal coherence are two sides of the same coin when considering how wave-like behaviors of thermal phonons can manifest and influence heat conduction.

\subsection{Shared theoretical frameworks \& computational models\label{sec:shared_frameworks}}

\begin{figure}
    \centering
    \includegraphics[width=\textwidth,height=0.8\textheight,keepaspectratio]{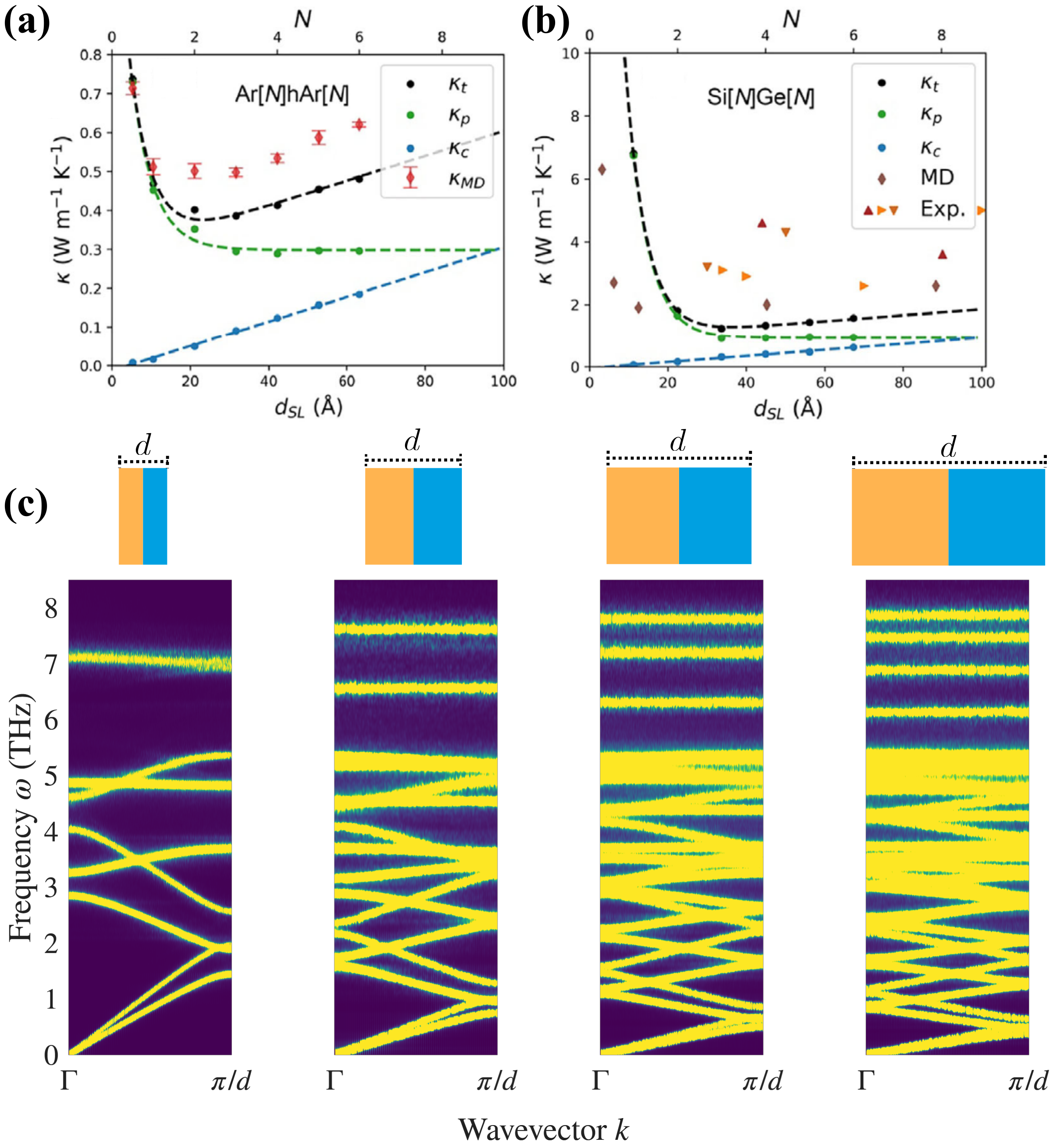}
    \caption{Application of the Wigner transport equation (WTE) for phonon temporal coherence to the minimum thermal conductivity phenomenon (see Sec.~\ref{sec:min_kappa_SL}) in periodic superlattices (SLs). Panels (a) and (b) show the WTE, when applied to the coherent phonon spectrum as defined by the SL's artificial periodicity, to accurately capture the crossover between coherent and incoherent phonon dominated transport in solid Argon and SiGe SLs, respectively. The data is reprinted with permission from Ref.~\cite{yang2026two}. Copyright (2026) John Wiley and Sons. Panel (c) plots the spectral energy density heat maps of solid Argon periodic SLs with progressively increasing period size $d$. }
    \label{fig:SL_SED_plots}
\end{figure}

Due to this conceptual link, several studies have accurately described aspects of one coherence effect with the theoretical frameworks of the other. For instance, Yang et al. \cite{yang2026two} demonstrated that the WTE when applied to periodic SLs can capture the minimum thermal conductivity trend detailed in Sec.~\ref{sec:min_kappa_SL}. For very small period sizes, the coherent modes belonging to the SL dispersion relation as defined by its artificial periodicity contribute more through the particle-like pathway. At larger period sizes, the increased zone-folding of the dispersion relation leads to (1) reduction in group velocities and (2) heightened linewidth overlap. The first effect reduces the particle contribution while the second effect increases the coherence contribution. The period size corresponding to the minimum thermal conductivity designates the crossover between particle and coherence dominated thermal conductivity. The transition is also visualized in the spectral energy density plots shown in Fig.~\ref{fig:SL_SED_plots} wherein for larger period sizes, the increasingly overlapping band structures resemble the diffuse spectra of Figs.~\ref{fig:dispersions_all}c-\ref{fig:dispersions_all}d. Simply, spatially coherent phonons become more diffuson-like as period size increases.

\begin{figure}
    \centering
    \includegraphics[width=\textwidth]{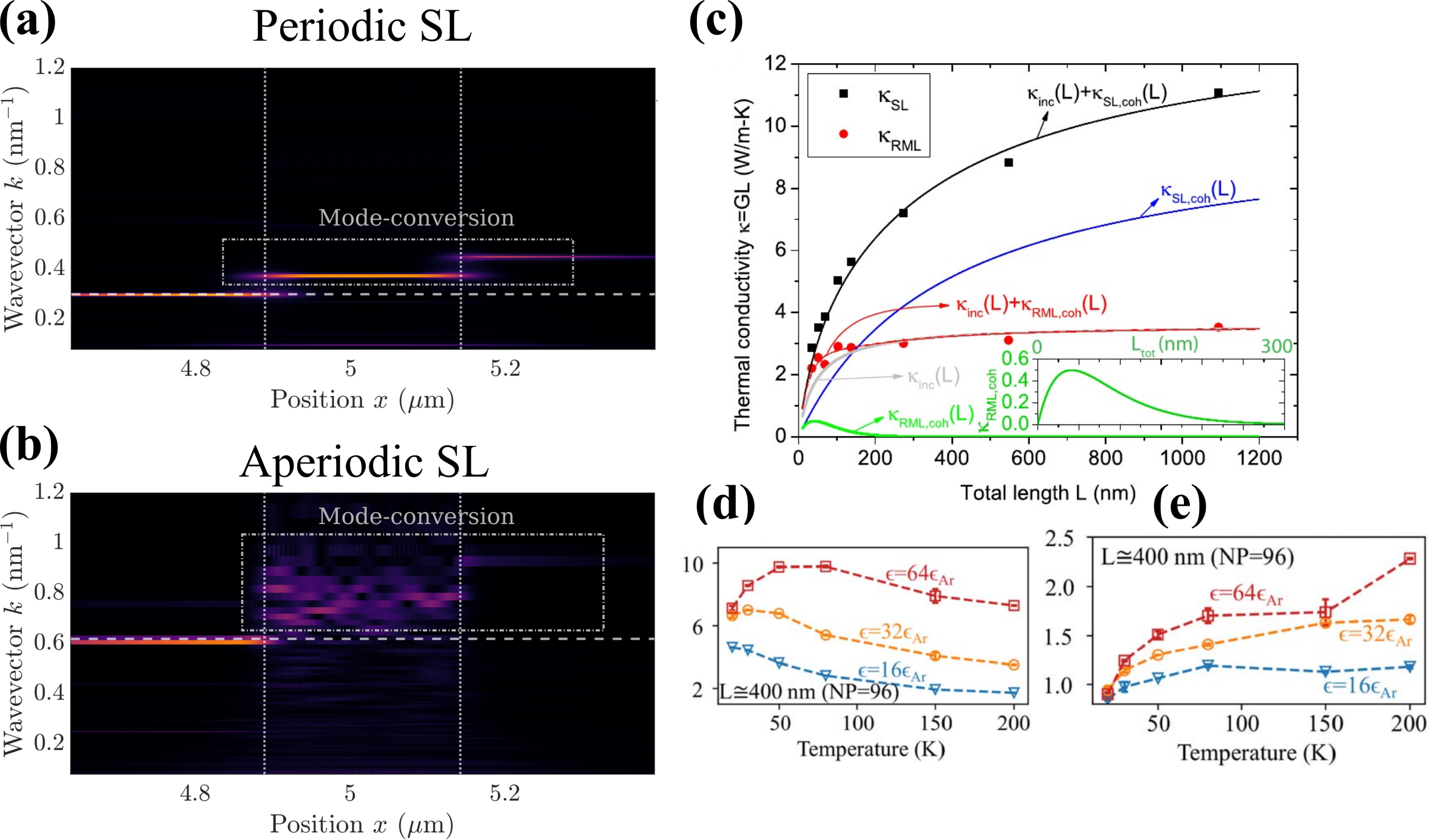}
    \caption{Coherent phonons manifesting propagon and diffuson behaviors in periodic and aperiodic SLs, respectively. Panels (a) and (b) present reciprocal-space wavelet transform calculations of atomistic wave-packet simulations demonstrating broadened wavevector linewidths and non-propagative states in the aperiodic SL. The data is reproduced with permission from Ref.~\cite{maranets2024prominent}. Copyright (2024) AIP Publishing. Panel (c) plots the length dependence of thermal conductivity as computed in the framework of Eqn.~\ref{eqn:coh_inc_equation}. The image is reprinted with permission from Ref.~\cite{wang2014decomposition}. Copyright (2014) from the American Physical Society. Panels (d) and (e) plot the temperature dependence of thermal conductivity, computed from non-equilibrium molecular dynamics simulations, for periodic and aperiodic SLs, respectively. The plots are reprinted with permission from Ref.~\cite{chakraborty2020complex}. Copyright (2020) IOP Publishing Ltd.}
    \label{fig:periodic_aperiodic_SL_kappa_dependence}
\end{figure}

A similar conclusion was made in our own studies elucidating contrasting coherent mode-conversion and thermal conductivity behaviors between periodic and aperiodic SLs \cite{maranets2024prominent,maranets2025phonon,maranets2025role}. Specifically, the characterization of coherent phonons as propagons and diffusons in periodic and aperiodic SLs, respectively, was found to more robustly explain the following trends in aperiodic SLs: (1) non-trivial thermal conductivity, (2) weak length dependence of thermal conductivity, (3) increase in thermal conductivity with temperature, and (4) increase in thermal conductivity with interface mixing. This analysis is based on analogizing aperiodic SLs to amorphous solids since the disruption of secondary periodicity is quite comparable to the disarrangement of crystal periodicity during amorphization. In Fig.~\ref{fig:periodic_aperiodic_SL_kappa_dependence}, we present several results strongly evidencing how coherent phonons interact like propagons and diffusons depending on the SL configuration. We specifically note comparable length and temperature dependencies of thermal conductivity to the data in Fig.~\ref{fig:length_temperature_propagon_diffuson}.

All in all, the findings establish that phonon spatial coherence and temporal coherence can manifest simultaneously when a material possesses the structural facets inducing both effects. Thus, spatial and temporal coherence are distinct but not mutually exclusive phenomena. Their overlap in certain cases signifies a shared wave nature of phonons and the relevance of phonon physical character in dictating heat conduction.

\section{Conclusions \& outlook\label{sec:conclusions}}

Here we conclude the review article with a summary of the developed concepts and a perspective on future research.

\subsection{Summary\label{sec:summary}}

In this review, we have rigorously dissected the scientific principles underlying thermal phonon coherence phenomena.

We first outlined in Sec.~\ref{sec:phonon_transport_overview} the foundational distinctions between spatial and temporal coherence, whose frequent conflation in the existing literature motivated the preparation of this review. Spatial coherence is a size effect of propagating phonon wave-packet transport, while temporal coherence is the intrinsic heat transfer mechanism of non-propagating phonon modes that are ill-defined as propagative wave-packets.

We then discussed heavily in Sec.~\ref{sec:spatial_coherence_theory} and Sec.~\ref{sec:temporal_coherence_theory} the theoretical frameworks for both coherence effects and explained how the unique thermal properties associated with phonon coherence fit within these theoretical contexts. Spatial and temporal coherence induce thermal conductivity magnitudes and dependencies that differ from the Peierls-BTE framework describing particle-like phonon transport, thereby necessitating the development of new models incorporating wave-like phonon dynamics. 

Following this dissection, in Sec.~\ref{sec:phonon_differences}, we clarified how thermal phonon coherence differs from several other non-Fourier phonon phenomena whose theories and observations share similar terminology. 

Lastly, in Sec.~\ref{sec:spatial_temporal_connection}, we offered analysis of the tangible conceptional connections between spatial and temporal coherence, providing a fruitful perspective on the wave nature of thermal phonons. Both spatial and temporal coherence fundamentally stem from the spatiotemporal quality of phonons, allowing for some overlap in microscopic properties to occur in specific materials.

\subsection{Future research directions\label{sec:Outlook}}

Our comprehensive review of the theories of thermal phonon coherence reveals several areas for further development in theoretical understanding, experimentation, computation, and materials design.

\subsubsection{Fundamental theory\label{sec:outlook_theory}}

Much of the fundamental knowledge of diffusons and temporal coherence is still lacking compared to that of propagons. Questions as to how heat conduction by diffusons is precisely affected by the multitude of different extrinsic scattering mechanisms \cite{hanus2021thermal} are currently unanswered. Additionally, it is unclear why certain diffuson-dominated materials have higher propagon contributions than others (see Sec.~\ref{sec:manipulate_temporal_coherence}). The ways in which interfacial thermal transport by diffusons might differ from propagons has also been minimally studied \cite{gordiz2015formalism}. 

Several concepts in phonon spatial coherence also warrant further dissection. For instance, more clarity is needed on why coherent phonon transport is observed to manifest at different length scales along opposing structural dimensions in phononic crystals \cite{anufriev2025phononic}. Furthermore, the role of spatial coherence length on the overall flow of heat, not just wave dynamics of individual coherent phonons, must be reconciled considering the emergence of spatial coherence can be observed across a finite characteristic length of spectral heat flux redistribution \cite{cui2025spectral}.

\subsubsection{Experimental measurement\label{sec:outlook_experiment}}

The spatial and temporal coherence of light can be directly resolved by measuring interference patterns in space and time, respectively of emission from optical sources \cite{hecht2012optics}. No such analog exists for heat since heat is facilitated by the energy dissipation of a broad spectrum of phonon modes possessing terahertz frequencies and wavelengths on the order of a few nanometers \cite{kaviany2014heat}. 

As of yet, there has not been any direct experimental observation of the thermal phonon wave dynamics, neither spatial nor temporal, discussed in this review. At best, indirect observation can be made through spectroscopy techniques that elucidate to varying degree, the proportional contributions of different phonon modes to thermal conductivity and important spectral quantities like vibrational density of states. Several spectroscopy experiments \cite{giri2018experimental,hoglund2022emergent,geng2024mapping} have elucidated many of the key concepts of phonon spatial coherence discussed in this review. However, further efforts to improve measurement apparatuses for resolving thermal phonon coherence experimentally are needed to validate increasingly sophisticated theories and computational models.

\subsubsection{Computational modeling\label{sec:outlook_computation}}

Most experiments that do not employ sophisticated spectroscopy methods to measure properties of individual phonon modes, typically measure thermal conductivity or thermal conductance/resistance. Subsequent comparison of measurements to computational models or vice versa can offer deep insights and strengthen fundamental understanding. A range of computational methods have been employed to study thermal phonon coherence \cite{bao2018review} and many key frameworks were discussed in this review. We point to several directions for advancement in modeling. 

Firstly, development a unified framework that deliberately models both phonon spatial and temporal coherence and their influence on heat conduction is a crucial endeavor. While Yang et al. \cite{yang2026two} and others \cite{huang2026intrinsic} have demonstrated the capabilities of the quantitative temporal coherence models reviewed in Sec.~\ref{sec:temporal_coherence_theoretical_framework} to capture some signatures of spatial coherence (owed to their intrinsic relationships discussed in Sec.~\ref{sec:spatial_temporal_connection}) like minimum thermal conductivity with period size, it is unclear whether these frameworks can predict the other effects such as wave localization, varied length dependencies of thermal conductivity, phononic crystal band gaps, and thermal conductivity anisotropy. 

Furthermore, the immense structural, geometric, and/or compositional complexity of materials exhibiting thermal phonon coherence has necessitated more efficient simulation approaches. For instance, calculation of third- and higher-order interatomic force constants (IFCs) are most imperative for accurately modeling thermal conductivity of materials exhibiting prominent temporal coherence \cite{lindsay2019perspective}. However, real-space finite-difference approaches for computing IFCs become prohibitively computational expensive for large unit cells, thereby requiring robust crystal symmetry analysis to reduce the number of displaced supercell calculations \cite{plata2017efficient,hicks2018aflow}. This problem is compounded when additional complexities like magnetic spin configurations affecting the force landscape are involved \cite{singh2025ultralow,singh2025v2se2o,singh2026phase}. Leveraging machine learning to efficiently generate first-principles-accurate IFCs \cite{eriksson2019hiphive,korotaev2019accessing,mortazavi2021accelerating,togo2024fly,srivastava2024accelerating} has demonstrated immense value in solving this scaling problem for pristine crystals. However, further challenges in efficiently constructing accurate IFCs adaptable to a wide range of temperatures, phases, levels of crystal defects and disorder remain.

\subsubsection{Materials design\label{sec:outlook_materials}}

The open problems in theory, experiment, and computation coalesce towards designing improved materials that harness the unique thermophysical properties and behaviors of thermal phonon coherence. Key directions for spatial coherence include exploring pathways to achieve stronger coherent mode-conversion by controlling temperature, pressure, stress, and strain \cite{maranets2025role}. Additionally, manifesting coherent phonon transport at higher temperatures and in more easily synthesizable materials compared to phononic crystals is a crucial focus (see Sec.~\ref{sec:nanocomposite}). 

Characterizing the role of phonon temporal coherence on thermal properties is becoming increasingly relevant as advanced materials with large tunable structural disorder and compositional complexity, like perovskites \cite{haeger2020thermal}, metal-organic frameworks \cite{nguyen2025thermal}, transition metal dichalcogenides \cite{liu2018thermal}, and high-entropy materials \cite{oses2020high}, proliferate in technologies. We believe that the detailed conceptual understanding of thermal phonon coherence developed in this review will be very useful for future materials engineering.

\section*{Credit author statement}

Theodore Maranets: Conceptualization (lead), Formal analysis (lead), Software (lead), Writing - Original Draft (lead), Writing - Review and Editing (equal). Haoran Cui: Writing - Original Draft (equal), Writing - Review and Editing (equal). Milad Nasiri: Writing - Original Draft (equal), Writing - Review and Editing (equal). Evan Doe: Writing - Original Draft (equal), Writing - Review and Editing (equal). Yan Wang: Conceptualization (equal), Formal analysis (equal), Supervision (lead), Funding Acquisition (lead), Writing - Original Draft (equal), Writing - Review and Editing (equal).

\section*{Declaration of competing interest}
The authors declare that they have no known competing financial interests or personal relationships that could have appeared to influence the work reported in this paper.

\section*{Data availability}
Data will be made available from the corresponding authors upon reasonable request.

\section*{Acknowledgments}

The authors gratefully acknowledge the financial support from the National Science Foundation Thermal Transport Processes program (CBET-2047109 and CBET-1953300). Cui and Wang also extend their thanks to the National Science Foundation EPSCoR Research Infrastructure Program (OIA-2033424). Maranets acknowledges the support of the Nuclear Power Graduate Fellowship from the Nuclear Regulatory Commission (31310021M0004) and the Graduate Research Opportunity Fellowship from the Nevada Space Grant Consortium. Doe acknowledges the support of the Undergraduate Research Opportunity Scholarship from the Nevada Space Grant Consortium. Maranets thanks Dr. Shubham Singh for helpful discussions.

\bibliography{references}% Produces the bibliography via BibTeX.

\end{document}